\documentclass[a4paper,twocolumn,10pt,aps,accepted=2023-11-21]{quantumarticle}
\pdfoutput=1

\usepackage[utf8]{inputenc}
\usepackage[english]{babel}
\usepackage[T1]{fontenc}
\usepackage{amsmath}
\usepackage{amsfonts}
\usepackage{amssymb}
\usepackage{graphicx}
\usepackage[caption=false]{subfig}
\graphicspath{{pics/}}
\usepackage[ruled, linesnumbered]{algorithm2e}
\usepackage{enumitem}

\usepackage[shortpaper]{aps_math}
\usepackage{qm}

\usepackage{ifthen}
\usepackage[numbers,sort&compress]{natbib}
\def\bibfile{My_Library}
\ifthenelse{\equal{\bibfile}{}}{%
  \def\myprintbibliography{}%
}{%
  \def\myprintbibliography{%
    \bibliographystyle{quantumjournal}%
    \bibliography{\bibfile}%
  }%
}%
\usepackage[%
]{hyperref}

\def\myprintglossary{%
}

\begin{document}
\title{Entanglement-efficient bipartite-distributed quantum computing}
\author{Jun-Yi Wu}
\email{junyiwuphysics@gmail.com}
\affiliation{Department of Physics and Center for Advanced Quantum Computing, Tamkang University, 151 Yingzhuan Rd., New Taipei City 25137, Taiwan, ROC}
\affiliation{Physics Division, National Center for Theoretical Sciences, Taipei 10617, Taiwan, ROC}
\orcid{0000-0001-8843-9269}
\author{Kosuke Matsui}
\affiliation{The University of Tokyo, Hongo 7-3-1, Bunkyo-ku, Tokyo 113-0033, Japan}
\author{Tim Forrer}
\affiliation{The University of Tokyo, Hongo 7-3-1, Bunkyo-ku, Tokyo 113-0033, Japan}
\orcid{0009-0007-1141-2755}
\author{Akihito Soeda}
\affiliation{The University of Tokyo, Hongo 7-3-1, Bunkyo-ku, Tokyo 113-0033, Japan}
\affiliation{Principles of Informatics Research Division, National Institute of Informatics, 2-1-2 Hitotsubashi, Chiyoda-ku, Tokyo 101-8430, Japan}
\affiliation{Department of Informatics, School of Multidisciplinary Sciences,
SOKENDAI (The Graduate University for Advanced Studies), 2-1-2 Hitotsubashi, Chiyoda-ku, Tokyo 101-8430, Japan}
\orcid{0000-0002-7502-5582}
\author{Pablo Andr\'{e}s-Mart\'{i}nez}
\affiliation{Quantinuum, Terrington House, 13-15 Hills Road, Cambridge CB2 1NL, UK}
\orcid{0000-0003-4456-7052}
\author{Daniel Mills}
\affiliation{Quantinuum, Terrington House, 13-15 Hills Road, Cambridge CB2 1NL, UK}
\orcid{0000-0001-5902-3774}
\author{Luciana Henaut}
\affiliation{Quantinuum, Terrington House, 13-15 Hills Road, Cambridge CB2 1NL, UK}
\orcid{0000-0002-9783-7450}
\author{Mio Murao}
\email{murao@phys.s.u-tokyo.ac.jp}
\affiliation{The University of Tokyo, Hongo 7-3-1, Bunkyo-ku, Tokyo 113-0033, Japan}
\orcid{0000-0001-7861-1774}

\begin{abstract}
  In noisy intermediate-scale quantum computing, the limited scalability of a single quantum processing unit (QPU) can be extended through distributed quantum computing (DQC), in which one can implement global operations over two QPUs by entanglement-assisted local operations and classical communication. To facilitate this type of DQC in experiments, we need an entanglement-efficient protocol. To this end, we extend the protocol in [Eisert et. al., PRA, 62:052317(2000)] implementing each nonlocal controlled-unitary gate locally with one maximally entangled pair to a packing protocol, which can pack multiple nonlocal controlled-unitary gates locally using one maximally entangled pair. In particular, two types of packing processes are introduced as the building blocks, namely the distributing processes and embedding processes. Each distributing process distributes corresponding gates locally with one entangled pair. The efficiency of entanglement is then enhanced by embedding processes, which merge two non-sequential distributing processes and hence save the entanglement cost. We show that the structure of distributability and embeddability of a quantum circuit can be fully represented by the corresponding packing graphs and conflict graphs. Based on these graphs, we derive heuristic algorithms for finding an entanglement-efficient packing of distributing processes for a given quantum circuit to be implemented by two parties. These algorithms can determine the required number of local auxiliary qubits in the DQC. We apply these algorithms for bipartite DQC of unitary coupled-cluster circuits and find a significant reduction of entanglement cost through embeddings. This method can determine a constructive upper bound on the entanglement cost for the DQC of quantum circuits.
\end{abstract}
\keywords{Keywords}
\maketitle

\section{\label{sec:introduction} Introduction}

In the noisy intermediate-scale quantum era \cite{Preskill2018-NISQ}, the noise in quantum processes and the decoherence of qubits are two bottlenecks of the scalability of quantum computing.
In a single  quantum processing unit (QPU), the scalability is constrained by the number, connectivity and coherence time of qubits, which determine the effective width and depth of a quantum circuit.
The effective size of a quantum processor can be quantified by its quantum volume \cite{MollEtAlTemme2018-QOptmztnVarAlgoQVolume, CrossEtAlGambetta2019-QVolume}.
It is believed that the connectivity of qubits on a QPU is a key feature to scale up quantum volume \cite{CrossEtAlGambetta2019-QVolume}.
However, each platform has its intrinsic topological limits on the number of qubits and their connectivity on a QPU.

To overcome these limits, one can extend the qubit connectivity across QPUs employing flying qubits to establish entanglement between two remote QPUs. Several protocols for establishing remote qubit connectivity have been proposed \cite{BoseEtAlVedral1999-TeleportAtomicState, CabrilloEtAlZoller1999-EntDstAtmByInterf, BrowneEtAlHuelga2003-RemoteEntBtwIons, DuanEtAlMonroe2004-TrpIonQcmpWthIonPhMppng,LimBeigeKwek2005-RUSLinOptDQC, DuanEtAlMonroe2006-RmtAtmQGtwthPh, YinEtAlDuan2015-QNetSpcndctQbtThrOptmchIntfc, KoshinoEtAlNakamura2017-EntGenRmtSpcndctAtm} and realized in the physical systems of trapped ions \cite{MoehringEtAlMonroe2007-RmtEntAtmQbt, SlodickaEtAlBlatt2013-EntRmtTrpIonBySnglPhDtctn}, neutral atoms \cite{RitterEtAlRempe2012-QNetSngleAtmInOptCav, HofmannEtAlWeinfurter2012-HeraldedEntNtrlAtm}, NV centers \cite{BernienEtAlHanson2013-HeraldedEntNV}, quantum dots \cite{DelteilEtAlImamoglu2015-HeraldedEntRmtHoleSpn, StockillEtAlAtature2017-EntRmtSpnQbt}, and superconducting qubits \cite{NarlaEtAlDevoret2016-RmtEntSpcndctQbt}.
One can even establish entanglement between remote trapped-ion qubits with a fidelity comparable with local entanglement generation \cite{StephensonEtAlBallance2020-HghRtFdltyHeraldedRmtEntTrpIon}.
In particular, one can even established entanglement between remote trapped-ion modules of servers \cite{HuculEtAlMonroe2014-MdlEntAtmIonTrp}.

All of these technologies facilitate the development of quantum internet \cite{Kimble2008-QNet} in different physical systems.
Benefiting from the topology of a quantum internet, one can extend the connectivity of local qubits to build up a hybrid quantum computing system with its QPUs distributed over a quantum network, e.g. quantum computation over the butterfly network \cite{SoedaKTMurao2011-QCmpBttflNtwk}.
To distribute universal quantum computing over a quantum internet, one has to implement the universal set of quantum gates over local servers.
The only global gate in the universal set that needs to be distributed is the CNOT gate. It is locally implementable with the assistance of entangled pairs according to the protocols in \cite{GottesmanChuang1999-UniQCmpTlpt1QbitOp, ZhouLeungChuang2000-QGtCnstrct}.
It shows the feasibility of distributed universal quantum computing through local operations and classical communication (LOCC), if a sufficient amount of entanglement is provided. However, the distribution of entanglement over a quantum network is probabilistic and time-consuming with respect to the coherence time of local qubits. To make distributed quantum computing (DQC) feasible, one has to therefore find an efficient way to exploit the costly entanglement resources.

There are two methods for nonlocal gate handling in DQC with entanglement-assisted LOCC.
One is based on quantum state teleportation \cite{GottesmanChuang1999-UniQCmpTlpt1QbitOp};
the other is based on the remote implementation of quantum gates through entanglement-assisted LOCC
\cite{EisertEtAlPlenio2000-LclImplNonlclQGt,  HuelgaEtAlPlenio2001-TlptOfUnitary, HuelgaPlenioVaccaro2002-TlptOfAngles, JiangEtAlLukin2007-DQC}, which is also referred to as ``telegate''\cite{MeterEtAlItoh2008-DQC, CaleffiEtAlCacciapuoti2022-DQCSurvey}.
In the former scheme, qubits associated with nonlocal gates are teleported forward and backward across parties. In the latter scheme,
one implements a nonlocal gate with entanglement-assisted LOCC in a way such that all associated qubits of the gate are kept functional locally without sending them to the other parties.
It has been experimentally implemented in \cite{ChouEtAlSchoelkopf2018-GateTeleportation, WanEtAlLeibfried219-QGtTeleptTrapIon}. Several telegate protocols are employed for compiling multipartite DQC \cite{MartinezHeunen2019-DQCPartition, SundaramGuptaRamakrishnan2021-EffDQC, CuomoEtAlCacciapuoti2023-DQCCompiler}.

In particular, the telegate protocol in \cite{EisertEtAlPlenio2000-LclImplNonlclQGt} referred to as the EJPP protocol in this paper can be employed to implement arbitrary nonlocal control unitaries.
The DQC schemes based on the state teleportation and the EJPP protocol are compared and combined in a general quantum network \cite{SundaramHimanshuRamakrishnan2022-DQCOverQNet}.
The EJPP protocol consumes only one ebit, that is one maximally entangled pair, which is optimum in entanglement efficiency.
Based on this protocol, one can find the optimum partition for a circuit consisting of control-Z gates and single-qubit gates through hypergraph partitioning \cite{MartinezHeunen2019-DQCPartition}.
The EJPP protocol employed in the partitioning can be extended for sequential control-Z gates and hence reduces entanglement cost \cite{SundaramGuptaRamakrishnan2021-EffDQC}.
In \cite{SundaramGuptaRamakrishnan2021-EffDQC}, the EJPP protocol is terminated by each single-qubit gate, and restarted for control-Z gates with a new entangled pair. This leads to an entanglement cost much higher than the theoretical lower bound given by the operator Schmidt rank of a general circuit \cite{StahlkeGriffiths2011-EntCostForUnitary}.
To further facilitate DQC in experiments, we develop a DQC protocol with a higher entanglement efficiency, in which we introduce a notion of entanglement-assisted packing processes for bipartite DQC through an extension of the EJPP protocol.
Note that this method is extended to multipartite DQC with some additional treatments and is employed as the core nonlocal-gate-handling subroutine for modular quantum computing under network architectures \cite{MartinezEtAlDuncan2023-DQCinQNet}.
A similar application of this method can be also employed for other modular compilation architectures \cite{BakerEtALChong2020-DQCPartition, DiAdamoEtAlCruise2021-DQCNetCtrl4VQE, FerrariEtAlCaleffi2021-DQCCompiler, CaleffiEtAlCacciapuoti2022-DQCSurvey, OvideEtAlFeld2023-DQCNetArch, CuomoEtAlCacciapuoti2023-DQCCompiler, FerrariCarrettaAmoretti2023-ModularCompilationDQC}.

As two particular types of entanglement-assisted packing processes, we introduce distributing processes and embedding processes as the two fundamental building blocks for DQC.
In each distributing process, one ebit of entanglement is employed to achieve an operation equivalent to a global unitary, implemented by packing local gates using entanglement-assisted LOCC.
The number of distributing processes equals the entanglement cost for DQC. To reduce the entanglement cost, we introduce embedding processes of intermediate gates to merge two non-sequential distributing processes into a single distributing process. Each embedding process can therefore save one ebit without changing the distributability of the circuit. The distributability and embeddability of gates in a circuit reveal the underlying entanglement cost for DQC employing packing processes. Such a packing method provides a tighter constructive upper bound on the entanglement cost for DQC.
It can significantly enhance the entanglement efficiency of DQC.

The rest of the paper is structured as follows.
In Section \ref{sec:ejpp_protocol}, we review the EJPP protocol.
In Section \ref{sec:EA_packing}, we introduce the entanglement-assisted packing processes and two special types of processes, namely, the distributing and embedding. The conflicts between embeddings are addressed. The packing graph and conflict graph are introduced to represent the distributing and embedding structure.
In Section \ref{sec:id_packing_conflict_graphs}, we derive the algorithms for the identification of packing graphs and conflict graphs for quantum circuits consisting of one-qubit and two-qubit gates.
In Section \ref{sec:ucc}, we demonstrate the enhancement of entanglement efficiency in unitary coupled-cluster (UCC) circuits \cite{TaubeBartlett2006-UCC,PeruzzoEtAlOBrien2014-VQE} employing our method.
In Section \ref{sec:conclusion}, we conclude the paper.

\section{\label{sec:ejpp_protocol} EJPP protocol}

\begin{figure}[htbp]
    \centering
    \includegraphics[width=0.48\textwidth,height=4.2cm,keepaspectratio]{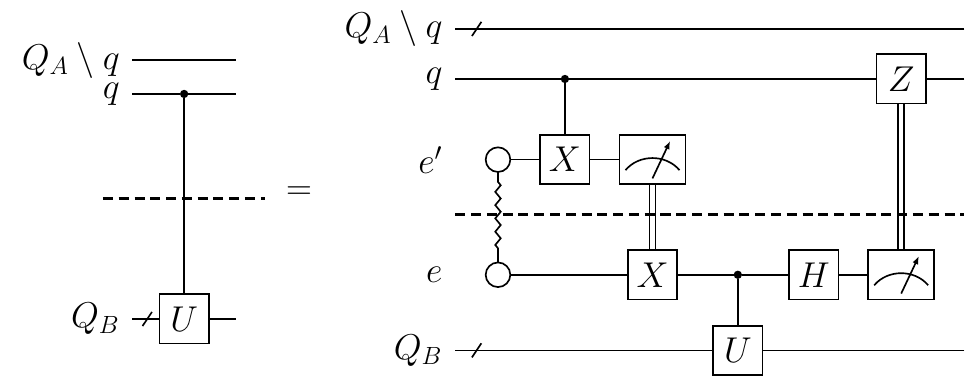}
    \caption{The quantum circuit on the left side is a controlled-unitary gate.
    The quantum circuit on the right side is a distributed implementation of the controlled-unitary gate according to the EJPP protocol, where $Q_{A}\cup\{e'\}$ and $Q_{B}\cup\{e\}$ are qubits in the local quantum processor $A$ and $B$, respectively. The qubits $\{e',e\}$ are the auxiliary memory qubits that store a pre-shared maximally entangled pair, namely one ebit entanglement.}
    \label{fig:ejpp_protocol}
  \end{figure}

  As shown in Fig. \ref{fig:ejpp_protocol}, a controlled-unitary gate $C_{U}$ on the left-hand side can be implemented through entanglement-assisted LOCC \cite{JonathanPlenio1999-EALOCCs} with only one maximally entangled state on the right-hand side according to the EJPP protocol presented in \cite{EisertEtAlPlenio2000-LclImplNonlclQGt}.
  The controlled unitary on the left side has a controlling qubit $q$ on the local quantum processor $A$ and a target unitary $U$ acting on the multi-qubit subsystem $Q^{B}$ on the local quantum processor $B$.
  In the EJPP protocol, there is a pre-shared maximally-entangled pair $\{e,e'\}$, on which one implements local controlled unitaries and global measurement-controlled gates through classical communication.
  It consumes only one ebit, i.e. a maximally-entangled pair, which is the minimum necessary amount of entanglement for a distributed implementation of a controlled unitary determined by its operator Schmidt rank \cite{StahlkeGriffiths2011-EntCostForUnitary}.
  It means that the EJPP protocol is most entanglement-efficient for a controlled unitary.

  The EJPP protocol can be divided into three parts, namely the start, kernel, and end. The starting and ending are constructed with fixed gates, while the kernel can be tailored for the implementation of particular unitaries.
  Here, we define the two fixed parts as the \emph{starting process}, and the \emph{ending process}, which correspond to the \emph{cat-entangler} and the \emph{cat-disentangler} in \cite{YimsiriwattanaLomonaco2004-GHZnDQC}, respectively.

  \begin{figure}[htbp]
    \centering
    \hfill
    \subfloat[]{\includegraphics[width=0.4\textwidth,height=2.5cm,keepaspectratio]{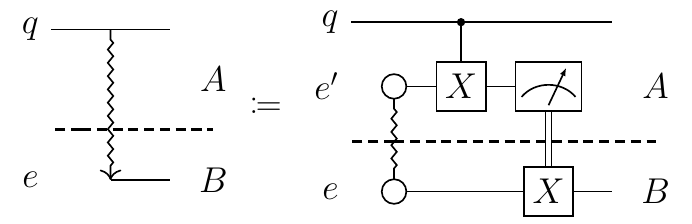}}
    \hfill{ }
    \\
    \hfill
    \subfloat[]{\includegraphics[width=0.4\textwidth,height=2.5cm,keepaspectratio]{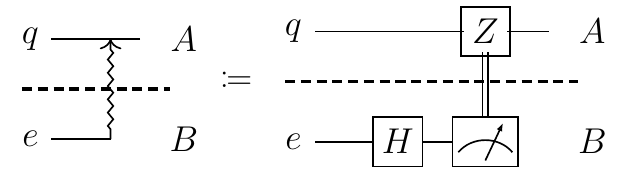}}
    \hfill{ }
    \\
    \hfill
    \subfloat[]{\includegraphics[width=0.4\textwidth,height=2.5cm,keepaspectratio]{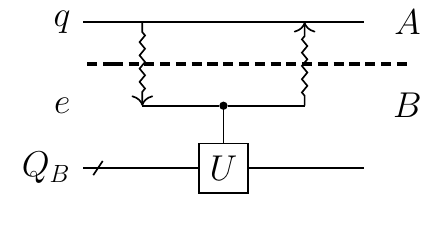}}
    \hfill{ }
    \caption{The starting and ending processes: the symbol on the left side represents the operation given by the quantum circuit on the right side. A working qubit $q$ and an auxiliary qubit $e'$ belong to a local QPU $A$, while and an auxiliary qubit $e$ belongs to another local QPU $B$.
    (a) The starting process.
    (b) The ending process.
    (c) The protocol in \cite{EisertEtAlPlenio2000-LclImplNonlclQGt} for distributed implementation of controlled unitary gates.
    }
    \label{fig:start_end_process}
  \end{figure}

  \begin{definition}[Starting process and ending process]\label{def:st_end_proc}
    The quantum circuit on the right-hand side of Fig.~\ref{fig:start_end_process} (a) and (b) are referred to as a \emph{starting process} and an \emph{ending process} rooted on $q$, respectively, which are denoted by the symbols on the left-hand side.
    The auxiliary qubits $e$ and $e'$ are initialized as a Bell state $(\ket{00}+\ket{11})/\sqrt{2}$ in the starting process.
  \end{definition}

  The starting process duplicates the computational-basis components of an input state $\ket{\psi}$ on $q$ to an entangled state
  \begin{align}
    S_{q,e}(\ket{\psi_{q}})
    & =
    C_{q,X_{e}}\ket{\psi_{q},0_{e}}
    \nonumber \\ &
    =
    \braket{0_{q}|\psi_{q}} \ket{0_{q},0_{e}}
    +
    \braket{1_{q}|\psi_{q}} \ket{1_{q},1_{e}}   ,
    \label{eq:result_starting}
  \end{align}
  where $C_{q,X_{e}}$ is a control-$X$ gate with the control qubit $q$ and target qubit $e$, and $S$ is a linear isometry operator $S \in \mathcal{L}(\mathcal{H}^\mathrm{q} \rightarrow \mathcal{H}^\mathrm{q} \otimes \mathcal{H}^\mathrm{e})$,
  \begin{align}
  \label{eq:starting_isometry}
    S_{q,e} :
    \ket{i}_q \mapsto
    \ket{i}_q \ket{i}_e,
    \text{ for all }
    i \in \{0,1\}.
  \end{align}

  On the other hand, by applying an ending process to the state $\ket{\Psi_{q,e}} = S_{q,e}(\ket{\psi_{q}})$, one obtains the partial trace of $C_{q,X_{e}}\ket{\Psi_{q,e}}$
  \begin{equation}
    E_{q,e}(\ket{\Psi_{q,e}})
    =
    \tr_{e}\left(C_{q,X_{e}} \projector{\Psi_{q,e}} C_{q,X_{e}}\right).
  \end{equation}
  The ending process of $\ket{\psi}$ is therefore the inverse of the starting isometry $S$, which restores the original state
  \begin{equation}
    E_{q,e}\circ S_{q,e} (\ket{\psi_{q}})
    =
    \ket{\psi_{q}}.
  \end{equation}
  Inserting a local unitary $C_{e,U_{Q_{B}}}$ acting on $\{e\}\cup Q_{B}$ between the starting and ending processes, one can implement a global controlled unitary $C_{q,U_{Q_{B}}}$ with local gates according to the following equality
  \begin{align}
    C_{q,U_{Q_{B}}}
    \projector{\psi_{q}}
    C_{q,U_{Q_{B}}}
    & =
    E_{q,e}\circ C_{e,U_{Q_{B}}} \circ S_{q,e} (\ket{\psi_{q}})
    \nonumber \\ &
    =
    \tr_{e}\left(
      \projector{\widetilde{\Psi}_{q,e}}
    \right),
  \end{align}
  where
  \begin{equation}
    \ket{\widetilde{\Psi}_{q,e}} = C_{q,X_{e}}C_{e,U_{Q_{B}}}C_{q,X_{e}}\ket{\psi_{q},0_{e}}.
  \end{equation}
  Employing the squiggly arrows for starting and ending processes, the protocol in \cite{EisertEtAlPlenio2000-LclImplNonlclQGt} can be represented as Fig. \ref{fig:start_end_process} (c).

\section{\label{sec:EA_packing} Packing distributable unitaries}
\subsection{Entanglement-assisted packing processes}

  \begin{figure*}[htbp]
    \centering
    \includegraphics[width= 0.9\textwidth]{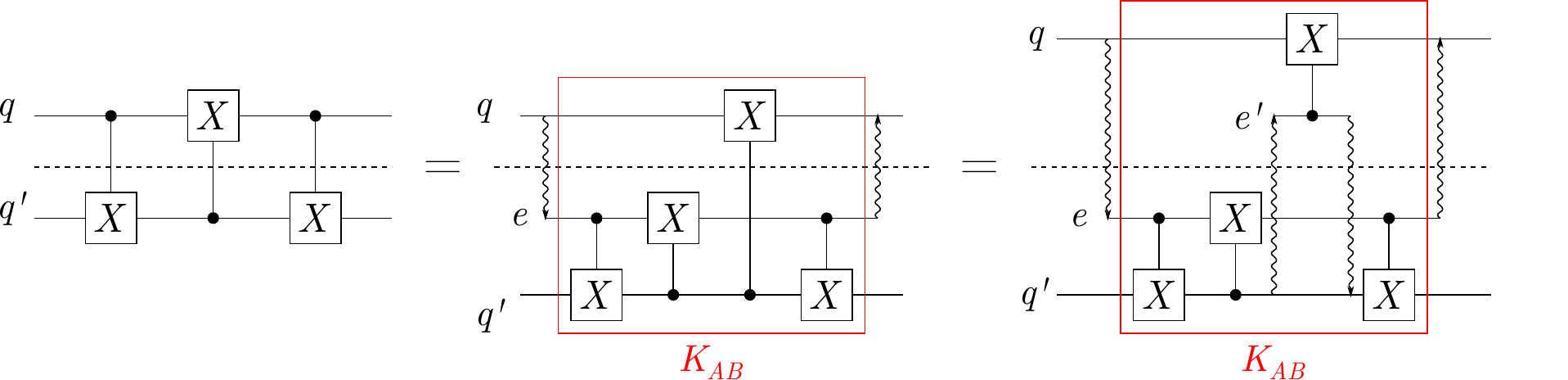}
    \caption{Distributed implementation of swapping with two entanglement-assisted LOCCs}
    \label{fig:swapping}
  \end{figure*}

  Beyond controlled-unitary gates, one can extend the EJPP protocol \cite{EisertEtAlPlenio2000-LclImplNonlclQGt} to locally implement a swapping gate with the assistance of two-ebit of entanglement, which is shown in Fig. \ref{fig:swapping}.
  There are two steps to construct the distributed implementation.
  In the first step, we extend the protocol through the replacement of the local control unitary $C_{e,U_{Q_{B}}}$ in Fig. \ref{fig:start_end_process} (c) by a global unitary $K_{AB}$.
  In the second step, we implement the EJPP protocol for the remaining global controlled-X gate starting and ending on the qubit $q'$.

  In this example, the process in the first step is implemented with a kernel $K_{AB}$ sandwiched by the starting and ending processes.
  The kernel $K_{AB}$ is a global unitary acting on the working qubit $\{q,q'\}$ and the auxiliary qubit $e$.
  In general, $K_{AB}$ can be either local or global.
  For a global kernel $K_{AB}$, one needs further steps to locally implement the global gates in the kernel with the starting and ending processes.
  Once all the global gates are distributed, then the distribution of the quantum circuit is finished.
  Such an extension of the EJPP protocol through the replacement of local controlled-unitary gates by a global kernel $K_{AB}$ is therefore a general building block for the distributed quantum computing based on the starting and ending processes.
  In this paper, we refer to these building blocks as \emph{entanglement-assisted packing processes}.
  \begin{definition}[\small{Entanglement-assisted packing process}]\label{def:EA_process}
    An \emph{entanglement-assisted packing process} $\mathcal{P}_{q,e}[K]$ with a kernel $K$ rooted on a qubit $q$ assisted by two auxiliary qubits $(e,e')$ is a quantum circuit that packs a unitary $K$ by the entanglement-assisted starting and ending processes rooted on $q$ (see Fig. \ref{fig:EA_process}(a)),
    \begin{equation}
    \label{eq:ent-assist_proc}
      \mathcal{P}_{q,e}[K]
      :=
      E_{q,e} \circ K \circ S_{q,e},
    \end{equation}
    In short, we call $\mathcal{P}_{q,e}[K]$ a \emph{$q$-rooted $e$-assisted packing process}.
  \begin{figure}[htbp]
  \label{fig:EA-packing_process}
      \centering
      \hfill
      \subfloat[]{\includegraphics[width=0.4\columnwidth]{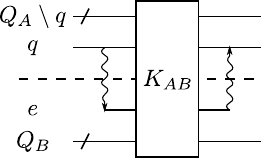}}
      \hfill
      \subfloat[]{\includegraphics[width=0.4\columnwidth]{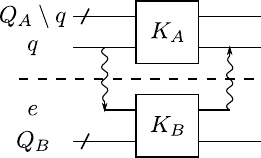}}
      \hfill{ }
      \caption{Entanglement-assisted packing process. (a) A $q$-rooted $e$-assisted packing process with a kernel $K_{AB}$. (b) A $q$-rooted $e$-assisted $A|B$-distributing process with a local kernel $K_{A}\otimes K_{B}$.}
      \label{fig:EA_process}
  \end{figure}
  \end{definition}\noindent%
  In the starting process, one needs a maximally entangled pair, namely one ebit, and two auxiliary memory qubits $e'$ and $e$.
  We refer to the first auxiliary qubit $e'$ as the \emph{root auxiliary qubit} and the second auxiliary qubit $e$ as the \emph{packing auxiliary qubit}.
  The memory space of a root auxiliary qubit can be released immediately after the starting process, while a packing auxiliary qubit is occupied for the kernel $K_{AB}$ until the ending process.
  The required memory space on each local QPU is therefore one root auxiliary qubit for initializing a starting process and several packing auxiliary qubits for parallel packing processes.
  In DQC based on entanglement-assisted packing processes, the external resources that can be optimized are therefore the entanglement cost and the number of packing auxiliary qubits.
  By definition, an entanglement-assisted packing process consumes one-ebit entanglement and occupies one packing auxiliary qubit $e$.
  A packing auxiliary qubit $e$ implies one-ebit preshared entanglement between $(e,e')$.
  The number of packing processes in a compilation of a unitary $U$ is therefore exactly the number of maximally entangled pairs and the number of packing auxiliary memory qubits that are required for the DQC implementation of $U$.
  This number is always a constructive upper bound on the entanglement cost for the DQC implementation of $U$ based on entanglement-assisted LOCC.
  A tighter bound can be determined if one finds a more entanglement-efficient compilation of a circuit.

  In general, a packing process is a completely positive and trace-preserving map (CPTP) represented by a set of two Kraus operators.
  For certain types of kernels, the two Kraus operators are equivalent up to a global phase $\varphi$.
  In this case, a packing process is equivalent to a unitary acting on the working qubits.
  We call $\mathcal{P}_{q,e}[K]$ a \emph{canonical unitary-equivalent packing process}, if the global phase $\varphi$ is equal to $\varphi=0$.
  The condition for a packing process being canonically unitary-equivalent is discussed in Appendix \ref{sec:EA_proc_Kraus}, where the packing processes with general kernels that leads to Kraus operators are also discussed.
  In the rest of this paper, we consider this ideal case and treat all packing processes as canonically unitary-equivalent.

  The canonical unitary-equivalent packing processes are the building blocks of DQC.
  They have an important property that two sequential packing processes can be merged into one packing process according to the following theorem.
  \begin{theorem}[Merging of packing processes]\label{theorem:EJPP_packing}{\ }\\
  Let $\mathcal{P}_{q,e}[K_{1,2}]$ be two $q$-rooted $e$-assisted packing processes that implement two unitaries $U_{1,2}$, respectively,
  \begin{equation}
    U_{i} = \mathcal{P}_{q,e}[K_{i}] \;\;\text{ for } i=1,2.
  \end{equation}
  The product of unitaries $U_{2}U_{1}$ can be then implemented by a single $q$-rooted $e$-assisted packing process with the kernel $K_{2}K_{1}$
  \begin{equation}
    U_{2}U_{1}
    =
    \mathcal{P}_{q,e}[K_{2}]
    \mathcal{P}_{q,e}[K_{1}]
    =
    \mathcal{P}_{q,e}[K_{2}K_{1}].
  \end{equation}
  \begin{proof}
  See Appendix \ref{proof:EJPP_packing}.
  \end{proof}
  \end{theorem}
  This theorem allows us to implement the unitary $U_{2}U_{1}$ through the packing of the two kernels $K_{2}K_{1}$ into one packing process and hence save one ebit.
  To decompose a circuit in DQC, two types of packing processes are needed, namely the distributing processes and embedding processes, which will be introduced in Section \ref{sec:distributing} and \ref{sec:embedding}, respectively.

\subsection{Distributing process\label{sec:distributing}}
  Our ultimate goal is to find an efficient implementation of a quantum circuit with packing processes that contain only local kernels.
  To this end, we need to decompose a circuit into distributing processses, which are defined as follows.
  \begin{definition}[Distributing process]\label{def:ejpp-distr_U}
  A unitary $U$ is $q$-rooted distributable (or distributable on $q$) over two local systems $A|B$, if there exists a $q$-rooted entanglement-assisted process with a local kernel with respect to $A|B$ (see Fig. \ref{fig:EA_process} (b)),
  \begin{equation}
    U = \mathcal{P}_{q,e}[K_{A}\otimes K_{B}].
  \end{equation}
  The kernel describes a \emph{$(A|B)$-distributing rule} $\mathcal{D}_{q}$ of $U$,
  \begin{equation}
    \mathcal{D}_{q}(U) = K_{A}\otimes K_{B}.
  \end{equation}
  The process $\mathcal{P}_{q,e}[K_{A}\otimes K_{B}]$ is called a \emph{$q$-rooted $(A|B)$-distributing process} of $U$.
  \end{definition}\noindent
  Explicitly, the EJPP protocol in Fig. \ref{fig:ejpp_protocol} is a distributing process, while the DQC implementation of the swapping gate in Fig. \ref{fig:swapping} involves two nested distributing processes.

  Distributing processes are the backbones of distributed quantum computing. The solution for entanglement-assisted quantum computation is to reveal the distributability structure of a quantum circuit, namely to find a decomposition of quantum circuits that consists of distributable unitaries.

  \begin{figure}[htbp]
      \centering
      \includegraphics[width=0.9\columnwidth]{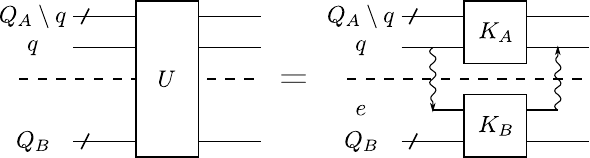}
      \caption{A $q$-rooted distributable unitary.}
      \label{fig:distributable_uni}
  \end{figure}


  In general, a unitary $U$ on the left-hand side of Fig. \ref{fig:distributable_uni} can be written in the following form
  \begin{align}
  \label{eq:unitary_general_form}
    U = \sum_{i,j} \ket{i}\bra{j}_{q} \otimes U_{ij}^{(\bar{q})},
  \end{align}
  where $\bar{q}$ denotes the set of qubits complement to $\{q\}$.
  The $q$-rooted $A|B$-distributability can be determined with the following necessary and sufficient condition.
  \begin{theorem}[Distributability condition]\label{theorem:EJPP_distr}{\ }\\
    A unitary $U$ is $q$-rooted distributable over $A|B$, if and only if $U$ is diagonal or anti-diagonal on $q$ and given in the following form,
    \begin{equation}
    \label{eq:EJPP-impl_unitary_form}
      U = \sum_{i,j} \Delta_{ij}
      \ket{i}\bra{j}_{q}
      \otimes
      U_{j}^{(\bar{q})}
      \;\text{ with }\;
      \Delta_{ij} = \delta^{i}_{j} \text{ or } \delta^{i \oplus 1}_{j},
    \end{equation}
    where $U^{(\bar{q})}_{j}$ is local with respect to $A|B$
    \begin{equation}
      U^{(\bar{q})}_{j} = V_{j}^{(Q_{A}\setminus q)}\otimes W_{j}^{(Q_{B})}.
    \end{equation}
    The unitary $V_{j}$ and $W_{j}$ act on the qubits $Q_{A}\setminus q$ and $Q_{B}$, respectively.
    Without loss of generality, we assume that $q\in Q_{A}$ is on the QPU $A$.
    The corresponding distributing rule described by the kernel $K_{A}\otimes K_{B}$ can be constructed as
    \begin{align}
      K_{A} & = \sum_{i,j\in\{0,1\}} \Delta_{ij} \ket{i}\bra{j}_{q}\otimes V_{j}^{(Q_{A}\setminus q)},
      \\
      K_{B} & = \sum_{i,j\in\{0,1\}} \Delta_{ij} \ket{i}\bra{j}_{e}\otimes W_{j}^{(Q_{B})}.
    \end{align}
  \begin{proof}
    See Appendix \ref{proof:EJPP_distr}
  \end{proof}
  \end{theorem} \noindent %
  The $q$-rooted distributability of a unitary $U$ depends on its representation in the computational basis of the root qubit.
  Note that the distributability condition can also be formulated in another basis of the root qubit, if one transforms the control qubit in the starting process and the correction gate $Z$ in the ending process into the corresponding basis.
  However, such variant starting processes and ending processes can not be merged with the standard ones defined in Definition \ref{def:st_end_proc}.
  We therefore stick to the definition of distributability with respect to the standard starting and ending processes, and determine the distributability in the computational basis of the root qubit with Theorem \ref{theorem:EJPP_distr}.

  As a result of Theorem \ref{theorem:EJPP_distr}, some trivial distributable gates can be determined as follows.
  \begin{corollary}
  \label{coro:EJPP-impl_gates}
    The following unitaries are all $q$-rooted distributable.
    \begin{enumerate}
      \item Single-qubit unitaries on $q$ that are diagonal or anti-diagonal.
      \item Two-qubit controlled-unitary gates that have $q$ as their control qubit.
    \end{enumerate}
    The distributability of these gates does not depend on the bipartition $A|B$.
  \end{corollary}\noindent%
  As a result of this corollary, a two-qubit control-phase gate $C_{V}(\theta)$ acting on $q_{1,2}$ is $q_{1,2}$-rooted distributable,
  \begin{align}
  \label{eq:control-phase_gate}
    C_{V}(\theta) & = \projector{0}_{q_{1}}\otimes \id_{q_{2}} + \projector{1}_{q_{1}}\otimes V_{q_{2}}(\theta)
    \nonumber \\
    & = \id_{q_{1}} \otimes \projector{0}_{q_{2}} + V_{q_{1}}(\theta) \otimes \projector{1}_{q_{2}},
  \end{align}
  where $V(\theta)$ is a phase gate
  \begin{equation}
    V(\theta):=\projector{0} + \exp( \imI \theta)\projector{1}.
  \end{equation}

  According to Theorem \ref{theorem:EJPP_packing}, one can implement two sequential $q$-rooted distributable unitaries $U_{2}U_{1}$ through the packing of their distributing kernels $\mathcal{D}_{q}(U_{2})\mathcal{D}_{q}(U_{1})$ within a single packing process consuming only one ebit.
  Such packing can be even possible for two non-sequential distributable unitaries through the another type of packing processs, namely embeddings.

\subsection{Embedding process\label{sec:embedding}}

  \begin{figure*}
    \centering
    \hfill
    \subfloat[]{\includegraphics[width=0.45\textwidth,height=2cm,keepaspectratio]{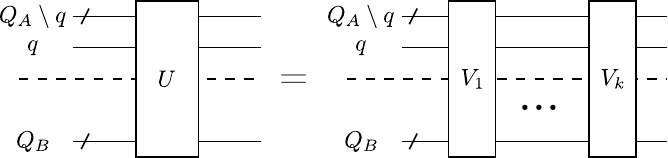}}
    \hfill
    \subfloat[]{\includegraphics[width=0.45\textwidth,height=2cm,keepaspectratio]{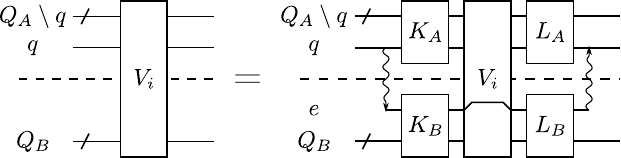}}
    \hfill{ }
    \\
    \subfloat[]{\includegraphics[width=0.8\textwidth,height=2.2cm,keepaspectratio]{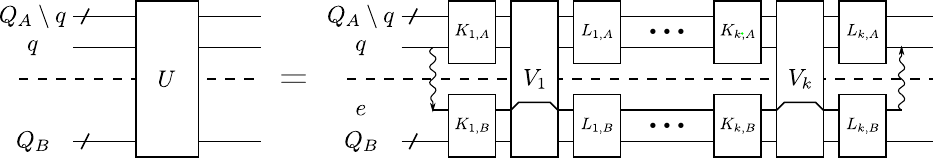}}
    \caption{Embeddable unitary:
      (a) The decomposition of $U$ as a product of embedding units.
      (b) The embeddability of each embedding unit.
      (c) The embedding of $U$.
    }\label{fig:embed_unitary}
  \end{figure*}

  In this section, we introduce embedding processes, which enable the packing of non-sequential distributing processes through the embedding of the intermediate unitary between them.
  \begin{definition}[Embedding process]\label{def:ejpp-embed_U}
    A unitary $U$ is \emph{$q$-rooted embeddable} with respect to a bipartition $A|B$, if there exists a decomposition of $U$, $U=\prod_{i}V_{i}$ (Fig.\ref{fig:embed_unitary} (a)), and corresponding kernels $\mathcal{B}_{q}(V_{i})$,
    \begin{equation}
      \mathcal{B}_{q}(V_{i}) = (L_{i,A}\otimes L_{i,B})\,V_{i}\,(K_{i,A}\otimes K_{i,B}),
    \end{equation}
    where $L_{i,A}$ and $K_{i,A}$ act on local qubits $Q_{A}$, and $L_{i,B}$ and $K_{i,B}$ act on local qubits $Q_{B}\cup\{e\}$ (Fig.\ref{fig:embed_unitary} (b)), such that
    \begin{equation}
      V_{i} = \mathcal{P}_{q,e}[\mathcal{B}_{q}(V_{i})].
    \end{equation}
    The unitary $V_{i}$ is an \emph{embedding unit} of $U$.
    The kernel $\mathcal{B}_{q}(V_{i})$ describes a \emph{$q$-rooted embedding rule} of $V_{i}$.
    The process $\mathcal{P}_{q,e}[\mathcal{B}_{q}(V_{i})]$ is a \emph{$q$-rooted $e$-assisted embedding process} of $V_{i}$.
  \end{definition}%
  According to Theorem \ref{theorem:EJPP_packing}, a $q$-rooted embeddable unitary can be packed into a packing process with additional local kernels $K_{i,A}\otimes K_{i,B}$ and $L_{i,A}\otimes L_{i,B}$ sandwiching the original embedding units $V_{i}$ (see Fig. \ref{fig:embed_unitary} (c)),
  \begin{equation}
  \label{eq:embedding_proc}
    U = \mathcal{P}_{q,e}[\prod_{i}\mathcal{B}_{q}(V_{i})].
  \end{equation}
  Note that $V_{i}$ can be a nonlocal unitary that still needs to be distributed.

\begin{figure}[t]
    \centering
    \includegraphics[width=0.45\textwidth,height=2.5cm,keepaspectratio]{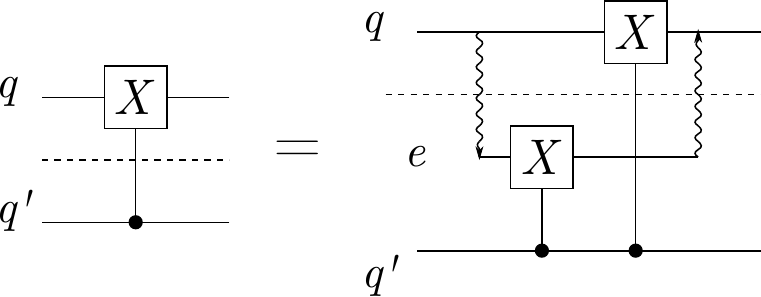}
    \caption{The embedding process of a CNOT gate}
    \label{fig:embedding_CX}
  \end{figure}

\begin{figure*}[t]
    \centering
    \subfloat[]{\includegraphics[width=\textwidth,height=2.5cm,keepaspectratio]{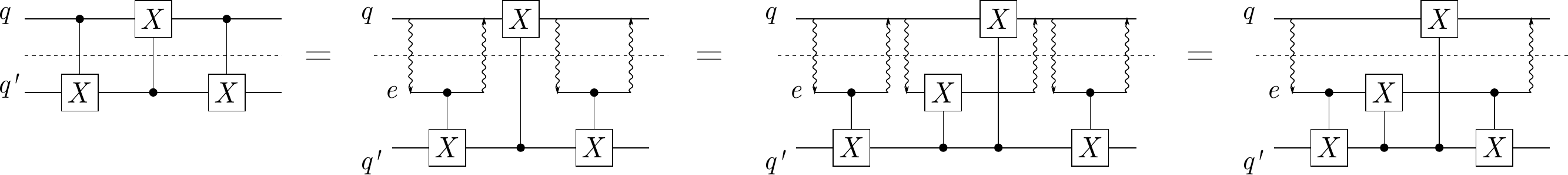}}
    \\
    \subfloat[]{\includegraphics[width=0.8\textwidth,height=2cm,keepaspectratio]{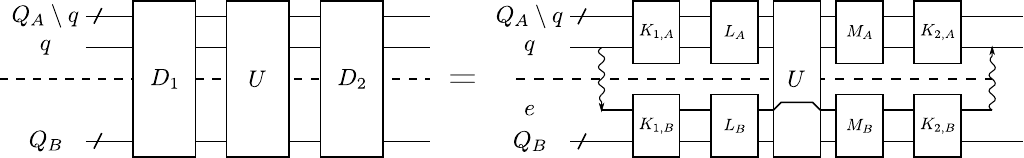}}
    \caption{(a) The embedding process between two distributing processes in the DQC of swapping. (b) Packing of two distributing processes through embedding: $D_{1,2}$ are $q$-rooted distributable unitaries. $U$ is $q$-rooted embeddable with the rule $\mathcal{B}_{q}(U) = M_{A}\otimes M_{B} \,U\, L_{A}\otimes L_{B}$.}
    \label{fig:merging_by_embedding}
  \end{figure*}

 An example of embedding processes can be found in the first step of DQC implementation of swapping in Fig. \ref{fig:swapping}, where a CNOT gate is packed into a $q$-rooted embedding process. As shown in Fig.\ref{fig:embedding_CX}, an additional local CNOT gate is introduced in the kernel of the embedding process, while the nonlocal CNOT gate remains unaltered in the circuit. Such an embedding process in the swapping circuit can merge two non-sequential distributing processes into one packing process according to Theorem \ref{theorem:EJPP_packing}, as illustrated in Fig. \ref{fig:merging_by_embedding} (a).


  Formally, let $U$ be a $q$-rooted embeddable unitary with the embedding rule $\mathcal{B}_{q}(U) = (M_{A}\otimes M_{B}) U (L_{A}\otimes L_{B})$ and placed between two $q$-rooted distributable unitaries $D_{1,2}$, which is shown in Fig. \ref{fig:merging_by_embedding} (b).
  According to Theorem \ref{theorem:EJPP_packing}, one can pack the two non-sequential distributable unitaries $D_{1,2}$ into a packing process through the embedding of $U$,
  \begin{align}
    D_{2} \, U \, D_{1}
    & =
    \mathcal{P}_{q,e}[\mathcal{D}_{q}(D_{2})\mathcal{B}_{q}(U)\mathcal{D}_{q}(D_{1})]
    \nonumber\\
    & =
    \mathcal{P}_{q,e}[\widetilde{K}_{2} U \widetilde{K}_{1}].
  \end{align}
  where $\mathcal{D}_{q}(D_{i}) = K_{i,A}\otimes K_{i,B}$ are the distributing rules for $D_{1,2}$ and $\widetilde{K}_{1} = L_{A}K_{1,A}\otimes L_{B}K_{1,B} $. In the end, the embeddable unitary $U$ is embedded into a merged packing process with the kernel $\widetilde{K}_{2} U \widetilde{K}_{1}$, where $\widetilde{K}_{2} = K_{2,A} M_{A}\otimes K_{2,A} M_{B}$ are both local unitaries and the distributability of $U$ remains untouched.

  \bigskip

  \begin{figure*}
    \centering
    \subfloat[]{\includegraphics[width=0.9\textwidth,height=3.5cm,keepaspectratio]{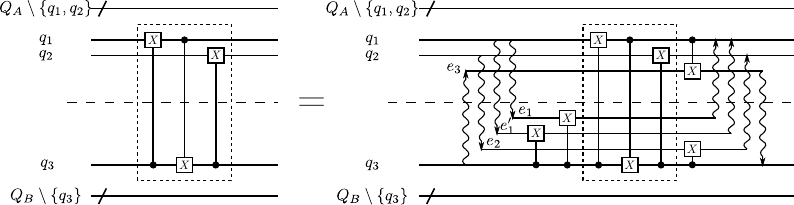}}    \\
    \subfloat[]{\includegraphics[width=0.9\textwidth,height=5cm,keepaspectratio]{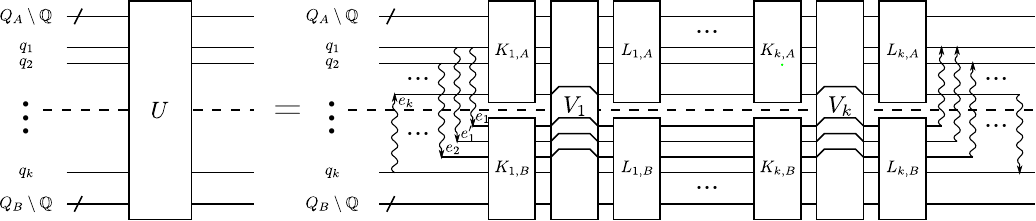}}
    \caption{Joint embedding.
    (a) A circuit that can be jointly embedded on the qubits $\{q_{1},q_{1},q_{2},q_{3}\}$.
    (b) A joint embedding rooted on $\mathbb{Q}=\{q_{1},q_{1},q_{2},...,q_{k}\}$ assisted with the auxiliary qubits $\mathbb{A} = \{e_{1},e'_{1},e_{2},...,e_{k}\}$.}
    \label{fig:joint_embedding}
  \end{figure*}

  The utility of embedding is not limited to packing processes rooted on a single qubit.
  It can be extended to simultaneous embedding processes rooted on multiple qubits. As shown in Fig. \ref{fig:joint_embedding} (a), the unitary implemented by three CNOT gates can be jointly embedded in four packing processes which start simultaneously with four starting processes rooted on $\{q_{1},q_{2},q_{3}\}$ assisted by $\{e_{1},e'_{1},e_{2},e_{3}\}$ and end simultaneously with four ending processes.
  In general, one can simultaneously embed a unitary into a joint packing process $\{\mathcal{P}_{q_{i},e_{i}}[K_{i}]\}_{i}$ rooted on multiple qubits $\{q_{1},...,q_{k}\}$, which is referred to as a $\{q_{1},...,q_{k}\}$-rooted joint embedding and illustrated in Fig. \ref{fig:joint_embedding} (b).
  \begin{definition}[Joint embedding]
  \label{def:joint_embedding}
    A unitary $U$ is jointly embeddable on a set of root qubits $\mathbb{Q} = \{q_{1},...,q_{k}\}$, if there exists a decomposition $U=\prod_{i}V_{i}$ and corresponding joint embedding rules
    \begin{equation}
      \mathcal{B}_{\mathbb{Q}}(V_{i})
      =
      (L_{i,A}\otimes L_{i,B}) V_{i} (K_{i,A}\otimes K_{i,B}),
    \end{equation}
    such that each embedding unit can be implemented by a joint entanglement-assisted process $\mathcal{P}_{\mathbb{Q},\mathbb{A}}$ rooted on a set of qubits $\mathbb{Q}$ and assisted with a set of auxiliary qubits $\mathbb{A} = \{e_{1},...,e_{k}\}$,
    \begin{equation}
      V_{i} = \mathcal{P}_{\mathbb{Q},\mathbb{A}}[\mathcal{B}_{\mathbb{Q}}(V_{i})]
      \text{ with }
      \mathcal{P}_{\mathbb{Q},\mathbb{A}}
      :=
      \mathcal{P}_{q_{k},e_{k}}\circ \cdots \circ \mathcal{P}_{q_{1},e_{1}}.
    \end{equation}
  \end{definition}
  With joint embedding, one can simultaneously merge non-sequential distributing processes on multiple qubits.
  Note that a joint embedding can have packing processes on the same root qubit multiple times.
  As shown in Fig. \ref{fig:joint_embedding}, the joint embedding of $U$ on $\{q_{1},q_{2},...,q_{k}\}$ has two packing processes rooted on the same qubit $q_{1}$ and assisted with two auxiliary qubits $e_{1}$ and $e'_{1}$, respectively.

  \bigskip

  In general, it is non-trivial to derive a necessary condition for the embeddability of $U$, since one has to explore all the possible decompositions of $U$.
  However, one can still derive a primitive embedding rule $\mathcal{B}_{\mathbb{Q}}(U)$ as a sufficient condition for the $\mathbb{Q}$-rooted embeddability of $U$ as follows.
  \begin{corollary}[Primitive embedding rule]\label{coro:primitive_cond_embed}{\ }\\
    A unitary $U$ is $\mathbb{Q}$-rooted embedabble with an embedding rule $\mathcal{B}_{\mathbb{Q}}(U) =  (L_{A}\otimes L_{B})\, U\, (K_{A}\otimes K_{B})$, if 
    \begin{equation}
    \label{eq:coro_embed_rule_derive}
      C_{\mathbb{Q},X_{\mathbb{A}}}\,U\,C_{\mathbb{Q},X_{\mathbb{A}}}
      =
      (L_{A}\otimes L_{B})\, U\, (K_{A}\otimes K_{B}),
    \end{equation}
    where $C_{\mathbb{Q},X_{\mathbb{A}}}$ is a sequence of control-X gates on $\{(q_{i},e_{i}): q_{i}\in\mathbb{Q},e_{i}\in\mathbb{A}\}$,
    \begin{equation}
      C_{\mathbb{Q},X_{\mathbb{A}}}
      =
      \prod_{q_{i}\in\mathbb{Q}, e_{i}\in\mathbb{A}}C_{q_{i},X_{e_{i}}}.
    \end{equation}
    This embedding rule is called the \emph{primitive $\mathbb{Q}$-rooted embedding rule} of $U$.
  \begin{proof}
    This corollary is a result of Corollary \ref{coro:U-eq_suff_cond} in Appendix \ref{sec:EA_proc_Kraus}.
  \end{proof}
  \end{corollary}

  Due to the diagonal and anti-diagonal condition for distributability in Theorem \ref{theorem:EJPP_distr}, one can show that a $q$-rooted distributable is always $q$-rooted embeddable.
  \begin{theorem}[Distributability and embeddability]
  \label{theorem:distr_and_embed}
    If a unitary $U$ is $q$-rooted distributable, then $U$ is also $q$-rooted embeddable with the following embedding rules,
    \begin{enumerate}
      \item for $U$ being diagonal on $q$,
        \begin{equation}
        \label{eq:emb_rule_diag}
          U = \mathcal{P}_{q,e}[U];
        \end{equation}
      \item for $U$ being anti-diagonal on $q$,
        \begin{equation}
        \label{eq:emb_rule_anti-diag}
          U = \mathcal{P}_{q,e}[U X_{e}] = \mathcal{P}_{q,e}[X_{e} U],
        \end{equation}
        where $X_{e}$ is an $X$ gate on the auxiliary qubit $e$.
    \end{enumerate}
  \begin{proof}
    According to Theorem \ref{theorem:EJPP_distr}, a $q$-rooted distributable unitary $U$ is diagonal or anti-diagonal on $q$.
    One can show that, for $U$ being diagonal on $q$,
    \begin{equation}
      C_{q,X_{e}} U C_{q,X_{e}} = U,
    \end{equation}
    while for $U$ being anti-diagonal on $q$,
    \begin{equation}
      C_{q,X_{e}} U C_{q,X_{e}} = UX_{e} = X_{e}U.
    \end{equation}
    As a result of Corollary \ref{coro:primitive_cond_embed}, $U$ is embeddable on $q$ with the embedding rules $\mathcal{B}_{q}(U) = C_{q,X_{e}} U C_{q,X_{e}}$.
%
  \end{proof}
  \end{theorem}



  In general, the $q$-rooted embeddability is not a sufficient condition for $q$-rooted distributability, since there are some unitaries which are $q$-rooted embeddable but not $q$-rooted distributable.
  However, if the unitary we consider is a local unitary, one can show that  embeddability is equivalent to distributability.
  \begin{corollary}[Embeddability of local unitaries]\label{coro:local_embed_unitary}{\ }\\
    A local unitary is $q$-rooted embeddable, if and only if it is $q$-rooted distributable.
    \begin{proof}
      Without loss of generality, we assume that $q$ is on the system $A$.
      If a local unitary $U_{A} \otimes \id_{B}$ is $q$-rooted embeddable, there exists an embedding rule,
      \begin{align}
        U_{A} \otimes \id_{B}
        & = \mathcal{P}_{q,e}[(M_{A}\otimes M_{B}) \, (U_{A} \otimes \id_{B}) \, (L_{A}\otimes L_{B})],
        \nonumber \\
        & = \mathcal{P}_{q,e}[(M_{A}U_{A}L_{A}) \otimes (M_{B}L_{B})].
      \end{align}
      The unitary $U_{A}$ is therefore $q$-rooted distributable with the local kernel $(M_{A}U_{A}L_{A}) \otimes (M_{B}L_{B})$, which shows that $q$-rooted embeddability is a sufficient condition for $q$-rooted distributability of a local unitary.
      According to Theorem \ref{theorem:distr_and_embed}, $q$-rooted embeddability is also a necessary condition for $q$-rooted distributability.
    \end{proof}
  \end{corollary}
  According to Theorem \ref{theorem:EJPP_distr}, this corollary implies that a local unitary is embeddable on $q$, if and only if it is diagonal or anti-diagonal on $q$.

\subsection{Compatibility among embeddings}
\label{sec:compatible_emb}
  \begin{figure*}[t]
    \centering
    \subfloat[]{\includegraphics[width=0.95\textwidth, height=3.5cm,keepaspectratio]{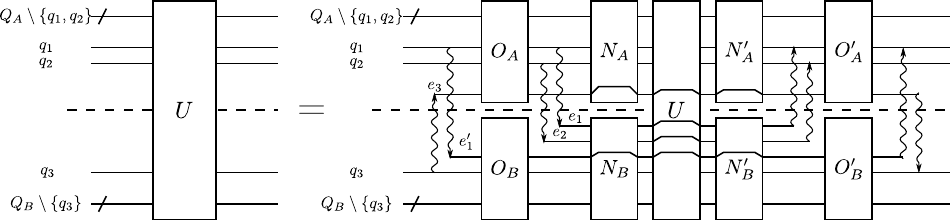}}
    \\
    \subfloat[]{\includegraphics[width=0.95\textwidth, height=3.5cm,keepaspectratio]{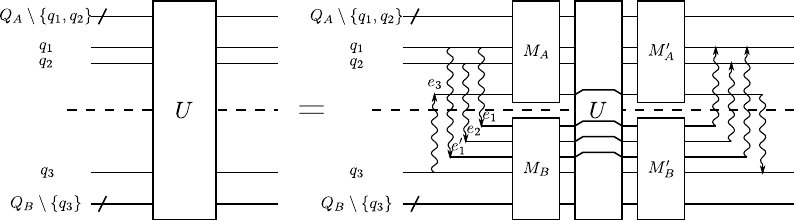}}
    \caption{A unitary, which is embeddable on both $\mathbb{Q}=\{q_{1},q_{2}\}$ and $\mathbb{Q}'=\{q_{1},q_{3}\}$
    (a) A $\mathbb{Q}$-rooted embedding followed by a $\mathbb{Q}'$-rooted embedding.
    (b) The $\mathbb{Q}$-rooted embedding and $\mathbb{Q}'$-rooted embedding are compatible, if $U$ is $\mathbb{Q}\uplus\mathbb{Q}'$-rooted embeddable.
    }
    \label{fig:recursive_embedding}
  \end{figure*}

  Let $U$ be embeddable on two different sets of root qubits $\mathbb{Q}$ and $\mathbb{Q}'$ with the following embedding rules, respectively
  \begin{align}
    \mathcal{B}_{\mathbb{Q}}(U)
    & =
    N_{A}\otimes N_{B}\,U\,N'_{A}\otimes N'_{B},
    \\
    \mathcal{B}_{\mathbb{Q}'}(U)
    & =
    O_{A}\otimes O_{B} \, U\, O'_{A}\otimes O'_{B}.
  \end{align}
  Suppose there are two $\mathbb{Q}$-rooted packing processes $\mathcal{P}_{\mathbb{Q},\mathbb{A}}[L]$ and $\mathcal{P}_{\mathbb{Q},\mathbb{A}}[L']$ sandwiching the unitary $U$. The intermediate unitary $U$ can be packed through its $\mathbb{Q}$-rooted embedding
  \begin{equation}
    \mathcal{P}_{\mathbb{Q},\mathbb{A}}[L']
    \,U\,
    \mathcal{P}_{\mathbb{Q},\mathbb{A}}[L]
    =
    \mathcal{P}_{\mathbb{Q},\mathbb{A}}[L'\,\mathcal{B}_{\mathbb{Q}}(U)\,L].
  \end{equation}
  The same packing holds for $\mathbb{Q}'$-rooted packing processes $\mathcal{P}_{\mathbb{Q}',\mathbb{A}'}[K]$ and $\mathcal{P}_{\mathbb{Q}',\mathbb{A}'}[K']$ that sandwich the unitary $U$,
  \begin{equation}
    \mathcal{P}_{\mathbb{Q}',\mathbb{A}'}[K']
    \,U\,
    \mathcal{P}_{\mathbb{Q}',\mathbb{A}'}[K]
    =
    \mathcal{P}_{\mathbb{Q}',\mathbb{A}'}[K'\,\mathcal{B}_{\mathbb{Q}'}(U)\,K].
  \end{equation}
  Although one can always implement the $\mathbb{Q}$-rooted embedding of $U$ followed by the $\mathbb{Q}'$-rooted embedding recursively,
  \begin{equation}
  \label{eq:recurs_embed_eg}
    U = \mathcal{P}_{\mathbb{Q}',\mathbb{A}'}\left[
      \mathcal{B}_{\mathbb{Q}'}\left(
        \mathcal{P}_{\mathbb{Q},\mathbb{A}}\left[
          \mathcal{B}_{\mathbb{Q}}\left(U\right)
        \right]
      \right)
    \right],
  \end{equation}
  it is not guaranteed that one can still pack the $\mathbb{Q}$-rooted processes $\mathcal{P}_{\mathbb{Q},\mathbb{A}}[L_{1,2}]$ and $\mathbb{Q}'$-rooted processes $\mathcal{P}_{\mathbb{Q}',\mathbb{A}'}[K_{1,2}]$ simultaneously.

  \begin{figure*}[t]
    \centering
    \subfloat[]{\includegraphics[width=0.95\textwidth, height=3.5cm,keepaspectratio]{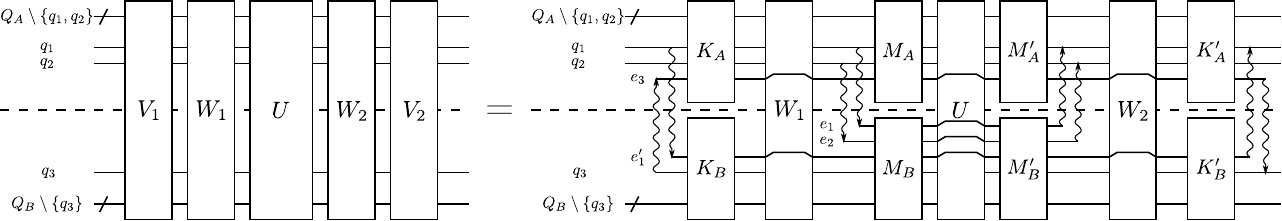}}
    \\
    \subfloat[]{\includegraphics[width=0.95\textwidth, height=3.5cm,keepaspectratio]{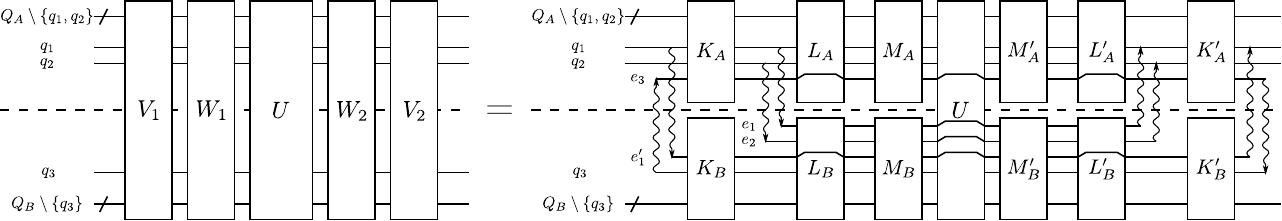}}
    \caption{Simultaneous packing through compatible recursive embeddings
    (a) The packing of the $\mathbb{Q}'$-rooted processes including the embedding of $W_{1,2}$, embedding of $U$ and distributing of $V_{1,2}$.
    (b) The simultaneous packing of $\mathbb{Q}$-rooted processes and $\mathbb{Q}'$-rooted processes that merges the $\mathbb{Q}'$-rooted embedding of $W_{1,2}$, $\mathbb{Q}'$-rooted distributing of $V_{1,2}$, and $\mathbb{Q}$-rooted distributing of $W_{1,2}$ through the $\mathbb{Q}\uplus\mathbb{Q}'$-rooted embedding of $U$.
    }
    \label{fig:recursive_embedding_eg}
  \end{figure*}

  Fig. \ref{fig:recursive_embedding} (a) shows a $\{q_{1},q_{2}\}$-rooted embedding of $U$ followed by a recursive $\{q_{1},q_{3}\}$-rooted embedding. In this example, the root qubits are $\mathbb{Q} = \{q_{1},q_{2}\}$ and $\mathbb{Q}' = \{q_{1},q_{3}\}$ associated with the auxiliary qubits $\mathbb{A} = \{e_{1},e_{2}\}$ and $\mathbb{A}' = \{e'_{1},e_{3}\}$, respectively.
  One can simultaneously pack $\mathbb{Q}$-rooted processes and $\mathbb{Q'}$-rooted processes through the joint embedding of $U$, if $U$ is $\mathbb{Q}\uplus\mathbb{Q}'$-rooted embeddable, as it is shown in Fig. \ref{fig:recursive_embedding} (b),
  \begin{equation}
  \label{eq:joint_embedding_proc}
    U = \mathcal{P}_{\mathbb{Q}\uplus\mathbb{Q}', \mathbb{A}\cup\mathbb{A}'}[M'_{A}\otimes M'_{B} \, U \, M_{A}\otimes M_{B}].
  \end{equation}
  Note that $\mathbb{Q}\uplus\mathbb{Q}'$ is the disjoint union of $\mathbb{Q}$ and $\mathbb{Q}'$, as the root qubits $\mathbb{Q}$ and $\mathbb{Q}'$ may share the same qubits. For example, the circuit on the left-hand side of Fig. \ref{fig:joint_embedding} (b) is embeddable on both $\mathbb{Q}=\{q_{1},q_{2}\}$ and $\mathbb{Q}'=\{q_{1},q_{3}\}$. The $\mathbb{Q}$-rooted embedding and $\mathbb{Q}'$-rooted embedding of the circuit are compatible, since one can simultaneously implement these two embeddings in one joint embedding process.

  As an embedding process can merge two non-sequential distributing processes, the compatibility of multiple embeddings allows us to merge multiple non-sequential distributing processes on multiple qubits. Fig. \ref{fig:recursive_embedding_eg} demonstrates the utility of two compatible embeddings of $U$ to merge the distributing process of $V_{1,2}$ and $W_{1,2}$.
  Suppose the unitary $U$ in Fig. \ref{fig:recursive_embedding_eg} is embeddable on both $\mathbb{Q}=\{q_{1},q_{2}\}$ and $\mathbb{Q}=\{q_{1},q_{3}\}$, which can be jointly embedded as shown in Fig. \ref{fig:recursive_embedding}.
  Let $V_{1,2}$ and $W_{1,2}$ be distributable on $\mathbb{Q'}$ and $\mathbb{Q}$, respectively, while $W_{1,2}$ be $\mathbb{Q'}$-rooted embeddable with the embedding rule $\mathcal{B}_{\mathbb{Q'}}(W_{1,2}) = W_{1,2}$.
  As shown in Fig. \ref{fig:recursive_embedding_eg} (a), one can firstly pack the $\mathbb{Q}'$-rooted distributing processes of $V_{1,2}$ through the $\mathbb{Q}'$-rooted embedding of $W_{1,2}$ and $\mathbb{Q}\uplus\mathbb{Q}'$-rooted embedding of $U$.
  Afterward, one can pack the $\mathbb{Q}$-rooted distributing processes of $W_{1,2}$ through the $\mathbb{Q}\uplus\mathbb{Q}'$-rooted embedding of $U$.
  In the end, the $\mathbb{Q}\uplus\mathbb{Q}'$-rooted embedding of $U$ allows the simultaneous merging of the $\mathbb{Q}$-rooted processes and the $\mathbb{Q}'$-rooted processes.

  In this example, the recursive embeddings of $U$ rooted on $\mathbb{Q}$ and $\mathbb{Q}'$ are compatible for a simultaneous packing on $\mathbb{Q}\uplus\mathbb{Q}'$.
  For simultaneous packing, two embeddings must possess a recursive compatibility defined as follows.
  \begin{definition}[Compatible recursive embeddings]\label{def:recursive_compatibility}
    Let $\mathcal{B}_{\mathbb{Q}}(U)$ and $\mathcal{B}_{\mathbb{Q}'}(U)$ be two embedding rules of a unitary $U$ rooted on $\mathbb{Q}$ and $\mathbb{Q}'$, respectively.
    They are \emph{compatible} for a recursive embedding on $\mathbb{Q}\uplus\mathbb{Q}'$, if $U$ is jointly embeddable on $\mathbb{Q}\uplus\mathbb{Q}'$ with an embedding rule $\mathcal{B}_{\mathbb{Q}\uplus\mathbb{Q}'}$
    \begin{equation}
      \mathcal{B}_{\mathbb{Q}\uplus\mathbb{Q}'}(U)
      =
      (M'_{A}\otimes M'_{B})
      \;U\;
      (M_{A}\otimes M_{B}),
    \end{equation}
    so that
    \begin{equation}
      U =
      \mathcal{P}_{\mathbb{Q}\uplus\mathbb{Q}',\mathbb{A}\cup\mathbb{A}'}
      \left[
        \mathcal{B}_{\mathbb{Q}\uplus\mathbb{Q}'}(U)
      \right].
    \end{equation}
  \end{definition}
  Since the two sets of root qubits $\mathbb{Q}$ and $\mathbb{Q}'$ can share some common elements, the compatible merging of the two embeddings $\mathcal{B}_{\mathbb{Q}}$ and $\mathcal{B}_{\mathbb{Q}'}$ takes the disjoint union set $\mathbb{Q}\uplus\mathbb{Q}'$ as its root qubits. For the example in Fig. \ref{fig:recursive_embedding} (b), the root qubits after the compatible merging are $\mathbb{Q}\uplus\mathbb{Q}' = \{q_{1},q_{2},q_{1},q_{3}\}$ associated with the auxiliary qubits $\mathbb{A}\cup\mathbb{A}' = \{e_{1},e_{2},e'_{1},e_{3}\}$.

  As a result of Corollary \ref{coro:primitive_cond_embed}, the compatibility of two recursive embeddings can be determined with the following sufficient condition.
  \begin{lemma}[Condition for recursive compatibility]
  \label{lemma:cond_recur_compat}
    Let $\mathcal{B}_{\mathbb{Q}'}$ be a $\mathbb{Q}'$-rooted embedding rule of $U$,
    \begin{equation}
      \mathcal{B}_{\mathbb{Q}'}(U) =
      (K'_{A}\otimes K'_{B})\, U\, (K_{A}\otimes K_{B}).
    \end{equation}
    It is recursively compatible with another $\mathbb{Q}$-rooted embedding of $U$, if
    \begin{equation}
    \label{eq:recurs_embed_commut}
      [K'_{A}\otimes K'_{B}, C_{\mathbb{Q},X_{\mathbb{A}}}] = [K_{A}\otimes K_{B}, C_{\mathbb{Q},X_{\mathbb{A}}}] = 0.
    \end{equation}
    The joint embedding rule is $\mathcal{B}_{\mathbb{Q}\uplus\mathbb{Q}'} = \mathcal{B}_{\mathbb{Q}'}\circ\mathcal{B}_{\mathbb{Q}}$.
%
    \begin{proof}
      See Appendix \ref{proof:cond_recur_compat}
    \end{proof}
  \end{lemma}

  As a result of Lemma \ref{lemma:cond_recur_compat}, the embedding rule for distributable unitaries are all recursively compatible.
  \begin{corollary}[{\small Recursive embedding of distributing}]
    The $q$-rooted embedding of a $q$-rooted distributable unitary is recursively compatible with any other embeddings.
  \begin{proof}
    As a result of Lemma \ref{lemma:cond_recur_compat}, the embedding rules in Theorem \ref{theorem:distr_and_embed} are compatible for any recursive embeddings, since the additional kernel is $\id$ or $X_{e}$, which commutes with any control-$X$ gate $C_{q',X_{e'}}$, $[\id, C_{q',X_{e'}}] = 0 \text{ and } [X_{e}, C_{q',X_{e'}}] = 0$.
  \end{proof}
  \end{corollary}

  \bigskip

  \begin{figure*}
    \centering
    \includegraphics[width=0.95\textwidth, height=3.5cm,keepaspectratio]{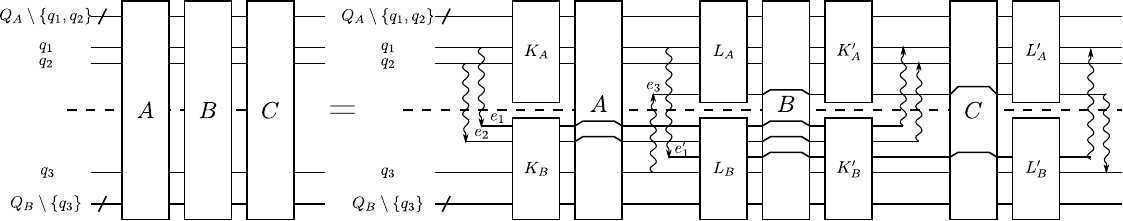}
    \caption{Compatible nested embeddings. The unitaries $U = BA$ and $U' = CB$ are embeddable on $\mathbb{Q} = \{q_{1},q_{2}\}$ and $\mathbb{Q}' = \{q_{1},q_{3}\}$ with the embedding rules $\mathcal{B}_{\mathbb{Q}}$ and $\mathcal{B}_{\mathbb{Q}'}$, respectively. If the equality holds, $\mathcal{B}_{\mathbb{Q}}$ and $\mathcal{B}_{\mathbb{Q}'}$ are compatible.
    }
    \label{fig:compatible_nested_embedding}
  \end{figure*}

  A more complicated situation of simultaneous embeddings is the nested embedding shown in Fig. \ref{fig:compatible_nested_embedding}.
  Suppose the unitaries $U = BA$ and $U' = CB$ are embeddable on $\mathbb{Q} = \{q_{1},q_{2}\}$ and $\mathbb{Q}' = \{q_{1},q_{3}\}$ with the embedding rules $\mathcal{B}_{\mathbb{Q}}$ and $\mathcal{B}_{\mathbb{Q}'}$, respectively. If we have a unitary decomposed as $CBA$, it is not guaranteed that one can implement the embeddings of $\mathcal{B}_{\mathbb{Q}}(BA)$ and $\mathcal{B}_{\mathbb{Q}'}(CB)$ simultaneously.
  If the equality in Fig. \ref{fig:compatible_nested_embedding} holds, the nested embeddings $\mathcal{B}_{\mathbb{Q}}(BA)$ and $\mathcal{B}_{\mathbb{Q}'}(CB)$ are compatible for simultaneous implementation.
  \begin{definition}[Compatible nested embedding]
    Let $U$ and $U'$ be embeddable on $\mathbb{Q}$ and $\mathbb{Q}'$ with the embedding rules $\mathcal{B}_{\mathbb{Q}}(U)$ and $\mathcal{B}_{\mathbb{Q}'}(U')$, respectively.
    For the decomposition of $U$ and $U'$,
    \begin{equation}
      U = BA, \;\; U' = CB.
    \end{equation}
    The embedding rules $\mathcal{B}_{\mathbb{Q}}(U)$ and $\mathcal{B}_{\mathbb{Q}'}(U')$ are \emph{compatible for the nested embedding} of $CBA$, if
    \begin{align}
      CBA
      = &
      E_{\mathbb{Q}',\mathbb{A}'} \circ
      L' C  \circ
      E_{\mathbb{Q}, \mathbb{A}} \circ
      K' B  L
      \nonumber \\ &
      \circ S_{\mathbb{Q}',\mathbb{A}'} \circ
      A K \circ
      S_{\mathbb{Q},\mathbb{A}}.
    \end{align}
     where $K = K_{A}\otimes K_{B}$, $K' = K'_{A}\otimes K'_{B}$, $L = L_{A}\otimes L_{B}$, and $L' = L'_{A}\otimes L'_{B}$ are all local unitaries (see Fig. \ref{fig:compatible_nested_embedding}).
  \end{definition}

\subsection{Packing of distributable packets}

  A quantum circuit of $\eta$ qubits and depth $\tau$ can be represented
  by a $\eta\times\tau$ grid with the nodes $\{(q,t): q\in\{q_{1},...,q_{\eta}\}, t\in\{1, ..., \tau\}\}$. At each depth $t$, we have a unitary $U_{t}$ implemented by a set of gates $g_{t,i}^{(\mathbb{Q}_{i})}$ acting on disjoint sets of qubits $\mathbb{Q}_{i}$, which satisfy $\cup_{i}\mathbb{Q}_{i} = \{q_{1},...,q_{\eta}\}$ and $\mathbb{Q}_{i}\cap\mathbb{Q}_{j} = \emptyset$.
  As a result of Theorem \ref{theorem:EJPP_distr}, the distributability of $U_{t}$ on a root qubit $q$ does not depend on the bipartition $A|B$.
  It means that one can easily identify all the distributable nodes of a circuit without paying any attention to the partitioning.
  The distributable nodes indicate the potential root points of distributing processes for global gates.
  In the packing procedure of the distributing processes, one can neglect the circuit nodes of single-qubit distributable gates and only consider the multi-qubit gates.
  According to Theorem \ref{theorem:EJPP_packing}, two subsequential distributable nodes $(q,t)$ and $(q,t+1)$ can be straightforwardly packed in one distributing process.

  Suppose there are two non-sequential distributable nodes $(q,t_{1})$ and $(q,t_{2})$ separated by a unitary $W^{(q)}_{t_{2};t_{1}}$, which is non-distributable but embeddable.
  One can still pack the two non-sequential distributing processes rooted on $(q,t_{1})$ and $(q,t_{2})$ into one packing process through the embedding of $W^{(q)}_{t_{2};t_{1}}$ according to Theorem \ref{theorem:EJPP_packing}.
  The embeddability of $W^{(q)}_{t_{2};t_{1}}$ therefore allows us to pack the two non-sequential distributing processes through the embedding of $W^{(q)}_{t_{2};t_{1}}$.
  Such a packing of non-sequential distributing processes reduces the entanglement cost in the distributed quantum computation.
  To explore all the possible packing of distributing processes in a quantum circuit, we need to identify its potential embedding processes and distributing processes.
  The embedding processes will then merge the root points of distributing processes into \emph{distributable packets}, which are introduced as follows.
  \begin{widetext}
  \begin{definition}[distributable packet]\label{def:distr_pack}{\ }\\
    Let $\mathcal{V}_{q,T}$ be a set of distributable nodes $(q,t)$ on the qubit $q$ at the depths $t\in T$ of a quantum circuit,
    \begin{equation}
    \label{eq:dist_pack_def}
      \mathcal{V}_{q,T} := \{(q,t): t\in T, \text{$U_{t}$ is $q$-rooted distributable, and $(q,t)$ is associated with a multi-qubit gate}\}.
    \end{equation}
    Let $\mathbb{B}$ be a set of embedding rules.
    The set $\mathcal{V}_{q,T}$ is $\mathbb{B}$-packable with respect to a bipartition $A|B$,
    if the unitary between each pair of depths $\{t_{1},t_{2}\}\subseteq T$ fulfills
    \begin{equation}
      \prod_{t:t_{1}<t<t_{2}}U_{t} \text{ is $q$-rooted embeddable}
    \end{equation}
    with respect to the bipartition $A|B$ according to the embedding rules in $\mathbb{B}$.
    The set $\mathcal{V}_{q,T}$ is called a \emph{distributable $\mathbb{B}$-packet}.
    A single distributable node $\{(q,t)\}$ is a \emph{trivial distributable $\mathbb{B}$-packet}.
    The set of distributable $\mathbb{B}$-packets of a circuit is denoted as $\mathbb{V}_{\mathbb{B}}$.
  \end{definition}
  \end{widetext}
  If there is no ambiguity in a context, we will employ the terminology ``distributable packet'' without the specification of the embedding rule $\mathbb{B}$ for conciseness.
  In this definition, we only consider the distributable nodes, which are associated with global gates in Eq. \eqref{eq:dist_pack_def}, since local gates do not need to be distributed.
  It is worth noting that the nodes in $\mathcal{V}_{q,T}$ are not necessarily contiguous.

  If two distributable packets have non-empty intersection, they can be merged into a larger set through the following corollary.
  \begin{lemma}[Merging of distributable packets]\label{lemma:merging_of_packing}{\ }\\
    Two distributable packets $\mathcal{V}_{q,T_{1}}$ and $\mathcal{V}_{q,T_{2}}$ can be merged into one packet $\mathcal{V}_{q,T_{1}\cup T_{2}}$, if they have non-empty intersection $T_{1} \cap T_{2} \neq \emptyset$,
    \begin{equation}
      \mathcal{V}_{q,T_{1}} \cup_{D} \mathcal{V}_{q,T_{2}}:=
      \mathcal{V}_{q,T_{1}\cup T_{2}}.
    \end{equation}
    The symbol $\cup_{D}$ represents the merging of distributable packets.
  \begin{proof}
    See Appendix \ref{proof:merging_of_packing}.
  \end{proof}
  \end{lemma}
  After the merging of these processes, one obtains a \emph{packing process on a distributable packet}, of which the kernels are given as follows.
  \begin{theorem}[Packing on distributable packets]
  \label{theorem:packing_dPacket}
    Given a distributable packet $\mathcal{V}_{q,T}$, the $q$-rooted distributing processes of $U_{t}$ with $t\in T$ can be packed into one packing process through embedding as follows,
    \begin{equation}
    \label{eq:packing_of_dist_pack}
      \prod_{\min T \le t \le \max T}
      U_{t} =
      \mathcal{P}_{q,e} [\prod_{\min T \le t \le \max T} K_{t}],
    \end{equation}
    where the kernel $K_{t}$ of the depth $t$ is
    \begin{equation}
      K_{t} =
      \left\{
        \begin{array}{ll}
          \mathcal{D}_{q,t}:=\mathcal{D}_{q}(U_{t}), & t\in T \\
          \mathcal{B}_{q,t}:=\mathcal{B}_{q}(U_{t}), & t\notin T
        \end{array}
      \right..
    \end{equation}
    \begin{proof}
      The unitary $\prod_{t\in[\min T, \max T]}U_{t}$ can be implemented with distributing processes for $t\in T$ and embedding processes for $t\notin T$
      \begin{equation}
        \prod_{t\in[\min T, \max T]}U_{t}
        =
        \prod_{t\in[\min T, \max T]} \mathcal{P}_{q,e_{t}} [K_{t}],
      \end{equation}
      where the kernel $K_{t}$ of the depth $t$ is the distributing rule $\mathcal{D}_{q}(U_{t})$ for $t\in T$ and the embedding rule $\mathcal{B}_{q}(U_{t})$ for $t\notin T$. As a result of Theorem \ref{theorem:EJPP_packing}, the product of unitary-equivalent packing processes can be packed into the single packing process given in Eq. \eqref{eq:packing_of_dist_pack}.
    \end{proof}
  \end{theorem}
  The set of distributable packets of a quantum circuit after all possible mergings contains only disjoint packets.
  The DQC of a quantum circuit can be implemented through the packing processes rooted on these distributable packets.
  The number of selected root packets is equal to the entanglement cost of the corresponding DQC implementation.

\subsection{Conflicts among packing processes}
  Although distributing processes and embedding processes can be packed in a distributable packet, it is not guaranteed that the packing processes rooted on two distributable packets can be simultaneously implemented.
  The packing process on a distributable packet contains the distributing processes rooted on its elements, and the embedding processes between them.
  The implementation of these distributing processes and embedding processes may introduce additional local kernels that change the distributablity and embeddability of another packet.
  This may cause incompatibility among distributable packets.

  To reveal the incompatibility among distributing processes and embedding processes on distributable packets, we can unpack a packing process on a packet $\mathcal{V}_{q,T}$ in Eq. \eqref{eq:packing_of_dist_pack} into a sequence of packing processes represented by distributing kernels and embedding kernels
  \begin{gather}
    \mathcal{V}_{q,T} \text{ with } T = \{t_{1},...,t_{k}\}
    \nonumber \\
    \downarrow \text{unpack}
    \nonumber \\
    (\mathcal{D}_{q;t_{1}}, \mathcal{B}_{q;t_{1},t_{2}}, \mathcal{D}_{q;t_{2}},
    ...,
    \mathcal{D}_{q;t_{k-1}}, \mathcal{B}_{q;t_{k-1},t_{k}}, \mathcal{D}_{q;t_{k}} ),
  \label{eq:dist_pack_unpack}
  \end{gather}
  where
  \begin{equation}
  \label{eq:embedding_kernel}
    \mathcal{B}_{q;t_{1},t_{2}}
    :=
    \prod_{t:t_{1}<t<t_{2}}\mathcal{B}_{q;t}.
  \end{equation}
  If there exists a trivial distributable packet $\mathcal{V}_{q;t}$ with $t_{1}<t<t_{2}$, then one can replace the embedding rules $\mathcal{B}_{q;t}$ in Eq. \eqref{eq:embedding_kernel} by the distributing rule $\mathcal{D}_{q;t}$
  \begin{equation}
    \mathcal{B}_{q;t_{1},t_{2}}
    \rightarrow
    \mathcal{B}_{q;t,t_{2}} \, \mathcal{D}_{q;t} \, \mathcal{B}_{q;t_{1},t}.
  \end{equation}
  It will add the node $(q,t)$ into the packet $\mathcal{V}_{q,T}$ in Eq. \eqref{eq:dist_pack_unpack} and form a larger packet $\mathcal{V}_{q,T\cup\{t\}}$.
  We call such embedding \emph{decomposable}, which is defined as follows.
  \begin{definition}[Indecomposable embedding]\label{def:indecomp_embedding}{\ }\\
    An embedding process $\mathcal{B}_{q;t_{1},t_{2}}$ is \emph{decomposable}, if there exists only a distributable node $\mathcal{V}_{q,t}$ of a multi-qubit gate with $t_{1} < t < t_{2}$, such that
    \begin{equation}
      \mathcal{P}_{q,e}[\mathcal{B}_{q;t_{1},t_{2}}] =
      \mathcal{P}_{q,e}[\mathcal{B}_{q;t,t_{2}} \, \mathcal{D}_{q;t} \, \mathcal{B}_{q;t_{1},t}],
    \end{equation}
    Otherwise, $\mathcal{B}_{q;t_{1},t_{2}}$ is \emph{indecomposable}.
    A distributable packet $\mathcal{V}_{q;\{t_{1},t_{2}\}}$ is indecomposable, if $\mathcal{B}_{q;t_{1},t_{2}}$ is indecomposable.
  \end{definition}
  By definition, the embedding $\mathcal{B}_{q;t}$ on a trivial distributable node $\mathcal{V}_{q,t}$ is also an indecomposable embedding.

  Given a circuit $Q$, the set of all indecomposable embeddings fully describes the embedding rules applied to the circuit $Q$,
  \begin{align}
    \mathbb{B}(Q)
    =
    \{\mathcal{B}_{q;T}: \text{indecomposable}\}.
  \end{align}
  On the other hand, the set of all distributing kernels on trivial distributable packets fully describes the distributing rules for $Q$
  \begin{equation}
    \mathbb{D}(Q)
    =
    \{\mathcal{D}_{q;t}:\mathcal{V}_{q,t} \text{ is a trivial distributable packet}\}.
  \end{equation}
  The incompatibility among distributing rules and embedding rules should be identified at the level of indecomposable embedding processes and trivial distributing processes, which forms a set of kernels $\mathbb{K}$ of the potential packing processes
  \begin{equation}
    \mathbb{K}(Q) := \mathbb{D}(Q)\cup\mathbb{B}(Q).
  \end{equation}
  We call such a set the \emph{indecomposable packing kernels} of $Q$.
  The incompatibility among indecomposable packing kernels can be identified with directed conflict edges
  \begin{definition}[Kernel-conflict edges]\label{def:kernel-conflict_edge}{\ }\\
    Let $K_{q_{1};T_{1}},K_{q_{2};T_{2}}\in\mathbb{K}(Q)$ be two indecomposable packing kernels of a circuit $Q$.
    The incompatibility between $K_{q_{1};T_{1}}$ and $K_{q_{2};T_{2}}$ is represented by a directed edge
    \begin{equation}
      K_{q_{1};T_{1}}\rightarrow K_{q_{2};T_{2}} \Leftrightarrow c(K_{1},K_{2})=1,
    \end{equation}
    if the implementation of $K_{1}$ prevents the implementation of $K_{2}$.
    If the two kernels are mutually incompatible, i.e. $c(K_{1},K_{2})=c(K_{2},K_{1})=1$, it forms an undirected edge
    \begin{equation}
      K_{q_{1};T_{1}}\leftrightarrow K_{q_{2};T_{2}}:=
      \{K_{q_{1};T_{1}}\rightarrow K_{q_{2};T_{2}}, K_{q_{2};T_{2}}\rightarrow K_{q_{1};T_{1}}\}.
    \end{equation}
    The matrix $\{c(K_{i},K_{j})\}_{i,j}$ is the adjacency matrix of conflict edges.
  \end{definition}

  A conflict caused by intrinsic incompatibility of packing processes due to the intrinsic quantum circuit structure, such as the incompatibility of embeddings, is called an \emph{intrinsic conflict}.
  In practice, two simultaneous packing processes need to compete for the necessary external resources.
  As shown in Fig. \ref{fig:start_end_process}, the external resources consumed in a packing process are one-ebit entangled pair, an auxiliary memory qubit $e'$ on the root QPU and an auxiliary memory qubit $e$ on the remote QPU, which are required in DQC implementation.
  The limitation on available external resources leads to conflicts among packing processes.
  Such a conflict caused by limited external resources is referred to as an \emph{extrinsic conflict}.

  Note that, in this paper, the auxiliary memory qubits in the root QPU are not counted as limited external resources, since they are only needed in the starting process (see Fig. \ref{fig:start_end_process} (a)), and can be reset and reused immediately after the starting processes.
  Besides, we ignore the limitation on the coherence time of auxiliary memory qubits, as we assume that their coherence time is not worse than the local working qubits and long enough for implementing packing processes.


  \bigskip

  \begin{figure*}[ht]
    \centering
    \subfloat[DD-type conflict (intrinsic)\label{fig:conflict_DD_intr}]{%
      \includegraphics[width=0.9\textwidth, height=3.5cm, keepaspectratio]{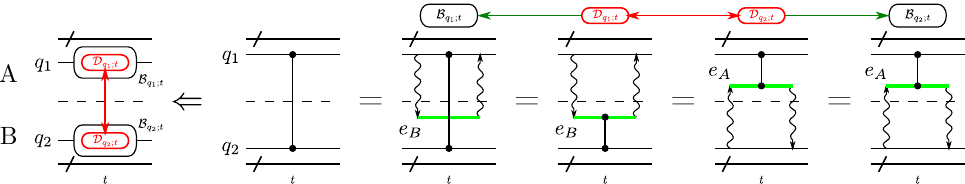}%
    }%
    \\
    \subfloat[DD-type conflict (extrinsic)\label{fig:conflict_DD_extr}]{%
      \includegraphics[width=0.9\textwidth, height=3.5cm, keepaspectratio]{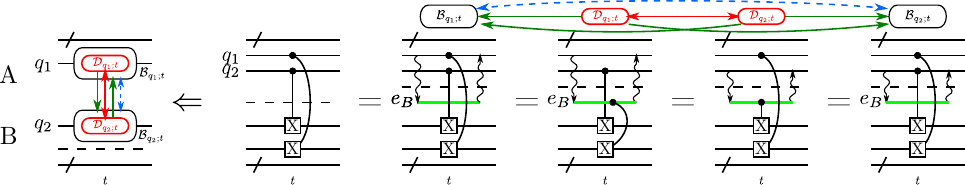}%
    }%
    \caption{Conflict between two distributing processes.
    \protect\subref{fig:conflict_DD_intr} Intrinsic conflict between two trivial distributable packets.
    \protect\subref{fig:conflict_DD_extr} Extrinsic conflict between two trivial distributable packets.
    }
    \label{fig:conflict_DD}
  \end{figure*}

  \begin{figure*}[ht]
    \centering
    \includegraphics[width=0.9\textwidth, height=3.5cm, keepaspectratio]{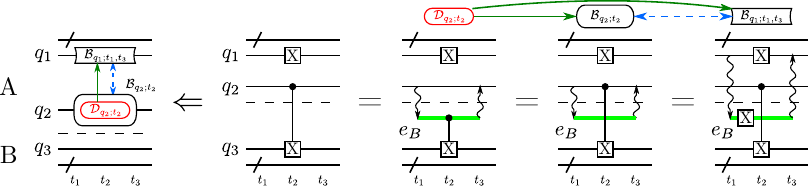}%
    \caption{Conflict between a distributing process on a trivial distributable packet and an indecomposable embedding, which compete for auxilliary memory qubits on the other local system.}
    \label{fig:conflict_DB}
  \end{figure*}

  \begin{figure*}[t]
    \centering
    \subfloat[Extrinsic BB-type conflict\label{fig:conflict_BB_extr}]{%
      \includegraphics[width=0.9\textwidth, height=3.5cm, keepaspectratio]{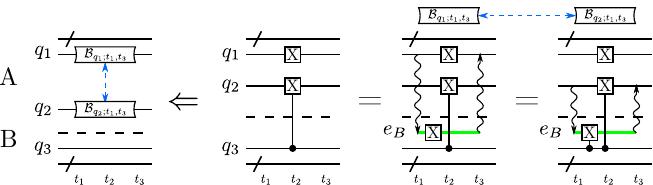}%
    }%
    \\
    \subfloat[Intrinsic BB-type conflict\label{fig:conflict_BB_intr}]{%
      \includegraphics[width=0.9\textwidth, height=3.5cm, keepaspectratio]{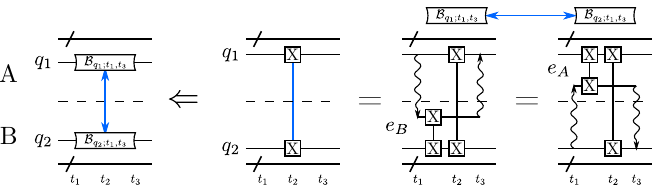}%
    }%
    \caption{Conflict between two embedding processes.
      \protect\subref{fig:conflict_BB_extr} Extrinsic conflict between two indecomposable embeddings competing for memory qubits.
      \protect\subref{fig:conflict_BB_intr} Intrinsic conflict between two indecomposable embeddings due to their intrinsice incompatibility.
    }
    \label{fig:conflict_BB}
  \end{figure*}

  In a circuit, an indecomposable packing kernel can be represented by a vertex placed at its root circuit nodes with its packing rule written on it.
  For example, in Fig. \ref{fig:conflict_DD}, the two trivial distributable packets $\mathcal{D}_{q_{1};t}$ and $\mathcal{D}_{q_{2};t}$ are represented as red vertices.
  According to Theorem \ref{theorem:distr_and_embed}, every distributable unitary is also embeddable. There must be an underlying indecomposable embedding process associated with each trivial distributable packet rooted on the same circuit nodes. On the same root circuit nodes, the root vertices of distributing kernels $\mathcal{D}_{q_{1};t}$ and $\mathcal{D}_{q_{2};t}$ are placed over their underlying indecomposable embedding vertex $\mathcal{B}_{q_{1};t}$ and $\mathcal{B}_{q_{2};t}$, respectively.
  A further example of indecomposable packing kernel $\mathcal{B}_{q_{1};t_{1},t_{3}}$ is shown in Fig. \ref{fig:conflict_DB} as a pincer-shaped vertex on its root circuit nodes, where its pincer-shaped geometry indicates its ability to merge two distributable packets before and after.

  At the level of indecomposable kernels, there are three types of incompatibility, namely the incompatibility between two distributing processes, between a distributing process and an embedding process, and between two embedding processes, which are demonstrated in Fig. \ref{fig:conflict_DD}, \ref{fig:conflict_DB} and \ref{fig:conflict_BB}, respectively.

  \paragraph{DD-type conflicts (Fig. \ref{fig:conflict_DD}).}
  The first type of conflict is the incompatibility between two distributing processes, which we call a \emph{DD-type} conflict.
  As shown in Fig. \ref{fig:conflict_DD}\subref{fig:conflict_DD_intr}, on the right-hand side, a control-Z gate can be distributed in a distributing process with the kernel $\mathcal{D}_{q_{1};t}$ or $\mathcal{D}_{q_{2};t}$ rooted on the node $(q_{1},t)$ or $(q_{2},t)$, respectively. If one of the distributing processes is implemented, the other process is not needed anymore.
  The DD-type conflict of this example is an intrinsic property of a nonlocal control-Z gate.
  It is therefore an instrinsic conflict.
  On the left-hand side of Fig. \ref{fig:conflict_DD}\subref{fig:conflict_DD_intr}, the distributability of the control-Z gate is represented by two vertices (trivial distributable packets), where the distributing rules $\mathcal{D}_{q_{1};t}$ and $\mathcal{D}_{q_{2};t}$ of the control-Z gate are rooted.
  The incompatibility of the distributability is then represented by a red bidirected edge $(\mathcal{D}_{q_{1};t} {\color{red}\leftrightarrow} \mathcal{D}_{q_{2};t})$ between these two red vertices.

  For the same CZ gate, besides the distributability, it can be also embedded with the embedding rule $\mathcal{B}_{q_{1};t}$ shown on the right-hand side of Fig. \ref{fig:conflict_DD} \subref{fig:conflict_DD_intr}.
  The embedding rule $\mathcal{B}_{q_{1};t}$ and the distributing rule $\mathcal{D}_{q_{1};t}$ compete for auxiliary memory qubits on system $B$, since each packing auxiliary qubit can only be assigned to a process at one time. If there is only one auxiliary memory qubit $e_{B}$ available, then only either $\mathcal{B}_{q_{1};t}$ or $\mathcal{D}_{q_{1};t}$ can be implemented, which is an extrinsic conflict.
  Since a distributing process has a higher priority over an embedding process, the conflict edge $(\mathcal{B}_{q_{1};t}{\color{green}\leftarrow}\mathcal{D}_{q_{1};t})$ is directed from the distributing rule represented by a green arrow in the right-hand side figure.

  As a whole, the incompatibility among the packing processes is represented by a graph of indecomposable packing root vertices. A red bidirected edge represents the intrinsic conflict between two trivial distributing processes, while a directed green edge represents the incompatibility between a trivial distributing process and its underlying embedding. Note that the directed green edge is omitted on the left-hand side graph by default.

  Other possible DD-type conflicts occur when two trivial distributable packets are placed at the same depth of a local QPU as shown in Fig. \ref{fig:conflict_DD} \subref{fig:conflict_DD_extr}. The conflict is caused by the competition among packing processes for external auxiliary memory qubits, which is extrinsic.
  The conflict between the embeddings $\mathcal{B}_{q_{1};t}$ and $\mathcal{B}_{q_{2};t}$ is automatically resolved if the conflicts associated with the trivial distributable packets are resolved.
  We therefore employ a dashed edge to represent it.

  \bigskip

  \paragraph{DB-type conflicts (Fig. \ref{fig:conflict_DB}).}
  The second type of conflict is the incompatibility between a trivial distributing process and an indecomposable embedding process, which we call a \emph{DB-type} conflict.
  As shown on the right side of Fig. \ref{fig:conflict_DB}, a circuit consisting of an $X$ gate and a CNOT gate on the depth $t_{2}$ can be distributed in a distributing process with the rule $\mathcal{D}_{q_{2};t_{2}}$ rooted on $(q_{2},t_{2})$, which is also embeddable with the rule $\mathcal{B}_{q_{2};t_{2}}$ rooted on $(q_{2},t_{2})$. Another indecomposable embedding rule $\mathcal{B}_{q_{1};t_{1},t_{3}}$ can be applied on $(q_{1},t_{2})$, which merges the packing processes rooted on $(q_{1},t_{1})$ and $(q_{1},t_{3})$.

  All of the three packing processes compete for local auxiliary memory qubits, which leads to extrinsic conflicts.
  Since the distributing $\mathcal{D}_{q_{2};t_{2}}$ has a higher priority than the embeddings, we employ green directed edges  $\mathcal{D}_{q_{2};t_{2}}{\color{green}\rightarrow}\mathcal{B}_{q_{2};t_{2}}$ and $\mathcal{D}_{q_{2};t_{2}}{\color{green}\rightarrow}\mathcal{B}_{q_{1};t_{1}, t_{3}}$ from the distributing root vertex $\mathcal{D}_{q_{2};t_{2}}$ to the embedding root vertices to represent the DB-type conflicts.

  The conflict between the indecomposable embeddings is represented by a blue bidirected edge $\mathcal{B}_{q_{2};t_{2}} {\color{blue}\leftrightarrow} \mathcal{B}_{q_{1};t_{1}, t_{3}}$. Here, a dashed line is employed, since this type of extrinsic conflicts will be automatically resolved, once the conflicts associated with trivial distributing process are resolved. For a detailed explanation please refer to the BB-type conflicts in the next paragraph.

  \paragraph{BB-type conflicts (Fig. \ref{fig:conflict_BB}).}

  The third type of conflict is the incompatibility between two indecomposable embeddings, which we call \emph{BB-type conflicts}.
  Fig \ref{fig:conflict_BB} \subref{fig:conflict_BB_extr} shows a circuit example of extrinsic BB-type conflicts, in which the embedding $\mathcal{B}_{q_{1};t_{1},t_{3}}$ of the $X$ gate on $(q_{1},t_{2})$ competes for auxiliary memory qubits with the embedding $\mathcal{B}_{q_{2};t_{1},t_{3}}$ of the CNOT gate on $(q_{2},t_{2})$.
  A blue bidirected edge $\mathcal{B}_{q_{1};t_{1},t_{3}} {\color{blue}\leftrightarrow} \mathcal{B}_{q_{2};t_{1},t_{3}}$ on the pincer-shaped vertices represents the extrinsic conflict.
  In DQC implementation, the utility of external resources for a packing process is meaningless unless a distributing process is involved. An indecomposable embedding must be packed inside a distributable packet. An extrinsic conflict between two indecomposable embeddings therefore implies a conflict between two distributable packets competing for the same extrinsic resource. It implies the existence of distributing processes that compete for the same resource and leads to DB-type or DD-type conflicts, which have a higher priority. As a result, such an extrinsic BB-type conflict is redundant, which is then represented by dashed lines.

  Besides extrinsic BB-type conflicts, there are some conflicts between indecomposable embeddings, which are intrinsic and not resolvable by adding external resources.
  For the circuit example shown in Fig \ref{fig:conflict_BB} \subref{fig:conflict_BB_intr}, the nonlocal gate is either embeddable on $(q_{1};t_{2})$ or $(q_{2};t_{2})$ with the embedding rules $\mathcal{B}_{q_{1};t_{1},t_{3}}$ or $\mathcal{B}_{q_{2};t_{1},t_{3}}$, respective.
  However, these two embedding rules are not compatible for joint embedding due to fundamental limits (see Theorem \ref{theorem:incompt_2qubit_circ} in Section \ref{sec:identify_edges}).
  One has to resolve this conflict by abandoning one of the embeddings.
  Such an intrinsic BB-type conflict is represented as a blue bidirected edge $\mathcal{B}_{q_{1};t_{1},t_{3}} {\color{blue}\leftrightarrow} \mathcal{B}_{q_{2};t_{1},t_{3}}$ connecting the embedding root vertices.
  Since an intrinsic conflict is never redundant, it is always represented with a solid line.

  \begin{figure*}[t]
    \centering
    \includegraphics[width=0.9\textwidth]{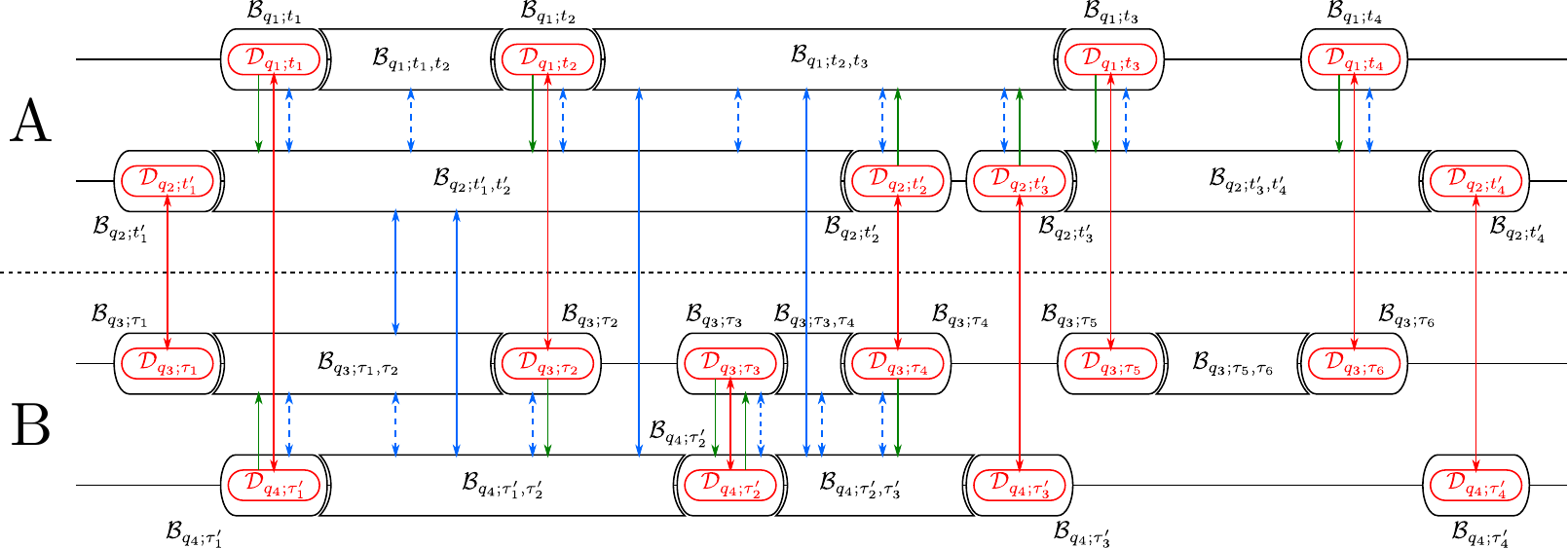}
    \caption{An example of incompatibility among distributing and embedding processes of a circuit represented by directed conflict edges.
    The red, green, and blue arrows represent the DD-type, DB-type, and BB-type conflicts, respectively.
    A dashed line indicates the redundancy of the conflict due to the existence of other conflicts associated with trivial distributable packets.
    }\label{fig:conflict_structure}
  \end{figure*}

  \bigskip

  \paragraph{Conflict structure and its solution.}
  The distribution of a quantum circuit can possess all these three types at the same time. Fig. \ref{fig:conflict_structure} shows the conflict structure of a circuit example.
  The kernels of trivial distributing processes and indecomposable embedding processes of the circuit are listed as follows,
  \begin{widetext}
  \begin{align}
  \label{eq:eg_EAK_rules}
    \mathbb{K} =
    \left\{
    \begin{tabular}{cccccc}
      $\mathcal{D}_{q_{1};t_{1}}$, &   & $\mathcal{D}_{q_{1};t_{2}}$, &   & $\mathcal{D}_{q_{1};t_{3}}$, & \\
      $\mathcal{B}_{q_{1};t_{1}}$, & $\mathcal{B}_{q_{1};t_{1},t_{2}}$, & $\mathcal{B}_{q_{1};t_{2}}$, & $\mathcal{B}_{q_{1};t_{2},t_{3}}$, & $\mathcal{B}_{q_{1};t_{3}}$, &...\\
      $\mathcal{D}_{q_{2};t'_{1}}$, &   & $\mathcal{D}_{q_{2};t'_{2}}$, & ...  &   & \\
      $\mathcal{B}_{q_{2};t'_{1}}$, & $\mathcal{B}_{q_{2};t'_{1},t'_{2}}$, & $\mathcal{B}_{q_{2};t'_{2}}$, & ... &  & \\
      ... & ... & ... & ... & ... & ...
    \end{tabular}%
    \right\}.
  \end{align}
  \end{widetext}
  With limited resources, one has to abandon conflicting processes to resolve the conflicts, the actions include the removal of distributing processes and the removal of embedding processes from a distributable packet.
  Let $\mathcal{V}_{q,T'}$ be a sub-packet of a distributable packet $\mathcal{V}_{q,T}$.
  The abandonment of the distributing processes associated with a sub-packet $\mathcal{V}_{q,T'}$ from a packet $\mathcal{V}_{q,T}$ does not affect the embeddability on $(q,T')$. It therefore leads to a new packet $\mathcal{V}_{q,T\setminus T'}$.
  The action of the removal of $\mathcal{V}_{q,T'}$ from $\mathcal{V}_{q,T}$ is defined as follows.
  \begin{definition}[Removal of distributing processes]
  \label{def:D-removal}
    After the removal of distributing processes associated with $\mathcal{V}_{q,T'}$ from $\mathcal{V}_{q,T}$, where $T'\subset T$, one obtains
    \begin{align}
      \mathcal{V}_{q,T}  \setminus_{D} \mathcal{V}_{q,T'}
      & :=
      \mathcal{V}_{q,T\setminus T'}.
    \end{align}
  \end{definition}
  On the contrary, the removal of embedding processes from a distributable packet splits the packet.
  \begin{definition}[Removal of embedding processes]
  \label{def:removal_B}
  Let $\mathcal{B}_{q;t_{1},t_{2}}$ be an indecomposable embedding in a distributable packet $\mathcal{V}_{q;T}$ with $\{t_{1},t_{2}\}\subseteq T$.
  After the removal of the embedding $\mathcal{B}_{q;t_{1},t_{2}}$, the packet $\mathcal{V}_{q;T}$ is split into two packets
  \begin{align}
    \mathcal{V}_{q,T}  \setminus_{B} \mathcal{B}_{q;t_{1},t_{2}}
    & :=
    \{\mathcal{V}_{q,T_{1}}, \mathcal{V}_{q,T_{2}}\}
  \end{align}
  where $T_{1,2}$ are the two split regions of depths,
  \begin{align}
    T_{1} &:= \{t\in T:t\le t_{1}\},\;\;
    T_{2} &:= \{t\in T:t\ge t_{2}\}.
  \end{align}
  \end{definition}

\subsection{Packing graphs and conflict graphs}
\label{sec:graphs}
  The conflict between packing processes can be solved through the removal of distributing processes or embedding processes.
  Since distributing processes have higher priority, we will first solve the $DD$-type conflicts among distributing processes prior to the $DB$-type and $BB$-type conflicts, which involve embedding processes.
  In a quantum circuit, the conflicts between two distributing processes are associated with global gates.
  For general multi-qubit gates, one needs hyperedges to describe the conflict.
  Since all multi-qubit gates can be decomposed into two-qubit gates, in this paper, we consider the circuits solely consisting of one-qubit and two-qubit gates.
  For a two-qubit gate which is distributable on both of its gate nodes $(q_{1},t)$ and $(q_{2},t)$, each gate node belongs to a distributable packet $\mathcal{V}_{q_{1},T_{1}}$ and $\mathcal{V}_{q_{2},T_{2}}$, respectively.
  One has the freedom to choose a root packet from $\mathcal{V}_{q_{1,2},T_{1,2}}$ to anchor a distributing process.
  Once a root packet is selected, the distributable node on the other packet has to be removed according to Definition \ref{def:D-removal}.
  One can assign a packing edge to each two-qubit gate and form a packing graph.
  \begin{definition}[Packing edge]\label{def:packing-edge}{\ }\\
    Given a set of distributable $\mathbb{B}$-packets $\mathbb{V}_{\mathbb{B}}$, two distributable $\mathbb{B}$-packets $\mathcal{V}_{q_{1},T_{1}}$ and $\mathcal{V}_{q_{2},T_{2}}$ are \emph{connected} by a two-qubit gate $g_{t}$ at the depth $t$, if the nodes $\{(q_{1},t),(q_{2},t)\}$ of $g_{t}$ belong to $\mathcal{V}_{q_{1},T_{1}}$ and $\mathcal{V}_{q_{2},T_{2}}$, respectively,
    \begin{equation}
      (q_{1},t)\in\mathcal{V}_{q_{1},T_{1}}
      \;\;\text{ and }\;\;
      (q_{2},t)\in\mathcal{V}_{q_{2},T_{2}}.
    \end{equation}
    The pair of connected distributable packets forms a \emph{packing edge} associated with the gate $g_{t}$
    \begin{equation}
      \mathcal{E}_{g_{t}} :=
      (\mathcal{V}_{q_{1},T_{1}} \longleftrightarrow \mathcal{V}_{q_{2},T_{2}})
      :=
      \{\mathcal{V}_{q_{1},T_{1}}, \mathcal{V}_{q_{2},T_{2}}\}.
    \end{equation}
    The gate $g_{t}$ is a \emph{packing-edge} gate.
    The packing edge set is denoted by $\mathbb{E}_{\mathbb{B}} = \{\mathcal{E}_{g_{t}}\}_{g_{t}}$.
    If $\mathcal{V}_{q_{1},T_{1}}$ and $\mathcal{V}_{q_{2},T_{2}}$ belong to the same local system, the edge $\mathcal{E}_{g_{t}}$ is \emph{local}, otherwise \emph{global}.
  \end{definition}
  \begin{definition}[Packing graph]
  A $\mathbb{B}$-packing graph of a circuit is formed by the set of distributable $\mathbb{B}$-packets and the set of packing edges,
  \begin{equation}
    G_{\mathbb{B}}
    :=
    (\mathbb{V}_{\mathbb{B}},\mathbb{E}_{\mathbb{B}}).
  \end{equation}
  \end{definition}
  To distribute all the global gates, one needs to select a set of distributable packets, which cover all packing edges.
  Note that the packing edges are defined at the level of distributable $\mathbb{B}$-packets, while the kernel-conflict edges in Definition \ref{def:kernel-conflict_edge} are defined at the level of indecomposable kernels of distributing and embedding processes.


  Besides the DD-type conflict, one also needs to take the conflicts associated with embeddings into account.
  Each conflict edge at the level of indecomposable kernels incident to an embedding kernel also defines a conflict edge at the level of distributable packets.
  \begin{widetext}
  \begin{align}
  \label{eq:packet-conflict_edge}
    \mathcal{V}_{q_{1},T_{1}}
    \underset{C}{\longrightarrow}
    \mathcal{V}_{q_{2},T_{2}}
    :\Leftrightarrow
    \left\{
      \begin{array}{ll}
    \exists \mathcal{D}_{q_{1};t_{1}}\rightarrow\mathcal{B}_{q_{2};\tau_{1},\tau_{2}}, & \text{s.t. } t_{1}\in T_{1}, \text{ and } \min T_{2} \le \tau_{1,2}\le\max T_{2}\\
    \text{or}
    \\
    \exists \mathcal{B}_{q_{1};t_{1},t_{2}}\rightarrow\mathcal{B}_{q_{2};\tau_{1},\tau_{2}}, & \text{s.t. } \min T_{1} \le t_{1,2}\le\max T_{1},\text{ and } \min T_{2} \le \tau_{1,2}\le\max T_{2}\\
      \end{array}
    \right.
  \end{align}
  \end{widetext}
  A quantum circuit has two sets of conflict edges at two different levels, one is the set of kernel-conflict edges at the level of indecomposable packing kernels, the other is the set of packet-conflict edges at the level of distributable packets.
  \begin{definition}[Conflict edge set]\label{def:conflict_edge_set}
    The set of kernel-conflict edges $\mathbb{C}(\mathbb{K})$ is defined at the level of indecomposable packing kernels $\mathbb{K}$,
    \begin{equation}
    \label{eq:process-conflit_embedding}
      \mathbb{C}(\mathbb{K}) := \{K_{1}\rightarrow K_{2}: K_{1,2}\in\mathbb{K}, K_{2} \text{ is an embedding}\},
    \end{equation}
    while the set of packet-conflict edges $\mathbb{C}(\mathbb{V})$ is defined at the level of distributable packets $\mathbb{V}$,
    \begin{equation}
      \mathbb{C}(\mathbb{V}) :=
      \{\mathcal{V}_{1}\rightarrow \mathcal{V}_{2}:  \text{defined in Eq. \eqref{eq:packet-conflict_edge}, and }\mathcal{V}_{1,2}\in\mathbb{\mathbb{V}}\}.
    \end{equation}
  \end{definition}

  The extrinsic conflict edges caused by the competition for external resource, e.g. auxiliary qubits, can be solved by supplying sufficient external resources. 
  The intrinsic conflict edges caused by the intrinsic incompatibility of embeddings is not solvable with external resources. 
  The extrinsic and intrinsic edge sets are denoted by subscripts $\mathbb{C}_{in}$ and $\mathbb{C}_{ex}$, respectively.
  We can employ the intrinsic and extrinsic conflict graphs to represent the conflicts of packing processes.
  \begin{definition}[Conflict graph]
  Packet-conflict graphs and kernel-conflict graphs are defined at the level of distributable $\mathbb{B}$-packets $\mathbb{V}_{\mathbb{B}}$ and indecomposable packing kernels $\mathbb{K}$, respectively, for intrinsic conflict
  \begin{equation}
    C_{in}(\mathbb{V}_{\mathbb{B}})
    :=
    (\mathbb{V}_{\mathbb{B}},\mathbb{C}_{in}(\mathbb{V}_{\mathbb{B}}))
    ,\;
    C_{in}(\mathbb{K})
    :=
    (\mathbb{K},\mathbb{C}_{in}(\mathbb{K})),
  \end{equation}
  and for extrinsic conflict
  \begin{equation}
    C_{ex}(\mathbb{V}_{\mathbb{B}})
    :=
    (\mathbb{V}_{\mathbb{B}},\mathbb{C}_{ex}(\mathbb{V}_{\mathbb{B}}))
    ,\;
    C_{ex}(\mathbb{K})
    :=
    (\mathbb{K},\mathbb{C}_{ex}(\mathbb{K})).
  \end{equation}
  \end{definition}

\subsection{Packing algorithm}
\label{sec:packing_algo}
  The packing graphs $G$ and conflict graphs $\kappa$ contain the full information of distributability, embeddability, and compatibility, which can be employed to determine the final packing of a quantum circuit.
  In this section, we provide heuristic algorithms for finding entanglement-efficient packing in the scenarios of unlimited and limited external resources, respectively.

  \begin{algorithm*}[t]
  \setcounter{algocf}{\value{theorem}}\addtocounter{theorem}{1}
  \caption{Packing algorithm without extrinsic limits}
  \label{algo:packing_no_ex_limit}
  $\{\mathbb{V}_{i}\}_{i} \gets $ minimum vertex covers of $G_{\mathbb{B}}$\;
  \ForEach{$\mathbb{V}_{i}$}{%
    $\mathbb{K}_{i} \gets $ the set of packing kernels $\mathbb{K}_{i}$ that have root packets in $\mathbb{V}_{i}$\;
    $\{\widetilde{\mathbb{K}}_{i,j}\}_{j} \gets $ minimum vertex covers of $\mathbb{K}_{i}$-induced intrinsic kernel-conflict graph $C_{in}(\mathbb{K}_{i})$\;

    \ForEach{$\widetilde{\mathbb{K}}_{i,j}$}{%
      $\mathbb{R}_{i,j} \gets \bigcup_{\mathcal{V}\in\mathbb{V}_{i},\mathcal{B}\in\widetilde{\mathbb{K}}_{i,j}} \mathcal{V} \setminus_{B} \mathcal{B}$:
      split the vertex cover $\mathbb{V}_{i}$ by removing its embedding processes included in $\widetilde{\mathbb{K}}_{i,j}$\;

      \SetKwFunction{destrHop}{extended\_embedding}
      \tcc{Find possible entanglement reduction with an addon function \destrHop based on an extended embedding. (See Appendix \ref{sec:destr_hopping} for the details of extended embeddings.)}
      \label{algo_line:destr_hopping_1}
      $\mathbb{R}_{i,j} \gets \destrHop(\mathbb{R}_{i,j})$ looks for possible extended embeddings in $\mathbb{R}_{i,j}$\;

      $\left(\mathbb{R}_{i,j}^{(A)},\mathbb{R}_{i,j}^{(B)}\right) \gets \mathbb{R}_{i,j}$: divided into two sets of local packets\label{algo_line:required_memory_1}\;

      $C_{ex}(\mathbb{R}_{i,j}^{(A,B)}) \gets (\mathbb{R}_{i,j}^{(A,B)},\mathbb{C}_{ex}(\mathbb{R}_{i,j}^{(A,B)}))$:
      update the extrinsic local packet-conflict graphs according to Eq. \eqref{eq:packet-conflict_edge}\label{algo_line:required_memory_2}\;

      $\chi_{i,j}^{(A,B)} \gets $ chromatic number of $C_{ex}(\mathbb{R}_{i,j}^{(A,B)})$\label{algo_line:required_memory_3}\;
    }

  }
  \KwResult{$\{(\mathbb{R}_{i,j}, \chi_{i,j}^{(A)}, \chi_{i,j}^{(B)})\}_{i,j}$}
  \end{algorithm*}

  \paragraph{Packing with unlimited auxiliary qubits.}
  Suppose that we have unlimited packing auxiliary qubits, one can determine the ultimate packing and the required amount on packing auxiliary qubits according to Algorithm \ref{algo:packing_no_ex_limit}.
  In Algorithm \ref{algo:packing_no_ex_limit}, one first searches all possible minimum vertex covers $\{\mathbb{V}_{i}\}_{i}$ of the packing graph $G_{\mathbb{B}}$.
  From each minimum vertex cover $\mathbb{V}_{i}$, one obtains a set of packing kernels $\mathbb{K}_{i}$ rooted on the packets in $\mathbb{V}_{i}$.
  The kernel set $\mathbb{K}_{i}$ induces a subgraph $C_{in}(\mathbb{K}_{i})$ of the intrinsic kernel-conflict graph.
  To solve the intrinsic conflicts in $C_{in}(\mathbb{K}_{i})$ with the least amount of additional entanglement, one needs to find a minimum vertex cover $\widetilde{\mathbb{K}}_{i,j}$ of $C_{in}(\mathbb{K}_{i})$.
  One will need to remove the embeddings in $\widetilde{\mathbb{K}}_{i,j}$ and split the corresponding packets in $\mathbb{V}_{i}$ to get a set of compatible root packets $\mathbb{R}_{i,j}$.
  The cardinality $|\mathbb{R}_{i,j}|$ is the number of packing processes and hence the entanglement cost for implementing the packing processes rooted on $\mathbb{R}_{i,j}$.
  One may further reduce the entanglement cost through the merging of the packets in $\mathbb{R}_{i,j}$ with an addon function ``\destrHop'' based on the extended embedding introduced in Appendix \ref{sec:destr_hopping}.
  The required number of packing auxiliary qubits on the local $A(B)$ system can be determined by the chromatic number $\chi_{i,j}^{(A,B)}$ of the extrinsic packet-conflict $C_{ex}(\mathbb{R}_{i,j}^{(A,B)})$.
  In the end, one obtains a set of compatible root packets $\mathbb{R}_{i,j}$ and their corresponding required amount of local auxiliary qubits $\chi_{i,j}^{(A,B)}$,
  \begin{equation}
  \label{eq:packing_algo_result}
    \{(\mathbb{R}_{i,j},\chi_{i,j}^{(A)}, \chi_{i,j}^{(B)})\}_{i,j}.
  \end{equation}

  Depending on explicit scenarios, one can choose either the set of root packets with least amount of ebits $\min_{i,j} |\mathbb{R}_{i,j}|$, or least number of local packing ancillas $\min_{i,j} |\chi_{i,j}^{(A,B)}|$, or some particular trade-off between the entanglement consumption and packing auxiliary qubits.

  Note that enumerating all minimum vertex covers is a hard problem, as the cardinality of the set of minimum vertex covers can exponentially increase\cite{Uno1997-EnumMtchBiGrph}. A heuristic solution is to search for a minimum vertex cover instead of enumerating all of them, which sacrifices the optimality for the entanglement efficiency.
  Besides, the determination of the chromatic number of a general conflict graph $C_{ex}$ is an NP-hard problem \cite{GareyJohnsonBook-NPComp}. Nevertheless, one can still efficiently estimate an upper bound on the chromatic number $\chi_{i,j}^{(A,B)}$, which implies a required number of auxiliary memory qubits that guarantees the implementation of the distributing processes.
  Overall, the time complexity of Algorithm \ref{algo:packing_no_ex_limit} is estimated as $\mathcal{O}( m^{2}\,3^{4m/3} 2^{2m} )$, while its heuristic implementation can be efficiently implemented with the complexity $\mathcal{O}(m^{7/2})$, where $m$ is the number of nonlocal 2-qubit control-phase gates in a circuit.
  (see Appendix \ref{proof:complexity_of_packing} for an explanation).

  \begin{algorithm*}[t]
  \setcounter{algocf}{\value{theorem}}\addtocounter{theorem}{1}
  \caption{Packing algorithm with extrinsic limits}
  \label{algo:packing_with_ex_limit}
  $\chi_{A},\chi_{B} \gets $ available amount of auxiliary qubits on each local systems\;
  $\{\mathbb{V}_{i}\}_{i} \gets $ minimum vertex covers of $G_{\mathbb{B}}$\;
  \ForEach{$\mathbb{V}_{i}$}{%
    $\mathbb{K}_{i} \gets $ the set of packing kernels $\mathbb{K}_{i}$ that have root packets in $\mathbb{V}_{i}$\;
    $\{\widetilde{\mathbb{K}}_{i,j}\}_{j} \gets $ minimum vertex covers of $\mathbb{K}_{i}$-induced intrinsic kernel-conflict graph $C_{in}(\mathbb{K}_{i})$\;

    \ForEach{$\widetilde{\mathbb{K}}_{i,j}$}{%
      $\mathbb{R}_{i,j} \gets \bigcup_{\mathcal{V}\in\mathbb{V}_{i},\mathcal{B}\in\widetilde{\mathbb{K}}_{i,j}} \mathcal{V} \setminus_{B} \mathcal{B}$:
      split the vertex cover $\mathbb{V}_{i}$ by removing its embedding processes included in $\widetilde{\mathbb{K}}_{i,j}$\;

      \SetKwFunction{destrHop}{extended\_embedding}
      \tcc{Find possible entanglement reduction with an addon function \destrHop based on an extended embedding. (See Appendix \ref{sec:destr_hopping} for the details of extended embeddings.)}
      \label{algo_line:destr_hopping_2}
      $\mathbb{R}_{i,j} \gets \destrHop(\mathbb{R}_{i,j})$ looks for possible extended embeddings in $\mathbb{R}_{i,j}$\;

      $(\mathbb{R}_{i,j}^{(A)},\mathbb{R}_{i,j}^{(B)}) \gets \mathbb{R}_{i,j}$: divided into two sets of local packets. \label{algo_line:intrinsic_conflict_end}

      $C_{ex}(\mathbb{R}_{i,j}^{(A,B)}) \gets (\mathbb{R}_{i,j}^{(A,B)},\mathbb{C}_{ex}(\mathbb{R}_{i,j}^{(A,B)}))$:
      update the extrinsic local packet-conflict graphs according to \eqref{eq:packet-conflict_edge} \label{algo_line:extrinsic_conflict_start}\;
      \ForEach{$k \in \{A,B\}$}{
        $\{\mathbb{R}_{i,j;c}^{(k)}\}_{c=1,...,\chi_{k}} \gets $ divide $\mathbb{R}_{i,j}^{(k)}$ into $\chi_{k}$ colors, $\mathbb{R}_{i,j;c}^{(k)}$ with $c = 1,...,\chi_{k}$, such that the number of all $\mathbb{R}_{i,j;c}^{(k)}$-induced conflict edges $\sum_{c} |\mathbb{C}_{ex}(\mathbb{R}_{i,j;c}^{(k)})|$ is minimum \label{algo_line:heuristic}\;

        \ForEach{color $c \in \{1, ...,\chi_{k}\}$}{
          \tcc{Solve the conflicts for each color $c$.}
          $\mathbb{K}_{i,j,c}^{(k)} \gets $ the set of packing kernels $\mathbb{K}_{i,j,c}^{(k)}$ that have root packets in $\mathbb{R}_{i,j;c}^{(k)}$\;

          $\{\widetilde{\mathbb{K}}_{i,j,c}^{(k)}\}_{c} \gets $ minimum vertex covers of $\mathbb{K}_{i,j,c}^{(k)}$-induced extrinsic kernel-conflict graph $C_{ex}(\mathbb{K}_{i,j,c}^{(k)})$\;


          $\widetilde{\mathbb{R}}_{i,j;c}^{(k)} \gets \bigcup_{\mathcal{V}\in\mathbb{R}_{i,j;c}^{(k)},\mathcal{B}\in\widetilde{\mathbb{K}}_{i,j,c}^{(k)}} \mathcal{V} \setminus_{B} \mathcal{B}$: Split the $c$-color vertex cover $\mathbb{R}_{i,j;c}^{(k)}$ by removing its embedding processes included in $\widetilde{\mathbb{K}}_{i,j,c}^{(k)}$\;
        }
      }
      $\widetilde{\mathbb{R}}_{i,j} \gets \bigcup_{k,c}\widetilde{\mathbb{R}}_{i,j;c}^{(k)}$.
    }
  }
  $\widetilde{\mathbb{R}}\gets \argmin_{\widetilde{\mathbb{R}}_{i,j}}|\widetilde{\mathbb{R}}_{i,j}|$\;
  \KwResult{$\{\widetilde{\mathbb{R}}, \chi_{A}, \chi_{B}\}$}
  \end{algorithm*}

  \bigskip

  \paragraph{Packing with limited auxiliary qubits.}
  \label{}
  For the packing with extrinsic limits of packing auxiliary qubits, one can employ Algorithm \ref{algo:packing_with_ex_limit} to find an entanglement-efficient packing given a fixed number of auxiliary qubits.
  In Algorithm \ref{algo:packing_with_ex_limit}, the steps up to line \ref{algo_line:intrinsic_conflict_end} are solving the intrinsic conflicts. They are identical to Algorithm \ref{algo:packing_no_ex_limit} for the packing without extrinsic limits.
  From line \ref{algo_line:extrinsic_conflict_start}, the steps for solving extrinsic conflicts under the extrinsic limits must be modified accordingly.
  Suppose the available amount of local packing auxiliary qubits are $(\chi_{A},\chi_{B})$.
  On line \ref{algo_line:heuristic}, one divides $\mathbb{R}^{A}_{i,j}$ into $\chi_{A}$ colors $\{\mathbb{R}^{A}_{i,j,c}\}_{c=1,...,\chi_{A}}$ on the system $A$ and same for the system $B$.
  Each color represents a packing auxiliary qubit.
  The goal is to find the optimum $\chi_{A(B)}$-color partition of $\mathbb{R}_{i,j}^{A(B)}$, such that the number of extrinsic $\mathbb{R}_{i,j}^{A(B)}$-conflict edges covered the same-color packets is minimum.
  A partitioning algorithm can be employed to find the minimum-conflict color partition \cite{BjorklunHusfeldtKoivisto2009-GraphPartition}, however, there is no efficient algorithm available yet.
  For a heuristic solution, one can simply find a $\chi_{A(B)}$-color coloring and continue to the next step.

  After obtaining the $c$-color extrinsic conflict graph $C_{ex}(\mathbb{R}_{i,j,c}^{A(B)})$, one extracts the process kernels $\mathbb{K}_{i,j,c}^{A(B)}$ from $\mathbb{R}_{i,j,c}^{A(B)}$ and finds the minimum vertex cover $\widetilde{\mathbb{K}}_{i,j,c}^{A(B)}$ of the extrinsic $\mathbb{K}_{i,j,c}^{A(B)}$-conflict graph $C_{ex}(\mathbb{K}_{i,j,c}^{A(B)})$.
  To solve the conflicts, one removes the embedding processes $\widetilde{\mathbb{K}}_{i,j,c}^{A(B)}$ from the packet $\mathbb{R}_{i,j,c}^{A(B)}$ and obtains a set of compatible $c$-color packets $\widetilde{\mathbb{R}}_{i,j,c}^{A(B)}$.
  In the end, one obtains the candidate sets of root packets $\widetilde{\mathbb{R}}_{i,j}$ through the union of the $c$-color root packets $\widetilde{R}_{i,j,c}^{A(B)}$
  The ultimate set of root packets is the one that has the minimum cardinality.
  \begin{equation}
  \label{eq:id_dPacket_algo_extr_result}
    \widetilde{\mathbb{R}} = \argmin_{\mathbb{R}_{i,j}} |\mathbb{R}_{i,j}|.
  \end{equation}

  The corresponding quantum circuit can be then locally implemented with the packing process rooted on the packets in $\widetilde{\mathbb{R}}$ with the assistance of $\chi_{A}$ and $\chi_{B}$ local packing auxiliary qubits and $|\widetilde{\mathbb{R}}|$ ebits of entanglement.
  Note that Algorithm \ref{algo:packing_with_ex_limit} has the same time complexity as Algorithm \ref{algo:packing_no_ex_limit} (see Appendix \ref{proof:complexity_of_packing} for an explanation).

\section{\label{sec:id_packing_conflict_graphs}Identification of packing graphs and conflict graphs}
  The packing algorithms in Section \ref{sec:packing_algo} determine the distributable packets based on the distributability and embeddability of a quantum circuit, which are fully described by the packing graphs and conflict graphs of a circuit. Therefore, one needs to identify these characteristic graphs before implementing the packing algorithms.
  Since single-qubit and two-qubit gates form a universal set of quantum circuits, all quantum circuits can be decomposed into single-qubit and two-qubit gates.
  We therefore consider the identification of packing graphs and conflict graphs of quantum circuits consisting of only single-qubit and two-qubit gates.
  For the vertices, we need to identify the set of distributable packets, which is equivalent to the identification of indecomposable embeddings.
  For the edges, we need to identify the packing edges associated with two-qubit gates, the intrinsic conflict edges associated with incompatible embeddings, and the extrinsic conflict edges associated with external resource limits.

\subsection{Identification of distributable packets}
\label{sec:identify_dPaket_algo}


  According to Definition \ref{def:distr_pack}, in a circuit consisting of $1$-qubit and $2$-qubit gates, the elements in a distributable packet are the gate nodes of $2$-qubit gates.
  According to Definition \ref{def:packing-edge}, a packing edge between two distributable packets is always associated with a two-qubit gate, which is referred to as a packing-edge gate.
  A packing-edge gate of particular interest is the control-phase gate $C_{V}(\theta)$ in Eq. \eqref{eq:control-phase_gate}.
  One can show that the connectivity of packing graphs can be identified solely with control-phase gates, since all packing-edge gates are locally equivalent to a control-phase gate up to single-qubit distributable gates.
  \begin{lemma}[The connecting $2$-qubit gate]\label{lemma:connecting_gate}{\ }\\
  A two-qubit gate $g$ acting on $q_{1}$ and $q_{2}$ is distributable on both of the grid points $(q_{1},t)$ and $(q_{2},t)$, if and only if it is equivalent to a control-phase gate up to single-qubit distributable gates,
  \begin{equation}
    g =
    (D_{q_{1}}\otimes D_{q_{2}})
    \; C_{V}(\theta)\;
    (\widetilde{D}_{q_{1}}\otimes \widetilde{D}_{q_{2}}),
  \end{equation}
  where $D_{q_{i}}$ and $\widetilde{D}_{q_{i}}$ are single-qubit distributable gates acting on $q_{i}$.
  \begin{proof}
    See Appendix \ref{proof::lemma_connecting_gate}
  \end{proof}
  \end{lemma}

  As a result of Lemma \ref{lemma:connecting_gate}, to reveal the connectivity of distributable packets, we need to convert control unitaries to control-phase gate through
  \begin{align}
  \label{eq:CV-conversion}
    C_{U_{0},U_{1}}
    = & \projector{0}\otimes U_{0} + \projector{1}\otimes U_{1}
    \nonumber \\
    = &
    (V(\theta_{0})\otimes U_{0}W)
    C_{V}(\theta_{1}-\theta_{0})
    (\id\otimes W^{\dagger}),
  \end{align}
  where $W$ is the transformation matrix for the diagonalization of $U^{\dagger}_{0}U_{1}$,
  \begin{equation}
    U^{\dagger}_{0}U_{1}
    =
    e^{\imI\theta_{0}}
    W\,V(\theta_{1}-\theta_{0})\,W^{\dagger},
  \end{equation}
  and $\theta_{i}$ are the eigenphases of $U^{\dagger}_{0}U_{1}$.

  \begin{figure*}[t]
    \centering
    \hfill
    \subfloat[]{\includegraphics[width=0.9\textwidth,height=5.4cm,keepaspectratio]{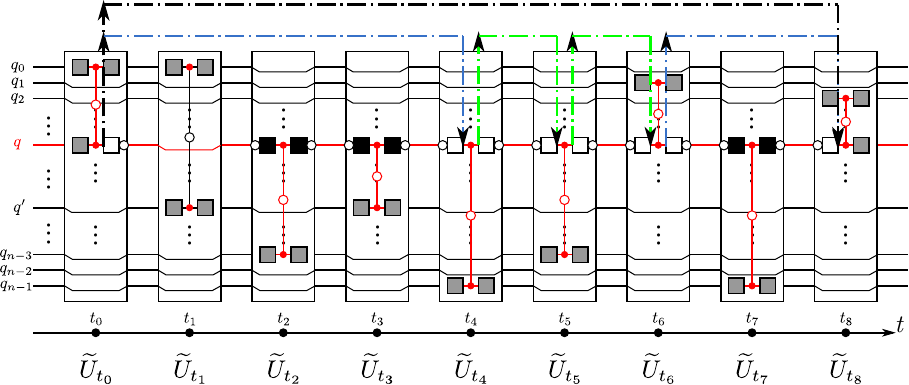}}
    \hfill{ }
    \\
    \hfill
    \subfloat[]{\includegraphics[width=0.2\textwidth,height=5.2cm,keepaspectratio]{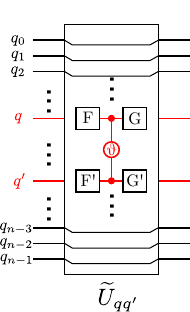}}
    \hfill{ }
    \subfloat[]{\includegraphics[width=0.4\textwidth,height=5.2cm,keepaspectratio]{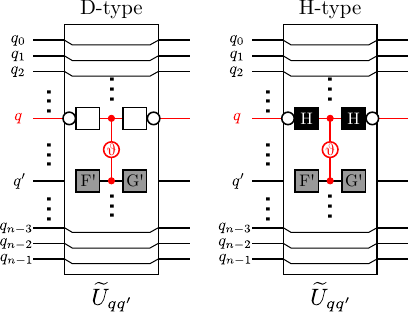}}
    \hfill{ }
    \caption{Identification of distributable packets: white circles represent distributable single-qubit gates, white squares represent identity gates, black square represents Hadamard gates, and gray square presents $q$-irrelevant gates.
    (a) A hopping embedding (black dot-dashed) decomposed into indecomposable hopping embeddings (blue dot-dashed) and neighbouring embeddings (green dash-dot).
    (b) A two-qubit block.
    (c) The D-type and H-type blocks.
    }
    \label{fig:pack_identify}
  \end{figure*}

  As it is shown in Fig. \ref{fig:pack_identify} (a), after the control-phase conversion , a circuit can be decomposed into a sequence of two-qubit blocks,
  \begin{equation}
  \label{eq:2q-blocks}
    U = \prod_{t} \widetilde{U}_{t},
  \end{equation}
  where the elemental two-qubit block $\widetilde{U}_{t}$ is shown in Fig. \ref{fig:pack_identify} (b) and given by
  \begin{equation}
  \label{eq:2-qubit_block}
    \widetilde{U}_{qq'} =
    (G_{q}\otimes G'_{q'})\;
    C_{V}(\theta)\;
    (F_{q}\otimes F'_{q'}).
  \end{equation}
  In Fig. \ref{fig:pack_identify} (a), we use the symbols white circles for single-qubit distributable gates, white squares for identity gates, black squares for Hadamard gates, gray squares for irrelevant gates.

  The control-phase nodes in Fig. \ref{fig:pack_identify} (a) are the nodes that we want to pack into packing processes through the embedding of the unitaries between them.
  If each two-qubit block $\widetilde{U}_{t}$ in Eq. \eqref{eq:2q-blocks} is $q$-rooted embeddable, then the total unitary $U$ is also $q$-rooted embeddable.
  One can employ this sufficient condition to determine the $q$-rooted embeddability of $U$ and identify the distributable packets in a circuit.
  Note that we do not exclude the possible embeddability of $U$ that possesses a non-embeddable block $\widetilde{U}_{t}$.

\begin{figure*}[t]
    \centering
    \hfill
    \subfloat[]{\includegraphics[width=0.45\textwidth,height=4.2cm,keepaspectratio]{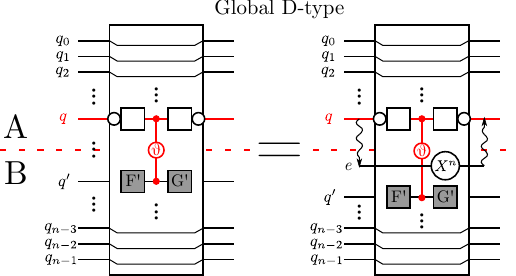}}
    \hfill{ }
    \subfloat[]{\includegraphics[width=0.45\textwidth,height=4.2cm,keepaspectratio]{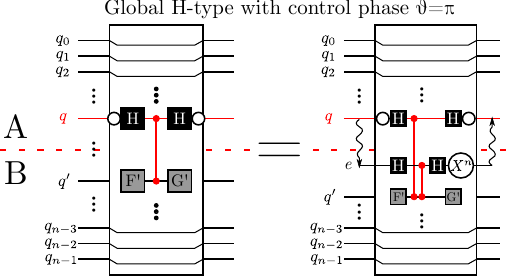}}
    \hfill{ }
    \\
    \hfill{ }
    \subfloat[]{\includegraphics[width=0.45\textwidth,height=4.2cm,keepaspectratio]{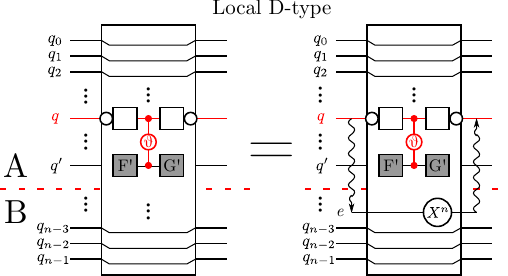}}
    \hfill{ }
    \subfloat[]{\includegraphics[width=0.45\textwidth,height=4.2cm,keepaspectratio]{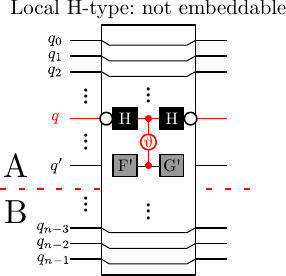}}
    \hfill{ }
    \caption{Embedding rules of two-qubit embedding units in Corollary \ref{coro:embed_rules_2q_block}.
    (a) Global D-type.
    (b) Global H-type with phase $\theta = \pi$.
    (c) Local D-type.
    (d) Local H-type (non-embeddable).}
    \label{fig:embedding_rules}
  \end{figure*}

  After the control-phase conversion, a quantum circuit consisting of one-qubit and two-qubit gates can be decomposed into two types of embedding units, which are shown in Fig. \ref{fig:pack_identify} (c).
  The embedding rules for such a quantum circuit can be summarized as the embedding rules for two-qubit blocks $\widetilde{U}_{qq'}$ as follows.
  \begin{corollary}[{Embedding rules for $2$-qubit blocks}]
  \label{coro:embed_rules_2q_block}
    A two-qubit block $\widetilde{U}_{qq'}$ in Eq. \eqref{eq:2-qubit_block} is $q$-rooted embeddable in the following two cases
    \begin{enumerate}
      \item \label{itm:embed_rule_1} (D-type, Fig. \ref{fig:embedding_rules} (a,b)) $F_{q}$ and $G_{q}$ are distributable, which means diagonal or antidiagoanl. The embedding rule is
          \begin{equation}
            \mathcal{B}_{q}(\widetilde{U}_{qq'})
            =
            \widetilde{U}_{qq'}X_{e}^{n},
          \end{equation}
          where $n$ is the number of anti-diagonal gates in $\{G_{q},F_{q}\}$.
      \item \label{itm:embed_rule_2} (Global $H$-type CZ, Fig. \ref{fig:embedding_rules} (c))
          The control phase gate $C_{V}(\theta)$  is global with a phase of $\theta=\pi$.
          The single-qubit gates $F_{q}$ and $G_{q}$ can be decomposed as
          \begin{align}
            F_{q} = H_{q}D_{q},
            \;\;
            G_{q} = \widetilde{D}_{q}H_{q},
          \end{align}
          where $D_{q}$ and $\widetilde{D}_{q}$ are diagonal or anti-diagonal. The embedding rule is
          \begin{align}
          \label{eq:emb_rule_ctrl-antidiag}
            \mathcal{B}_{q}(\widetilde{U}_{qq'})
            & =
            (G'_{q'}C_{q',X_{e}} X_{e}^{n} G_{q'}^{'\dagger}) \; \widetilde{U}_{qq'}
            \\
            & =
            \widetilde{U}_{qq'} \; (F_{q'}^{'\dagger}C_{q',X_{e}}X_{e}^{n} F'_{q'}),
          \end{align}
          where $n$ is the number of anti-diagonal gates in $\{D_{q},\widetilde{D}_{q}\}$.
    \end{enumerate}
    The block $\widetilde{U}_{qq'}$ is not embeddable in the following case,
    \begin{enumerate}[resume]
      \item \label{itm:embed_rule_3} (Local $H$-type, Fig. \ref{fig:embedding_rules} (d)) $\widetilde{U}_{qq'}$ is local $H$-type.
    \end{enumerate}
  \begin{proof}
    For a D-type block, $\widetilde{U}_{qq'}$ is distributable on $q$, and has embedding rules in Eq. \eqref{eq:emb_rule_diag} and \eqref{eq:emb_rule_anti-diag} according to Theorem \ref{theorem:distr_and_embed}.
    For a global H-type CZ block, one can derive the primitive embedding rule through

    \begin{align}
      C_{q,X_{e}}\widetilde{U}_{qq'}C_{q,X_{e}}
      & =
      (G'_{q'}C_{q',X_{e}}X_{e}^{k} G_{q'}^{'\dagger}) \; \widetilde{U}_{qq'}
      \\
      & =
      \widetilde{U}_{qq'} \; (F_{q'}^{'\dagger}C_{q',X_{e}}X_{e}^{k} F'_{q'}).
    \end{align}
    For a local H-type block, it is not distributable according to Theorem \ref{theorem:EJPP_distr}. As a result of Theorem \ref{theorem:distr_and_embed}, a $q$-rooted non-distributable local unitary can never be $q$-rooted embeddable.
  \end{proof}
  \end{corollary}%

  Now, we have the embedding rules for single-qubit gates $\mathbb{B}_{1}$ (Theorem \ref{theorem:distr_and_embed}) and two-qubit gates $\mathbb{B}_{2}$ (Corollary \ref{coro:embed_rules_2q_block}), which form a set of embedding rules $\mathbb{B}$ for the packing of gate nodes of control-phase gates,
  \begin{equation}
  \label{eq:emb_rules_B}
    \mathbb{B} = \{\text{$\mathbb{B}_{1}$:= Theorem \ref{theorem:distr_and_embed}, $\mathbb{B}_{2}$:=Corollary \ref{coro:embed_rules_2q_block}}\}.
  \end{equation}%
  Employing the set of embedding rules $\mathbb{B}$, one can identify the set of distributable $\mathbb{B}$-packet $\mathbb{V}_{\mathbb{B}}$ of a quantum circuit.

  As it is shown Fig. \ref{fig:pack_identify} (a), to determine the $\mathbb{B}$-packability of the nodes $(q,t_{0})$ and $(q,t_{8})$, one needs to determine the $q$-rooted embeddability of the unitary between them
  \begin{equation}
  \label{eq:packing_identification_eg}
    W_{t_{8};t_{0}} =
    F_{q,t_{8}}
    \widetilde{U}_{t_{7}}\widetilde{U}_{t_{6}}\widetilde{U}_{t_{5}}
    \widetilde{U}_{t_{4}}\widetilde{U}_{t_{3}}\widetilde{U}_{t_{2}}\widetilde{U}_{t_{1}}
    G_{q,t_{0}}.
  \end{equation}
  In this example, the embedding of $W_{t_{8};t_{0}}$ can be decomposed into four indecomposable embeddings $W_{i=1,...,4}$ between control-phase gates,
  \begin{equation}
    W_{t_{8};t_{0}} =
    W_{4}\,C_{q,V}^{(t_{6})}\,
    W_{3}\,C_{q,V}^{(t_{5})}\,
    W_{2}\,C_{q,V}^{(t_{4})}\,
    W_{1}.
  \end{equation}
  where
  \begin{align}
    W_{1} & =
    F_{q,t_{4}}
    \widetilde{U}_{t_{3}}\widetilde{U}_{t_{2}}\widetilde{U}_{t_{1}}
    G_{q,t_{0}},
    \\
    W_{2} & = F_{q,t_{5}}G_{q,t_{4}},
    \\
    W_{3} & = F_{q,t_{6}}G_{q,t_{5}},
    \\
    W_{4} & =
    F_{q,t_{8}}
    \widetilde{U}_{t_{7}}
    G_{q,t_{6}}.
  \end{align}

  The embeddings of $W_{2}$ and $W_{3}$ (green dot-dashed lines) do not include any embedding unit.
  Their embeddability can be easily determined with the embedding rules $\mathbb{B}_{1}$ for single qubit gates (Theorem \ref{theorem:distr_and_embed}).
  These two embeddings do not contain embedding units.
  They identify the packet of neighbouring control-phase nodes $\mathcal{V}_{q,\{t_{4},t_{5}\}}$ and $\mathcal{V}_{q,\{t_{5},t_{6}\}}$, respectively.
  We call such an embedding a \emph{neighbouring embedding}, and its corresponding distributable packet a \emph{neighbouring distributable packet}.

  On the other hand, the embeddings of $W_{1}$ and $W_{4}$ (blue dot-dashed lines) contain embedding units, which allow the packing of remote control-phase nodes $\mathcal{V}_{q,\{t_{0},t_{4}\}}$ and $\mathcal{V}_{q,\{t_{6},t_{8}\}}$, respectively.
  We call such an embedding a \emph{hopping embedding}, and its corresponding distributable packet a \emph{hopping distributable packet}.
  One need the $\mathbb{B}_{2}$ embedding rules (Corollary \ref{coro:embed_rules_2q_block}) to identify hopping embeddings.

  The embedding of $W_{t_{8};t_{0}}$ in Eq. \eqref{eq:packing_identification_eg} is also a hopping embedding, which forms the hopping distributable packet $\mathcal{V}_{q,\{t_{0},t_{8}\}}$.
  However, according to Definition \ref{def:indecomp_embedding}, $W_{t_{8};t_{0}}$ is decomposable.
  In this example, one can see that the indecomposable embeddings $\{W_{1},...,W_{4}\}$ lead to the largest distributable packet $\mathcal{V}_{q,\{t_{0}, t_{4},t_{5},t_{6}, t_{8}\}}$, which is merged according to Lemma \ref{lemma:merging_of_packing},
  \begin{align}
    \mathcal{V}_{q,\{t_{0}, t_{4},t_{5},t_{6}, t_{8}\}} = & \mathcal{V}_{q,\{t_{0},t_{4}\}}
    \cup_{D}
    \mathcal{V}_{q,\{t_{4},t_{5}\}}
    \nonumber \\ &
    \cup_{D}
    \mathcal{V}_{q,\{t_{5},t_{6}\}}
    \cup_{D}
    \mathcal{V}_{q,\{t_{6},t_{8}\}}.
  \end{align}
  It is therefore necessary to identify all indecomposable embeddings of a quantum circuit to fully explore its packability.
  The identification of distributable packets of a circuit is therefore equivalent to the identification of all the indecomposable embeddings.

  For neighbouring embeddings, the identification according to $\mathbb{B}_{1}$ (Theorem \ref{theorem:distr_and_embed}) always returns indecomposable embeddings.
  For indecomposable hopping embeddings constructed with the embedding units in Corollary \ref{coro:embed_rules_2q_block}, one can search them with the following condition.
  \begin{lemma}[{Indecomposable hopping embedding}]
  \label{lemma:indecomp_hopping_emb}
    Two distributable nodes $\{(q,t_{i}),(q,t_{j})\}$ with $i\le j-2$ form an indecomposable hopping distributable $\mathbb{B}$-packet $\mathcal{V}_{q,\{t_{i},t_{j}\}}$, if and only if the unitary $W_{t_{j};t_{i}}$ between $(q, t_{i})$ and $(q,t_{j})$ can be decomposed as
    \begin{equation}
    \label{eq:uninsertable_pSet}
      W_{t_{j};t_{i}}
      =
      D'_{q,t_{j}}
      \left(\prod_{k:i<k<j}  \widetilde{U}_{t_{k}}\right)
      D_{q,t_{i}},
    \end{equation}
    where $D'_{q,t_{j}}$ and $D_{q,t_{i}}$ are diagonal or anti-diagonal, and each embedding unit $\widetilde{U}_{t_{k}}$ is global H-type with the control phase $\theta = \pi$.
  \begin{proof}
    See Appendix \ref{proof:indecomp_hopping_emb}.
  \end{proof}
  \end{lemma}
  This lemma implies that a control-phase gate, which is local or has a control phase $\theta \neq \pi$, cannot be a part of indecomposable hopping embedding.
  \begin{corollary}
  \label{coro:local_2q_gate_division}
    Let $\mathcal{V}_{q,\{t_{i},t_{j}\}}$ be an indecomposable distributable packet.
    All the control-phase gates $C_{V}^{q,q_{t}}(\theta_{t})$ with $t_{i}<t<t_{j}$ between the nodes $(q,t_{i})$ and $(q,t_{j})$ must be global and $\theta_{t} = \pi$.
  \begin{proof}
    This is a direct result of Lemma \ref{lemma:indecomp_hopping_emb}.
  \end{proof}
  \end{corollary}

  Now, we have all the essential tools to develop an identification algorithm for distributable packets, which is summarized in Algorithm \ref{algo:dPacket_identify}.
  To implement this algorithm, one needs two functions, \mbox{\textit{neighbouring()}} and \mbox{\textit{hopping()}}, to determine the $q$-rooted neighbouring and hopping embeddings, respectively. 
  The neighbouring embeddings can be determined according to Algorithm \ref{algo:neighbouring_emb} through the embedding rules $\mathbb{B}_{1}$ in Theorem \ref{theorem:distr_and_embed}, which leads to a set of distributable $\mathbb{B}_{1}$-packets $\mathbb{V}_{\mathbb{B}_{1}}$.
  Meanwhile the indecomposable hopping embeddings can be determined according to Algorithm \ref{algo:hopping_emb} through the embedding rules $\mathbb{B}_{2}$ in Corollary \ref{coro:embed_rules_2q_block}.
  The indecomposable hopping embeddings determined by Algorithm \ref{algo:hopping_emb} remotely connect two neighbouring distributable packets in $\mathbb{V}_{\mathbb{B}_{1}}$.
  One can then merge two remote neighbouring distributable packets through the indecomposable hopping embeddings in $\mathbb{B}_{2}$ and obtain the ultimate set of distributable packets $\mathbb{V}_{\mathbb{B}}$.
  The complexity of the identification of distributable packets is estimated to be $\mathcal{O}(nd^{2})$, where $n$ and $d$ are the total qubit number and depth of a circuit (see Appendix \ref{proof:id_dPacket_complexity} for an explanation).

  \begin{algorithm*}[t]
  \setcounter{algocf}{\value{theorem}}\addtocounter{theorem}{1}
  \caption{Identification of distributable packets}
  \label{algo:dPacket_identify}
  \SetKwProg{Fn}{Function}{:}{}
    $Q \gets $ initialize a circuit with the control-phase conversion (Eq. \eqref{eq:CV-conversion})\;

    $\mathbb{V}_{0} \gets $ the set of trivial distributable packets of the circuit $Q$\;
    \begin{equation}
      \mathbb{V}_{0}:= \{\mathcal{V}_{q,t} = (q,t,\theta): (q,t) \text{ is a node of a control-phase gate } C_{V}(\theta).\}
    \end{equation}

    \ForEach{subsequential nodes $\{(q,t_{i},\theta_{i}),(q,t_{i+1},\theta_{i+1})\} \subseteq \mathbb{V}_{0}$ }{
      $\{W_{t_{i+1;t_{i}}}\}_{i} \gets $ conversion of single-qubit gates $W_{t_{i+1};t_{i}}$ \;
      \begin{equation}
      \label{eq:single-qubit_gate_conversion}
        W_{t_{i+1};t_{i}} =
        \left\{
          \begin{array}{ll}
            D, & \hbox{$D$ diagonal or anti-diagonal;} \\
            V(\gamma)\,H\, V(\alpha), & \hbox{$Z$-rotation-equivalent to $H$;} \\
            V(\gamma)\,H\, V(\beta)\, H\, V(\alpha), & \hbox{else, Euler decomposition.}
          \end{array}
        \right.\;
      \end{equation}
    \label{algo_line:dPacket_id_1qubut_conversion}
    }
    \ForEach{qubit $q$ in $Q$}{%
      \SetKwFunction{ngbr}{neighbouring}
      \SetKwFunction{hop}{hopping}
    \label{algo_line:dPacket_id_neighbour}
      $\{\mathbb{B}_{1}^{(q)}, \mathbb{V}_{\mathbb{B}_{1}}^{(q)}\}
      \gets $ \ngbr{$q$, $\mathbb{V}_{0}$, $\{W_{t_{i+1;t_{i}}}\}_{i}$}: identification of neighbouring distributable packets\;
    \label{algo_line:dPacket_id_hopping}
      $\{\mathbb{B}_{2}^{(q)}, \mathbb{H}^{(q)}\}
      \gets $ \hop{$q$, $\mathbb{V}_{0}$, $\{W_{t_{i+1;t_{i}}}\}_{i}$}: identification of hopping distributable packets\;

      $\mathbb{B}^{(q)} \gets \mathbb{B}^{(q)}_{1} \cup \mathbb{B}^{(q)}_{2}$\;
      $\mathbb{V}_{\mathbb{B}}^{(q)} \gets $ $\mathbb{V}_{\mathbb{B}_{1}}^{(q)}$ merged by $\mathbb{H}^{(q)}$ according to Lemma \ref{lemma:merging_of_packing}\;
    }
  \KwResult{$\{\mathbb{B}, \mathbb{B}_{1}, \mathbb{B}_{2}, \mathbb{V}_{\mathbb{B}}, \mathbb{V}_{\mathbb{B}_{1}}\}$}
  \end{algorithm*}

  \begin{algorithm*}[t]
  \setcounter{algocf}{\value{theorem}}\addtocounter{theorem}{1}
  \caption{Identification of neighbouring distributable packets on $q$}
  \label{algo:neighbouring_emb}
  \SetKwProg{Fn}{Function}{:}{}
  \setcounter{AlgoLine}{0}
  \Fn{\ngbr{$q$, $\mathbb{V}_{0}$, $\{W_{t_{i+1},t_{i}}\}_{i}$}}{
    $\mathbb{D}^{(q)} \gets \left\{\mathcal{V}_{q;t_{i},t_{i+1}} = \{(q,t_{i},\theta_{i}),(q,t_{i+1},\theta_{i+1})\}:  W_{t_{i+1};t_{i}} = D \right\}$: the set of neighbouring distributable packets\;

    $\mathbb{B}^{(q)}_{1} \gets \{(\mathcal{V}_{q;t_{i},t_{i+1}},\mathcal{B}_{q}(W_{t_{i+1};t_{i}})): W_{t_{i+1};t_{i}} = D\}$: the set of neighbouring embedding kernels determined by Theorem \ref{theorem:distr_and_embed}\;

    $\mathbb{V}^{(q)}_{\mathbb{B}_{1}} \gets$ the set of distributable packets merged from $\mathbb{D}^{(q)}$ according to Lemma \eqref{lemma:merging_of_packing}\;
    \KwRet{$\{\mathbb{B}^{(q)}_{1}, \mathbb{V}^{(q)}_{\mathbb{B}_{1}}\}$}\;
  }
  \end{algorithm*}

  \begin{algorithm*}[p]
  {\small
  \setcounter{algocf}{\value{theorem}}\addtocounter{theorem}{1}
  \caption{Identification of hopping distributable packets on $q$}
  \label{algo:hopping_emb}
  \SetKwProg{Fn}{Function}{:}{}
  \setcounter{AlgoLine}{0}
  \Fn{\hop{$q$, $\{W_{t_{i+1},t_{i}}\}_{i}$}}{
    $\{\mathbb{B}_{2}^{(q)}, \mathbb{H}^{(q)}\}\gets\{\emptyset,\emptyset\}$\;

    $\mathbb{S}^{(q)} \gets
      \left\{
        \{(q,t_{i},\theta_{i}),(q,t_{i+1},\theta_{j})\}:
        W_{t_{i+1};t_{i}}
        =
        V(\gamma'_{i})\,H\, V(\alpha'_{i})
        , \text{and }
        \theta_{i+1} = \pi
      \right\}
    $: the set of single-qubit starting-gate candidates.
    \label{algo_line:group_start}

    $\mathbb{E}^{(q)} \gets
      \left\{
        \{(q,t_{i},\theta_{i}),(q,t_{i+1},\theta_{j})\}:
        W_{t_{i+1};t_{i}}
        =
        V(\gamma'_{i})\,H\, V(\alpha'_{i})
        , \text{and }
        \theta_{i} = \pi
      \right\}
    $: the set of single-qubit ending-gate candidates.

    $\mathbb{M}^{(q)} \gets
      \left\{
        \{(q,t_{i},\theta_{i}),(q,t_{i+1},\theta_{i+1})\}:
        W_{t_{i+1};t_{i}}
        =
        V(\gamma_{i})\,H\, V(\beta_{i})\, H\, V(\alpha_{i})
        , \text{and }
        \theta_{i} = \theta_{i+1} = \pi
      \right\}$: the set of single-qubit intermediate-gate candidates.\;

    $\mathbb{D}^{(q)} \gets
      \left\{
        \{(q,t_{i},\theta_{i}),(q,t_{i+1},\theta_{i+1})\}:
        W_{t_{i+1};t_{i}}
        =
        D, \text{and }
        \theta_{i} = \theta_{i+1} = \pi
      \right\}$: the set of single-qubit intermediate-gate candidates, which are distributable.\;
    \label{algo_line:group_end}

    $\{\mathbb{S}_{k}^{(q)}, \mathbb{E}_{k}^{(q)}\}_{k} \gets $ division of $\mathbb{S}^{(q)}$ and $\mathbb{E}^{(q)}$ by local control-phase gates and global control-phase with $\theta\neq \pi$ (Corollary \ref{coro:local_2q_gate_division}).
    Let $T(\mathbb{P})$ be the set of node depths in $\mathbb{P}$, $T(\mathbb{P}):=\{t: \exists p\in\mathbb{P}, s.t. (q,t,\theta_{t})\in p\}$
    \begin{align}
      \left\{
        \mathbb{S}^{(q)}_{k}\subseteq\mathbb{S}:
        C_{V}^{(q,q'_{t})}(\theta_{t})
        \text{ is global and $\theta_{t}=\pi$ for all }
        \min T(\mathbb{S}^{(q)}_{k}) <t< \max T(\mathbb{S}^{(q)}_{k})
      \right\}_{k}.
    \end{align}
    \begin{align}
      \left\{
        \mathbb{E}^{(q)}_{k}\subseteq\mathbb{E}:
        C_{V}^{(q,q'_{t})}(\theta_{t})
        \text{ is global and $\theta_{t}=\pi$ for all }
        \min T(\mathbb{E}^{(q)}_{k}) <t< \max T(\mathbb{E}^{(q)}_{k})
      \right\}_{k}.
    \end{align}
    \label{algo_line:division_start}

    \ForEach{division $k$}{
      $\mathbb{P}_{k}^{(q)} \gets $ get the set of potential indecomposable distributable packets%
      \begin{align}
        \mathbb{P}_{k}^{(q)}
        =
        \Big\{
        \{(q,t_{i},\theta_{i}),(q,t_{j},\theta_{j})\}:
        &
        \{(q,t_{i},\theta_{i}), (q,t_{i+1},\theta_{i+1})\}\in\mathbb{S}_{k}^{(q)},
        \{(q,t_{j-1},\theta_{j-1}), (q,t_{j},\theta_{j})\}\in\mathbb{E}_{k}^{(q)}.
        \Big\}
      \end{align}%
      \label{algo_line:division_end}

      \ForEach{$\{(q,t_{i}),(q,t_{j})\}\in \mathbb{P}_{k}^{(q)}$
      \label{algo_line:indecomp_hop_start}
      }{
        \tcc{Check the $q$-rooted embeddability of $W_{t_{j};t_{i}}$ according to Lemma \ref{lemma:indecomp_hopping_emb}.}
        \BlankLine
        \label{algo_line:hopping_emb_check}
        $\widetilde{\mathbb{M}} \gets \emptyset$
        \ForEach{$k \in [i+1,j-2]$}{
          \Switch{$\{(q,t_{k}),(q,t_{k+1})\}$}{
            \label{algo_line:M-type_conv_start}
            \Case{$\{(q,t_{k}),(q,t_{k+1})\}\in\mathbb{S}^{(q)}$ or $\{(q,t_{k}),(q,t_{k+1})\}\in\mathbb{E}^{(q)}$}{
              Convert the $\mathbb{S}$-type or $\mathbb{E}$-type gates 
              into $\mathbb{M}$-type gates through
              \begin{equation}
              \label{eq:conversion_1H_to_2H}
                W_{t_{k+1};t_{k}}
                =
                V(\gamma'_{k})\,H\,V(\alpha'_{k})
                =
                V(\gamma_{k})\,H\,V(\beta_{k})\,H\, V(\alpha_{k}),
              \end{equation}
              where $(\alpha_{k},\beta_{k},\gamma_{k}) =
              (\alpha'_{k} + \frac{1}{2}\pi, \frac{1}{2}\pi, \gamma'_{k} + \frac{1}{2}\pi)$\;
            }
            \Case{$\{(q,t_{k}),(q,t_{k+1})\}\in\mathbb{D}^{(q)}$}{
              Convert $\mathbb{D}$-type gates 
              into $\mathbb{M}$-type gates through
              \begin{equation}\label{eq:conversion_D_to_2H}
                W_{t_{k+1};t_{k}}
                =
                H\,V(n\pi)\,H\, V(\alpha'_{k})
                =
                V(\gamma_{k})\,H\,V(\beta_{k})\,H\, V(\alpha_{k}),
              \end{equation}
              where $(\alpha_{k}, \beta_{k}, \gamma_{k}) = (\alpha'_{k}, n\pi, 0)$ or
              $(0, n\pi, (-1)^{n}\alpha'_{k})$\;
              $\widetilde{\mathbb{M}} \gets \widetilde{\mathbb{M}}\cup\{(\alpha_{k},\gamma_{k})\}$\;
            }
            \Case{$\{(q,t_{k}),(q,t_{k+1})\}\in\mathbb{M}^{(q)}$}{
              $\widetilde{\mathbb{M}} \gets \widetilde{\mathbb{M}}\cup\{(\alpha_{k},\gamma_{k})\}$\;
            }
          }
        \label{algo_line:M-type_conv_end}}
        \tcc{Determine the $q$-embeddability of $W_{t_{j};t_{i}}$ with the following necessary and sufficient condition. For the proof, see Appendix \ref{proof:phase_cond_for_hopping_emb}.}
        \If{\label{algo_line:hopping_emb_phase_cond}%
        $\gamma_{k} + \alpha_{k+1} = n\pi$, for all $k\in[i,j-1]$ and $(\alpha_{k},\gamma_{k})\in\widetilde{\mathbb{M}}$}{
          $\mathbb{B}_{2} \gets \mathbb{B}_{2}\cup\{(\mathcal{V}_{q;t_{i},t_{j}}, \mathcal{B}_{q}(W_{t_{j};t_{i}}))\}$\;
          $\mathbb{H}^{(q)} \gets \mathbb{H}^{(q)}\cup\{\mathcal{V}_{q;t_{i},t_{j}}\}$\;
        }
      \label{algo_line:indecomp_hop_end}
      }
    }
    \KwRet{$\{\mathbb{B}_{2}^{(q)},\mathbb{H}^{(q)}\}$}\;
  }
  }
  \end{algorithm*}

\begin{figure*}[p]
  \centering
  \subfloat[]{\includegraphics[width=0.9\textwidth]{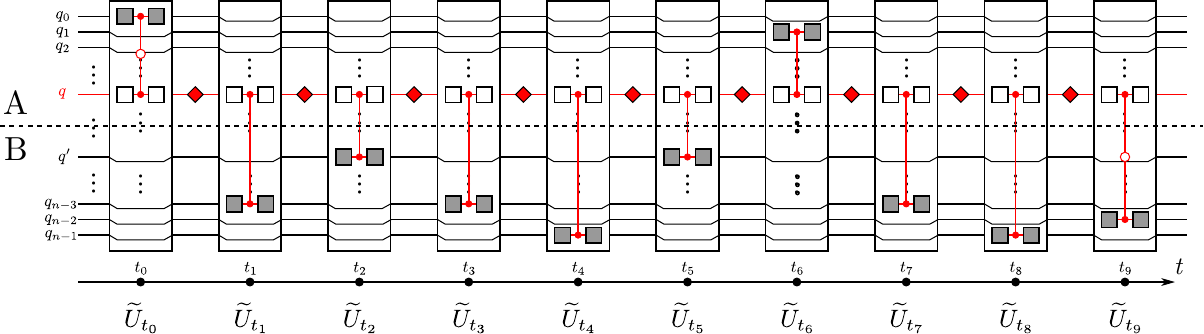}}
  \\
  \subfloat[]{\includegraphics[width=0.9\textwidth]{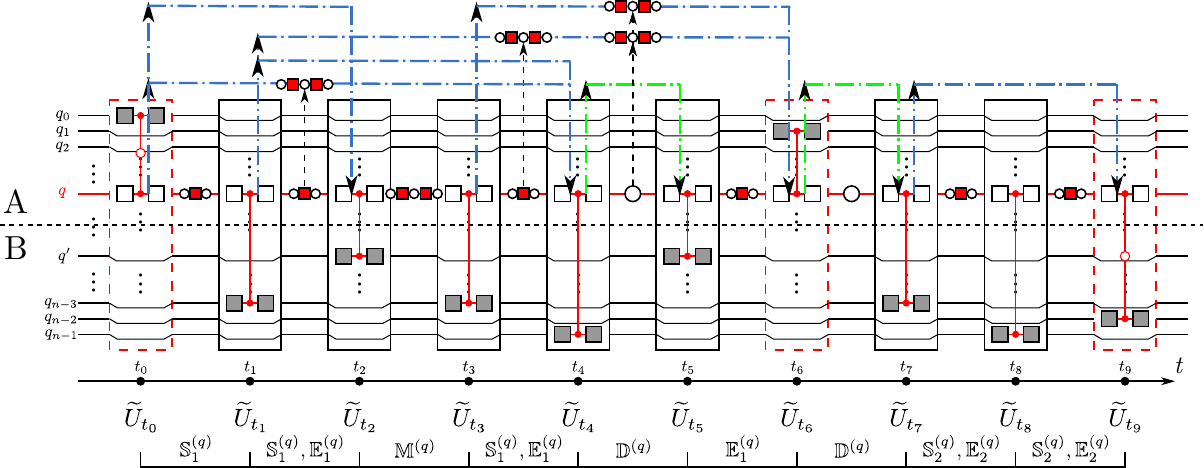}}
  \\
  \subfloat[]{\includegraphics[height=4.5cm,keepaspectratio]{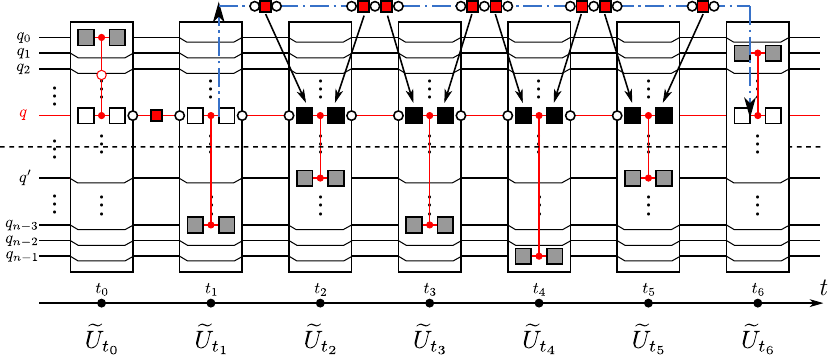}}
  \\
  \subfloat[]{\includegraphics[height=4.5cm,keepaspectratio]{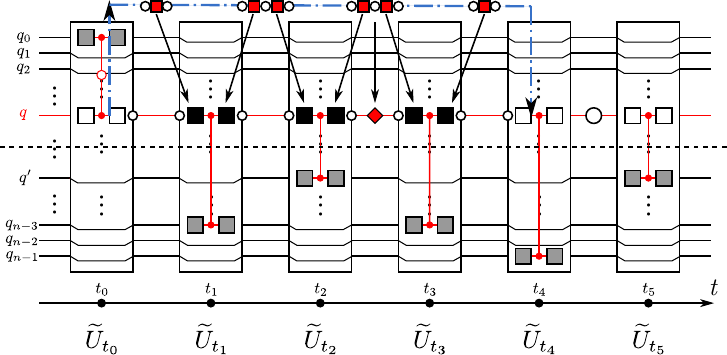}}
  \caption{Algorithm for the identification of distributable packets. (a) A circuit after control-phase conversion. The white squares are identity gates, the white circles are single-qubit distributable gates, and the red diamonds are the single-qubit non-distributable gates. (b) The circuit after the 1st step in Algorithm \ref{algo:dPacket_identify}. The red squares are Hadamard gates. (c) The determination of the packability of $\{(q,t_{1}),(q,t_{6})\}$. (d) The determination of the packability of $\{(q,t_{0}),(q,t_{4})\}$. }
  \label{fig:pSet_identify}
\end{figure*}

  In the identification algorithm for indecomposable hopping embeddings (Algorithm \ref{algo:hopping_emb}), from line \ref{algo_line:indecomp_hop_start} to \ref{algo_line:indecomp_hop_end}, the ultimate goal is to shift the Hadamard gates between two control-phase nodes
  into an embedding block through commuting them with the single-qubit distributable gates to form $H$-type embedding units, which is demonstrated by the examples in Fig. \ref{fig:pSet_identify}.
  In Fig. \ref{fig:pSet_identify} (a), it shows a quantum circuit after the control-phase conversion.
  In this example, we look at the packing on the qubit $q$. The white squares are identity gates.

  The single qubit gates (red diamonds) between the control phase nodes are converted into three types according to Eq. \eqref{eq:single-qubit_gate_conversion}.
  In Fig \ref{fig:pSet_identify} (b), these single-qubit gates are then grouped into $\mathbb{S}$ for starting-gate candidates, $\mathbb{E}$ for ending-gate candidates, $\mathbb{M}$ for intermediate-gate candidates, and $\mathbb{D}$ for distributable intermediate-gate candidates according to line \ref{algo_line:group_start} to \ref{algo_line:group_end} in Algorithm \ref{algo:hopping_emb}.
  The gates in the groups $\mathbb{S}$ and $\mathbb{E}$ are the potential starting and ending single-qubit gates of an indecomposable hopping embedding, since they can be decomposed into gates containing only one Hadamard gate.
  The gates in the group $\mathbb{M}$ can only be the intermediate single-qubit gates in an indecomposable hopping embedding, since they can only be converted into the gates containing two Hadamard gates.
  Note that all gates in $\mathbb{S}$, $\mathbb{E}$, $\mathbb{D}$ and $\mathbb{M}$ can serve as single-qubit intermediate gates in an indecomposable hopping emdedding, since they can be decomposed into a gate that contains two Hadamards.

  In Fig. \ref{fig:pSet_identify} (b), the red-dashed two-qubit blocks $\widetilde{U}_{t_{0}}$, $\widetilde{U}_{t_{6}}$, and $\widetilde{U}_{t_{9}}$ contain a local control-phase gate or a global control-phase gate with the phase $\theta\neq\pi$, which can never be involved in an indecomposable hopping embedding according to Lemma \ref{lemma:indecomp_hopping_emb}. One can therefore divide the starting-gate and ending-gate groups into different divisions, such as $\mathbb{S}^{(q)}_{1,2}$ and $\mathbb{E}^{(q)}_{1,2}$ .
  In each division, one can then construct a set of candidate $q$-rooted packets $\mathbb{P}^{(q)}_{k}$ from $\mathbb{S}^{(q)}_{k}$ and $\mathbb{E}^{(q)}_{k}$.
  These steps are summarized on line \ref{algo_line:division_start} to \ref{algo_line:division_end} in Algorithm \ref{algo:hopping_emb}.

  For each candidate packet $\{(q,t_{i}),(q,t_{j})\}\in \mathbb{P}_{k}^{(q)}$, one can then check the $q$-rooted embeddability following line \ref{algo_line:indecomp_hop_start} to \ref{algo_line:indecomp_hop_end}.
  For the example of the candidate packet $\{(q,t_{1}),(q,t_{6})\}$ in Fig. \ref{fig:pSet_identify} (b), the single-qubit gate $W_{t_{4};t_{3}}$ is an $\mathbb{S}$-type and $\mathbb{E}$-type gate, while $W_{t_{5};t_{4}}$ is $\mathbb{D}$-type.
  One first converts these gates into an $\mathbb{M}$-type gate following line \ref{algo_line:M-type_conv_start} to \ref{algo_line:M-type_conv_end}.
  In the next step, one checks the embeddability of $W_{t_{6};t_{1}}$ by shifting the red squares into the embedding units according to line \ref{algo_line:hopping_emb_phase_cond} to \ref{algo_line:indecomp_hop_end}, which is as shown in Fig. \ref{fig:pSet_identify} (c),
  If there are no non-embeddable gates remaining outside the embedding units, then $\{(q,t_{1}), (q,t_{6})\}$ forms a $\mathbb{B}$-distributable packet $\mathcal{V}_{q,\{t_{1},t_{6}\}}$.
  Explicitly, the condition for a successful construction of sequential embedding units is given on line \ref{algo_line:hopping_emb_phase_cond}.
  A counterexample of embeddability for the candidate $\{(q,t_{0}),(q,t_{4})\}$ is shown in Fig. \ref{fig:pSet_identify} (d), in which a non-embeddable gate (red diamond) is lying between the embedding units $\widetilde{U}_{t_{2}}$ and $\widetilde{U}_{t_{3}}$. Accordingly, $\{(q,t_{0}),(q,t_{4})\}$ is not embeddable.

  In the end, Algorithm \ref{algo:hopping_emb} returns the set of indecomposable hopping embeddings $\mathbb{B}_{2}$ and the set of their corresponding indecomposable distributable packets $\mathbb{H}$.

\subsection{Identification of packing and conflict edges}
\label{sec:identify_edges}

  The packing edges can be straightforwardly assigned to control-phase gates.
  A global control-phase gate $C_{V}^{(q,q')}(\theta_{t})$ acting on $\{q,q'\}$ at the depth $t$ defines an undirected edge between two distributable packets $\mathcal{V}_{q,T_{\mathbb{B}}}$ and $\mathcal{V}_{q',T'_{\mathbb{B}}}$ in $\mathbb{V}_{\mathbb{B}}$, where $t\in T_{\mathbb{B}}\cap T'_{\mathbb{B}}$,
  \begin{equation}
    \mathcal{E}_{q,q';t}^{(\mathbb{B})}
    :=\{\mathcal{V}_{q,T_{\mathbb{B}}}, \mathcal{V}_{q',T'_{\mathbb{B}}}\}
    \text{ with }
    t\in T_{\mathbb{B}}\cap T'_{\mathbb{B}}.
  \end{equation}
  These edges form an edge set of the $\mathbb{B}$-packing graph,
  \begin{equation}
    G_{\mathbb{B}} = (\mathbb{V}_{\mathbb{B}}, \mathbb{E}_{\mathbb{B}})
    \text{ with }
    \mathbb{E}_{\mathbb{B}}:= \{\mathcal{E}_{q,q';t}: \exists C_{V}^{(q,q')}(\theta_{t}) \}.
  \end{equation}

  \begin{figure*}[t]
    \centering
    \subfloat[]{\includegraphics[width=0.5\textwidth]{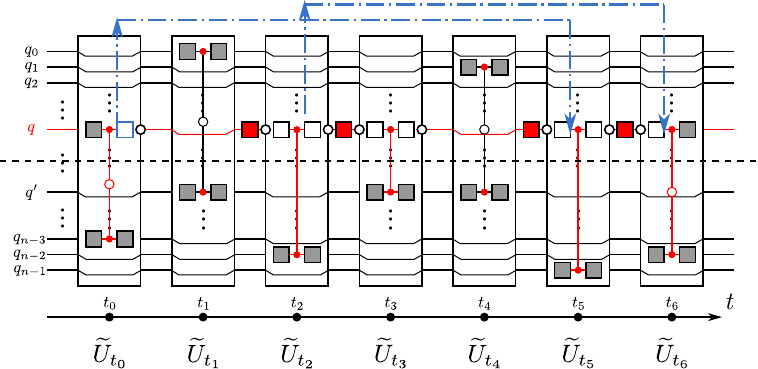}}
    \subfloat[]{\includegraphics[width=0.5\textwidth]{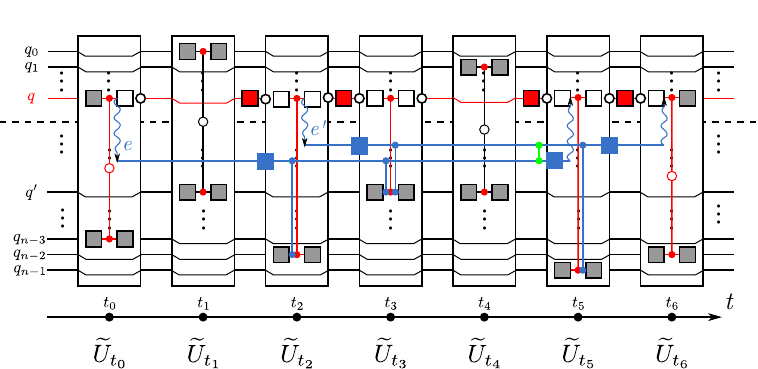}}
    \\
    \subfloat[]{\includegraphics[width=0.5\textwidth]{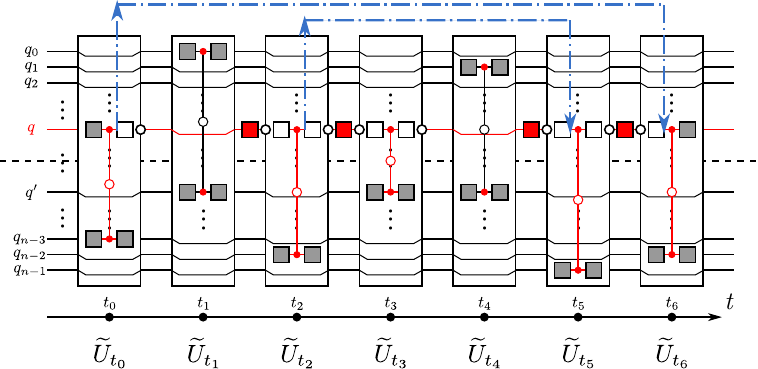}}
    \subfloat[]{\includegraphics[width=0.5\textwidth]{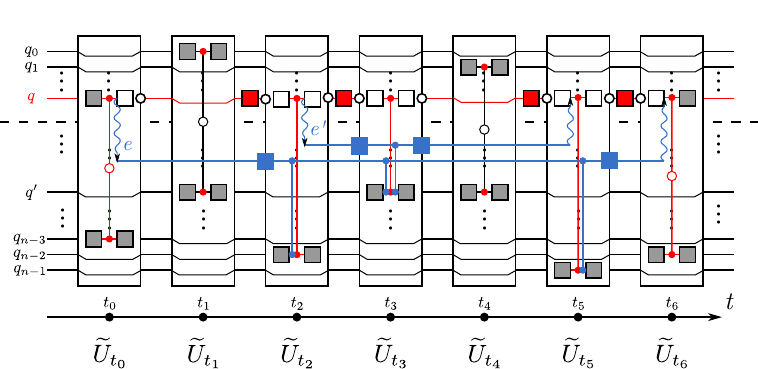}}
    \\
    \subfloat[]{\includegraphics[width=0.5\textwidth]{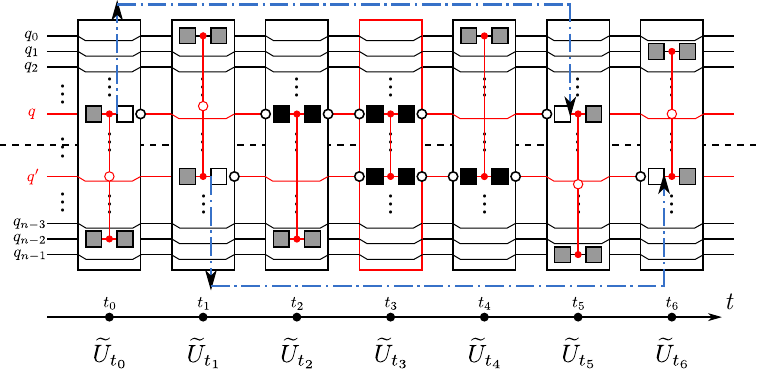}}
    \subfloat[]{\includegraphics[width=0.5\textwidth]{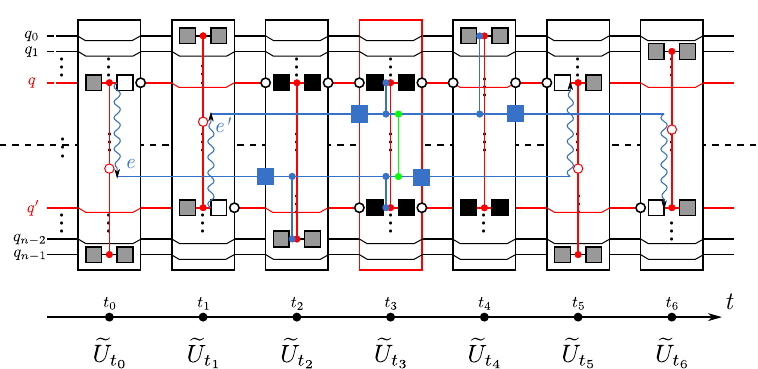}}
    \caption{Nested embeddings in elemental blocks.
    (a) Nested embeddings on the same qubit $q$.
    (b) Implementation of the embedding in (a).
    (c) Inclusive nested embeddings on the same qubit $q$.
    (d) Implementation of the embedding in (c).
    (e) Nested embeddings on two qubits $\{q,q'\}$ that belong to different local systems.
    (f) Incompatibility of the two nested embedding due to the additional global gate $C_{Z}^{(e,e')}$.
    }\label{fig:nested_embedding}
  \end{figure*}

  For the examples in Fig. \ref{fig:pSet_identify} (c,d), if $\mathcal{V}_{1}=\{(q,t_{0}), (q,t_{4})\}$ and $\mathcal{V}_{2}=\{(q,t_{1}), (q,t_{5})\}$ are both distributable $\mathbb{B}$-packets, they are nested and identified by two nested hopping embeddings.
  If the two nested hopping embeddings are incompatible, only one of them can be packed in one packing process.
  It means that one can not simultaneously pack two distributable packets identified by incompatible embeddings.
  The compatibility of two distributable packets can be determined through the following theorem.
  \begin{theorem}[Compatibility of embedding]\label{theorem:incompt_2qubit_circ}{\ }\\
    For a circuit consisting of single-qubit and two-qubit gates, two indecomposable embeddings $\mathcal{B}_{q; \tau_{1},\tau_{2}}$ and $\mathcal{B}_{q'; \tau'_{1},\tau'_{2}}$ are incompatible, if and only if there exists a global control-Z gate $C_{Z}(q,q';t)$ acting on $\{q,q'\}$ at the depth $t$ with $\tau_{1}<t<\tau_{2}$ and $\tau'_{1}<t<\tau'_{2}$.
    \begin{proof}
    See Appendix \ref{proof:incompt_2qubit_circ}
    \end{proof}
  \end{theorem}

  Fig. \ref{fig:nested_embedding} shows the three nested embeddings in a circuit consisting of single-qubit and two-qubit gates. The figures (b),(d),(f) are the implementation of the nested embeddings indicated by the blue dot-dashed lines in the figures (a),(c),(e), respectively. The nested embeddings in Fig. \ref{fig:nested_embedding} (e) are not compatible, since their implementation introduces a global control-Z gate (green).

  Theorem \ref{theorem:incompt_2qubit_circ} implies that an indecomposable neighbouring embedding is always compatible with any other indecomposable embeddings.
  The incompatibility of indecomposable embeddings exists only for the indecomposable hopping embeddings, which contain a global control-Z gate.
  Such incompatibility is intrinsic and independent from external resources.
  After the identification of the intrinsic incompatibility between two indecomposable hopping embeddings, one can represent the incompatibility associated with a global control-Z gate by a conflict edge at the level of a set of distributable packets $\mathbb{V}$.
  \begin{definition}[Intrinsic conflict edges in $\mathbb{V}$]{\ }\\
    In a set of distributable packets $\mathbb{V}$, a global control-Z gate $C_{Z}^{(q,q';t)}$ defines a conflict edge
    \begin{equation}
      \{\mathcal{V}_{q,T},\mathcal{V}_{q',T'}\},
    \end{equation}
    where $\mathcal{V}_{q,T},\mathcal{V}_{q',T'}\in\mathbb{V}$ are the distributable packets that cover the control-Z gate, namely $\min T<t<\max T$, $\min T'<t<\max T'$.
    The set of intrinsic conflict edges in $\mathbb{V}$ is denoted by
    \begin{align}
      \mathbb{C}_{in}(\mathbb{V}):=
      \{&
        \{\mathcal{V}_{q,T},\mathcal{V}_{q',T'}\}:
        \mathcal{V}_{q,T},\mathcal{V}_{q',T'}\in\mathbb{V},
      \nonumber \\
        & \exists C_{Z}^{(q,q';t)} \text{ covered by } \mathcal{V}_{q,T},\mathcal{V}_{q',T'}.
      \}
    \end{align}
  \end{definition}

  Since each indecomposable embedding in $\mathbb{B}_{2}$ corresponds to a unique distributable packet in $\mathbb{H}$, the intrinsic conflict edges between indecomposable embeddings introduced in Eq. \eqref{eq:process-conflit_embedding} are equivalent to the intrinsic conflict edges between the distributable packets in $\mathbb{H}$.
  One can therefore construct the intrinsic packet-conflict and kernel-conflict graphs as
  \begin{equation}
    C_{in}(\mathbb{V}_{\mathbb{B}})
    =
    (\mathbb{V}_{\mathbb{B}}, \mathbb{C}_{in}(\mathbb{V}_{\mathbb{B}}))
  \end{equation}
  and
  \begin{equation}
    C_{in}(\mathbb{H})
    =
    (\mathbb{H}, \mathbb{C}_{in}(\mathbb{H})),
  \end{equation}
  respectively.

  For the extrinsic kernel-conflict graphs, one will need to include the trivial distributable packets $\mathbb{V}_{0}$ in the vertices,  $\mathbb{K} = \mathbb{V}_{0}\cup\mathbb{H}$.
  The extrinsic kernel-conflict graph $C_{ex}(\mathbb{K})$ will be then constructed at the level of $\mathbb{K}$, from which one can construct the extrinsic packet-conflict graph $C_{ex}(\mathbb{V}_{\mathbb{B}})$ according to Eq. \eqref{eq:packet-conflict_edge}.

\section{\label{sec:ucc}Applications of embedding-enhanced distributed quantum computing}
\begin{figure*}[t]
  \centering
  \hfill
  \subfloat[]{\includegraphics[width=0.33\textwidth]{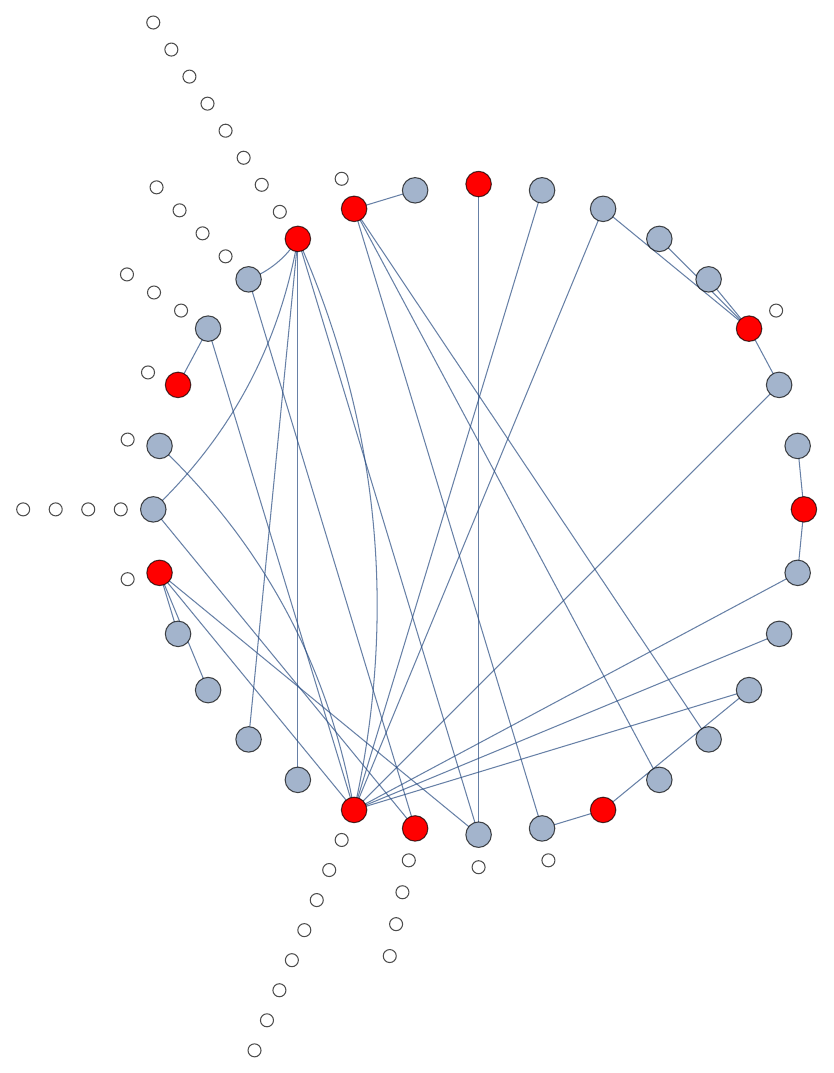}}
  \hfill
  \subfloat[]{\includegraphics[width=0.33\textwidth]{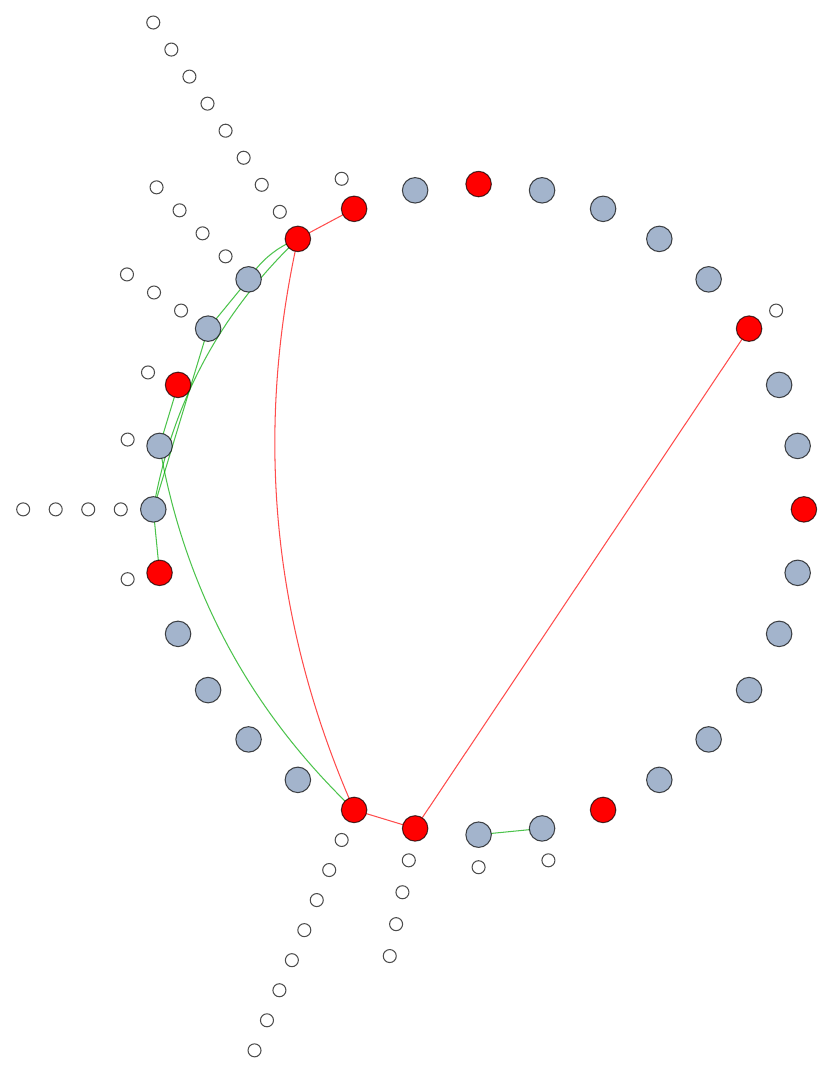}}
  \hfill
  \subfloat[]{\includegraphics[width=0.33\textwidth]{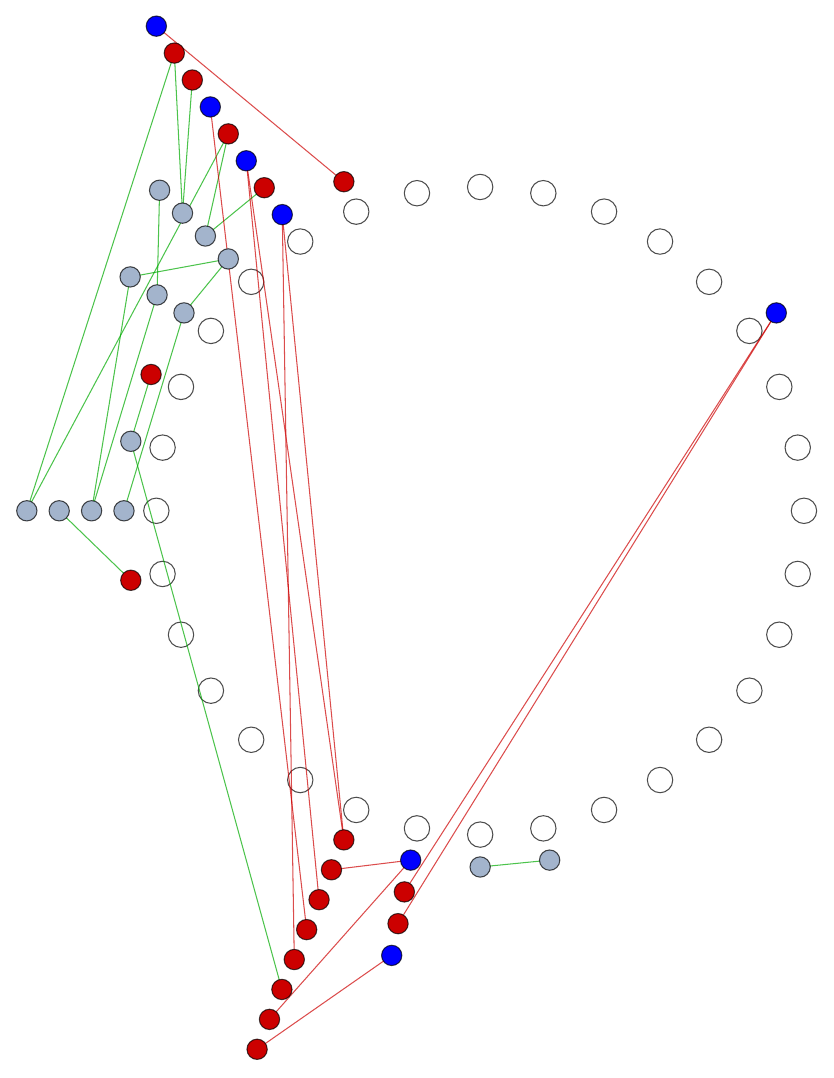}}
  \hfill{ }
  \caption{
  The distributability and embeddability structure of a $4$-qubit UCC circuit. (a) The packing graph: Each colored vertex is a distributable packet, which is formed by the merging of the neighbouring packets through indecomposable hopping packets. The indecomposable hopping packets are depicted as white vertices and aligned in a line stretching from each colored vertex.
  The edges are packing edges. The red vertices are a minimum vertex cover $\mathbb{V}_{\text{mvc}}$ of the packing graph. The order of $\mathbb{V}_{\text{mvc}}$ is $10$.
  (b) Intrinsic packet-conflict graph: the conflict graph defined at the level of distributable packets has the same vertex set as the packing graph. The conflict edges between packets are induced from the intrinsic conflict edges between kernels according to Eq. \eqref{eq:packet-conflict_edge}. The red edges are the conflict edges induced by the minimum vertex cover $\mathbb{V}_{\text{mvc}}$.
  (c) Intrinsic kernel-conflict graph defined at the level of kernels of packing processes. The colored vertices are the indecomposable hopping packets representing the kernels of hopping embeddings, while the white circles are the distributable packet formed from these embedding kernels. The red vertices $\mathbb{K}_{\text{mvc}}$ are the indecomposable hopping packets that included in the minimum vertex cover $\mathbb{V}_{\text{mvc}}$. The red edges are the conflict edges induced by $\mathbb{K}_{\text{mvc}}$. The blue vertices $\widetilde{\mathbb{K}}_{\text{mvc}}$ are a minimum vertex cover of the $\mathbb{K}_{\text{mvc}}$-induced conflict graph, which are the embeddings that needs to be removed. The order of $\widetilde{\mathbb{K}}_{\text{mvc}}$ is $7$.}
  \label{fig:UCC_appl}
\end{figure*}

In this section, we demonstrate our packing algorithms for DQC of unitary coupled-cluster (UCC) circuit \cite{TaubeBartlett2006-UCC,PeruzzoEtAlOBrien2014-VQE}, with which one implements a quantum variational eigensolver to find the eigenvalue of an observable that simulates a chemical system.
The implementation of our algorithm can be found in an open-source package pytket-dqc \footnote{https://github.com/CQCL/pytket-dqc}, which is developed for multipartite DQC based on our embedding-enhanced method \cite{MartinezEtAlDuncan2023-DQCinQNet}.

\subsection{DQC of a $4$-qubit UCC circuit}

A $4$-qubit UCC circuit after the control-phase conversion is shown in Fig. \ref{fig:ucc4q} in Appendix \ref{sec:ucc_circuit}. It contains $64$ global control-Z gates. The qubits are partitioned in two local systems $A = \{q_{0},q_{1}\}$ and $B = \{q_{2}, q_{3}\}$.
Employing Algorithm \ref{algo:dPacket_identify}, \ref{algo:neighbouring_emb} and \ref{algo:hopping_emb}, one obtains a packing graph $G_{\mathbb{B}}$ in Fig. \ref{fig:UCC_appl} (a).
The colored vertices are the $\mathbb{B}$-distributable packets, while the white vertices stretching from the colored vertices represent the hopping embeddings that merge the neighbouring packets to form a colored vertex.
According to Algorithm \ref{algo:packing_no_ex_limit}, one first determines a minimum vertex cover $\mathbb{V}_{\text{mvc}}$ of the packing graph $G_{\mathbb{B}}$, which are highlighted as red vertices.
Note that in this example, we only heuristically find a minimum vertex cover instead of searching for all minimum vertex covers.

According to Section \ref{sec:identify_edges}, one can obtain the corresponding packet-conflict graph in Fig. \ref{fig:UCC_appl} (b).
The conflict edges are defined at the level of distributable packets.
One has to solve the conflicts (red edges) induced by the minimum vertex cover $\mathbb{V}_{\text{mvc}}$.
To this end, we need to expand the packet-conflict graph to a kernel-conflict graph, as it is shown in Fig. \ref{fig:UCC_appl} (c).
According to Algorithm \ref{algo:packing_no_ex_limit}, we select the set of kernels $\mathbb{K}_{\text{mvc}}$ that are induced by the minimum vertex cover $\mathbb{V}_{\text{mvc}}$, and highlight them in red.
The $\mathbb{K}_{\text{mvc}}$-induced kernel-conflict graph $C_{in}(\mathbb{K}_{\text{mvc}})$ representing the conflicts that need to be resolved.
We then find the minimum vertex cover $\widetilde{\mathbb{K}}_{\text{mvc}}$ of the conflict graph $C_{in}(\mathbb{K}_{\text{mvc}})$, and highlight them in blue. To solve the conflicts, we remove the blue-highlighted hopping embeddings $\widetilde{\mathbb{K}}_{\text{mvc}}$ from the minimum-vertex-covering packets $\mathbb{V}_{\text{mvc}}$.

In total, we find $10$ distributable packets in $\mathbb{V}_{\text{mvc}}$ that cover all the packing edges and remove $7$ hopping embedding kernels  $\widetilde{\mathbb{K}}_{\text{mvc}}$ from the selected packets.
In the end, we need only $17$ distributing processes to implement a DQC of the $4$-qubit UCC circuit, which contains $64$ global control-Z gates.
Our method therefore saves $47$ ebits in total.

\subsection{Comparison with other protocols}
\label{sec:compare_protocol}
We use the heuristic implementation of the packing algorithm (Algorithm \ref{algo:packing_no_ex_limit}) to analyze the bipartite distribution of UCC circuits compared with the $G^{\ast}$-simple and $G^{\ast}$-LP algorithms based on the ``Migration Selection'' method introduced in \cite{SundaramGuptaRamakrishnan2021-EffDQC}.
In the “Migration Selection” method, neighboring nonlocal CZ gates can also be packed together, but its packing stops when the process encounters single-qubit gate or local CZ gates, no matter whether embeddable or not.
In our embedding-enhanced packing algorithm, one packs control-phase gates and uses embedding to merge non-sequential distributing processes over embeddable single-qubit and two-qubit gates. The $G^{\ast}$-algorithm can therefore be understood as a special packing algorithm for CZ-compiled circuits without embedding.

\begin{figure*}[t]
  \centering
  \subfloat[]{\includegraphics[width=\textwidth]{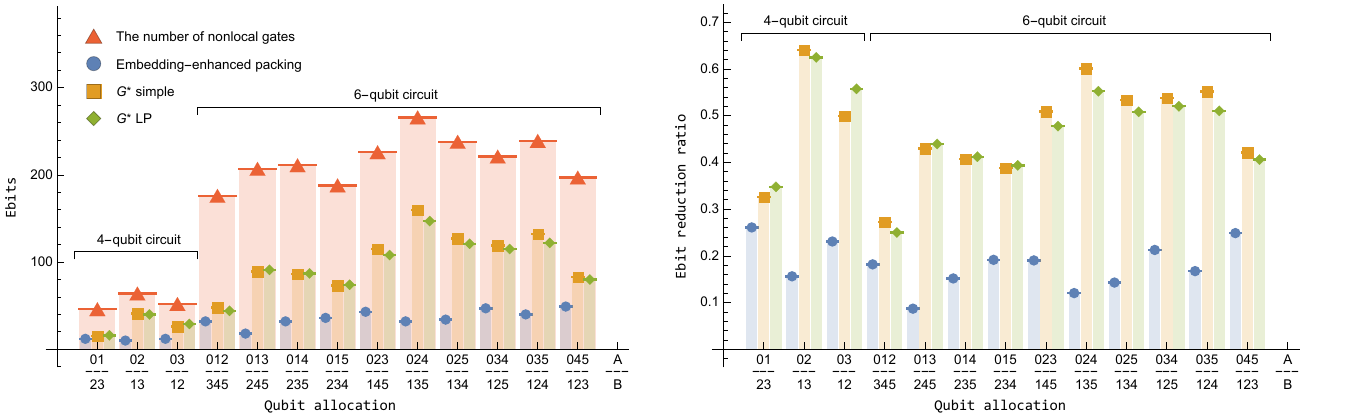}}
  \\
  \subfloat[]{\includegraphics[width=\textwidth]{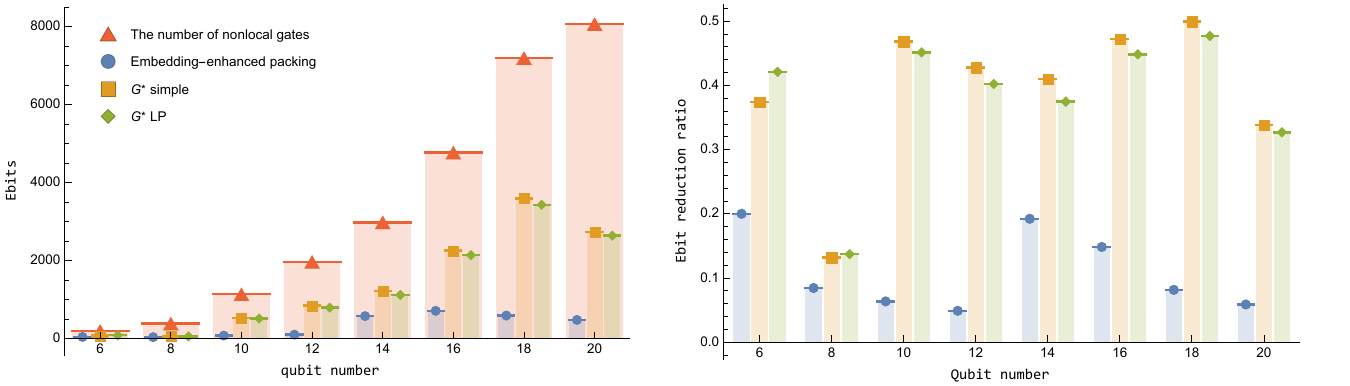}}
  \caption{The entanglement costs of different DQC protocols for UCC circuits (left column) and their reduction ratios (right column). (a) The entanglement cost for 4-qubit and 6-qubit UCC circuits under different bipartitions. (b) The entanglement cost for UCC circuits with a qubit number from $6$ up to $20$ under a fixed bipartition.}\label{fig:UCC_comparison}
\end{figure*}

Our benchmarking is performed on the UCC circuits with an even number of qubits that are uniformly distributed over two QPUs, each of which has the same specifications.
To account for the entanglement cost that depends on qubit allocation, we benchmark all possible bipartitions with an equal number of local qubits for 4-qubit and 6-qubit circuits.
The result is shown in Fig. \ref{fig:UCC_comparison} (a).
It shows that embedding can significantly reduce entanglement costs under every possible bipartition.
The entanglement efficiencies of these three protocols are further compared for UCC circuits up to a qubit number of 20 in Fig. \ref{fig:UCC_comparison} (b). Since the number of bipartitions increases exponentially with respect to the qubit number, in this comparison, we fixed a bipartition for each qubit number.
The result shows that the entanglement efficiency of embedding-enhanced distributing is also significantly improved for circuits with large qubit numbers.

\subsection{Embedding-enhanced multipartite-distributed quantum computing}
\label{sec:multipartite_DQC}
The extension of our embedding-enhanced packing protocol to multipartite DQC requires an extended definition of distributable packets (Definition \ref{def:distr_pack}) through an extension of the bipartite embedding rules to multipartite systems.
Once we get the extended packing graphs and conflict graphs constructed by the extended distributable packets as their vertices, one can follow Algorithms \ref{algo:packing_no_ex_limit} and \ref{algo:packing_with_ex_limit} to compile the DQC.
However, the identification of packing edges and conflict edges is challenging.
For multipartite embeddings, for instant in an $A|B|C$ system, their compatibility has to be considered under all possible bipartite subsystems, since the additional local correction gates in an embedding process of a nonlocal gate between two parties $A|B$ may destroy its embeddability between the other party $A|C$.
The incompatibility of embeddings is therefore not more describable by a bipartite conflict edge.

A straightforward but not optimal solution is to extend the notion of distributable packets to a set of gate nodes that are associated with two-qubit gates acting on the same local systems, such that each distributable packet is associated with only two parties.
With this solution, one can circumvent the multipartite conflict problem of embeddings and introduce the embedding method to existing compilers for multipartite DQC \cite{MartinezHeunen2019-DQCPartition} to allow entanglement-efficient DQC on multipartite quantum computing networks \cite{MartinezEtAlDuncan2023-DQCinQNet}.

\section{\label{sec:conclusion}Conclusion}

  In this paper, we have developed a theoretical architecture for entanglement-efficient distributed quantum computing over two local QPUs with the assistance of entanglement. The building blocks of the DQC architecture are the \emph{entanglement-assisted packing processes} (Definition \ref{def:EA_process}), which are extended from the EJPP protocol \cite{EisertEtAlPlenio2000-LclImplNonlclQGt}.
  An entanglement-assisted packing process packs a sequence of gates as its kernel sandwiched by the entanglement-assisted \emph{starting and ending processes} (Definition \ref{def:st_end_proc}).
  Two types of entanglement-assisted packing processes are essential, namely the \emph{distributing processes} (Definition \ref{def:ejpp-distr_U}) and the \emph{embedding processes} (Definition \ref{def:ejpp-embed_U}).
  The ultimate goal of distributed quantum computing with entanglement-assisted packing processes is to implement a quantum circuit with distributing processes, in which the kernels are all local gates. An embedding process allows the packing of two non-sequential distributing processes and hence save one ebit of entanglement.
  We therefore explore the embeddability of quantum circuits to save the entanglement cost in DQC.

  The distributability of a unitary can be determined by Theorem \ref{theorem:EJPP_distr}.
  Each distributing process consumes an ebit of entanglement.
  To save entanglement, we introduced the embedding processes to merge two non-sequential distributing processes into one single packing process.
  The embeddability of a unitary between two distributing processes can be determined with a sufficient condition, from which one can derive the primitive embedding rules (Corollary \ref{coro:primitive_cond_embed}, Theorem \ref{theorem:distr_and_embed}, and Corollary \ref{coro:local_embed_unitary}).
  These embedding rules join the nodes of distributable gates into \emph{distributable packets} (Definition \ref{def:distr_pack} and Lemma \ref{lemma:merging_of_packing}), on each of which one can implement the gates locally assisted with one-ebit entanglement in a single distributing process (Theorem \ref{theorem:packing_dPacket}).

  The \emph{trivial distributing processes} and \emph{indecomposable embedding processes} (Definition \ref{def:indecomp_embedding}) associated with different packets may be incompatible for simultaneous implementation. We therefore introduce \emph{packing edges} between distributable packets (Definition \ref{def:packing-edge}) and conflict edges incident to embedding processes (Definition \ref{def:kernel-conflict_edge} and \ref{def:conflict_edge_set}) to represent the incompatibility among distributing and embedding processes.
  With these edges, one can construct the \emph{packing graphs} and \emph{conflict graphs} of a circuit, which contain the full information of distributability, embeddability and incompatability in a circuit.
  Based on these graphs, one can determine the packing of a circuit with heuristic Algorithm \ref{algo:packing_no_ex_limit} (or Algorithm \ref{algo:packing_with_ex_limit}) for the scenario of unlimited (or limited) external resources.

  In particular, we consider circuits consisting of only one-qubit and two-qubit gates, which are universal for quantum computing.
  The distributability structure (packing edges) of such a circuit is completely represented by the control-phase gates (Lemma \ref{lemma:connecting_gate}) in its control-phase conversion.
  In a control-phase conversion, a circuit is decomposed as a sequence of two-qubit building blocks.
  We have derived the primitive embedding rules for these building blocks in Corollary \ref{coro:embed_rules_2q_block}.
  The embedding rules for single-qubit gates (Theorem \ref{theorem:distr_and_embed}) and two-qubit blocks (Corollary \ref{coro:embed_rules_2q_block}) identify the indecomposable neighbouring embeddings (Algorithm \ref{algo:neighbouring_emb}) and hopping embeddings (Lemma \ref{lemma:indecomp_hopping_emb} and Algorithm \ref{algo:hopping_emb}), respectively.
  Based on these embedding rules, one can then identify the set of distributable packets of a circuit with Algorithm \ref{algo:dPacket_identify}.
  Finally, one obtains the ultimate packing graph through the association of packing edges to global control-phase gates, and the ultimate conflict graphs through the association of conflict edges to incompatible embeddings (Theorem \ref{theorem:incompt_2qubit_circ}).

  These algorithms are demonstrated in the distributed implementation of a 4-qubit UCC circuit, which contains 64 global control-Z gates. The ultimate number of packing processes in the distributed implementation is $17$, hence requiring $17$ ebits of entanglement and saving $47$ ebits. %
  We benchmark the algorithms with further instants of the UCC circuits up to $20$ qubits. The comparison with other protocols shows a significant enhancement of entanglement efficiency by embeddings.

\bigskip

  The packing method in this paper can be summarized as ``distributing enhanced by embedding''. It incorporates the extrinsic limit of quantum resources, such as the number of available local memory qubits, in the conflict edges. It can be therefore employed to study the scalability of distributed quantum computing. One can further extend our method to multipartite scenarios and adopt quantum network topology to establish entanglement-efficient DQC over quantum internet \cite{MartinezEtAlDuncan2023-DQCinQNet}.


  The DQC architecture of a quantum circuit based on entanglement-assisted packing processes revealed by the packing algorithms is a constructive method to determine an upper bound on the entanglement cost of the entanglement-assisted LOCCs of a circuit.
  Benefit from embeddings, one can obtain a tighter upper bound on the entanglement cost approaching the lower bound, which is determined by the operator Schmidt rank \cite{StahlkeGriffiths2011-EntCostForUnitary}.
  One can therefore exploit the packing method to explore the unitaries that have the optimal entanglement cost implemented with entanglement-assisted packing processes.
  The packing protocol in this paper is therefore practical for finding an entanglement-efficient solution for distributed quantum computing, and also useful for the fundamental study of the optimality in entanglement-assisted LOCC.

\begin{acknowledgments}
JYW is supported by National Science and Technology Council, Taiwan, under Grant no. NSTC 110-2112-M-032-005-MY3, 111-2923-M-032-002-MY5, 111-2119-M-008-002, and 112-2112-M-032-008-MY3.
KM, TF, AS, and MM are supported by MEXT Quantum Leap Flag-ship Program (MEXT QLEAP) JPMXS0118069605, JPMXS0120351339, Japan Society for the Promotion of Science (JSPS) KAKENHI Grant No. 18K13467, 21H03394.
\end{acknowledgments}

\appendix

\section{The representation of an entanglement-assisted packing process}
\label{sec:EA_proc_Kraus}
  With a general kernel $K$, the quantum operations of a perfect packing process is described by
  \begin{equation}
    \rho = \tr_{e}\left( C_{q,X_{e}} K C_{q,X_{e}} \projector{\psi,0_{e}} C_{q,X_{e}} K^{\dagger} C_{q,X_{e}}  \right).
  \end{equation}
  A packing process is therefore a set of Kraus operators
  \begin{align}
    \mathcal{P}_{q,e}[K](\ket{\psi})
    =
    \frac{1}{2} \mathcal{K}_{+}\projector{\psi}\mathcal{K}_{+}^{\dagger}
    +
    \frac{1}{2} \mathcal{K}_{-}\projector{\psi}\mathcal{K}_{-}^{\dagger},
  \end{align}
  where
  \begin{align}
  \label{eq:Kraus_op_ejpp_proc}
    \mathcal{K}_{\pm}
    & =
    \sqrt{2}\braket{\pm_{e}|C_{q,X_{e}} K C_{q,X_{e}}|0_{e}}.
  \end{align}
  For some special kernels, the Kraus operators $\mathcal{K}_{\pm}$ are equivalent up to a global phase,
  \begin{equation}
  \label{eq:U-eq_cond}
    \mathcal{K}_{+} = e^{\imI \varphi}\mathcal{K}_{-}.
  \end{equation}
  In such a case, a packing process is equivalent to a unitary.
  \begin{lemma}[$U$-equivalent packing process]\label{lemma:U-eq_ejpp}{\ }\\
    Let $\mathcal{P}_{q,e}[K]$ be an EJPP process. It is equivalent to a unitary $U$,
    \begin{equation}
      U = \mathcal{P}_{q,e}[K],
    \end{equation}
    if and only if there exists a single-qubit $X$-rotation on the auxiliary qubit $e$,
    \begin{equation}
      R_{X}^{(e)}(\varphi):=\projector{+_{e}} + e^{\imI\varphi}\projector{-_{e}},
    \end{equation}
    such that $R_{X}^{(e)}(\varphi) K$ admits the following form in the computational basis $\{\ket{i_{q}i'_{e}}\}_{i,i'}$ of $q$ and $e$.
    \begin{align}
      R_{X}^{(e)}(\varphi)K
      & =
      \left\{\braket{i_{q}i'_{e}|R_{X}^{(e)}(\varphi)K|j_{q}j'_{e}}\right\}_{ii',jj'}
      \nonumber \\
      & =
      \left(
        \begin{array}{cccc}
          U^{(\bar{q})}_{00} & 0 & 0 & U^{(\bar{q})}_{01} \\
          0 & * & * & 0 \\
          0 & * & * & 0 \\
          U^{(\bar{q})}_{10} & 0 & 0 & U^{(\bar{q})}_{11} \\
        \end{array}
      \right),
    \end{align}
    where $U^{(\bar{q})}_{ij}:= \braket{i_{q}|U|j_{q}}$.
    The phase $\varphi$ is called the \emph{auxiliary phase}.
  \begin{proof}
    A packing process is $U$-equivalent, if and only if the Kraus operator $\mathcal{K}_{\pm}$ in Eq. \eqref{eq:Kraus_op_ejpp_proc} fulfills Eq. \eqref{eq:U-eq_cond}. Introducing a local phase calibration $R_{X}^{(e)}(\varphi)$ on the auxiliary qubit $e$ to eliminate the phase $\varphi$ in Eq. \eqref{eq:U-eq_cond}
    \begin{equation}
      \widetilde{K}:=R_{X}^{(e)}(\varphi)K,
    \end{equation}
    one obtains
    {\small
    \begin{equation}
      \braket{+_{e}|\,\widetilde{K}\,C_{q,X_{e}}\,|0_{e}}
      =
      \braket{-_{e}|\,Z_{q}\,\widetilde{K}\,C_{q,X_{e}}\,|0_{e}}.
    \end{equation}
    }
    This equality is equivalent to
    \begin{align}
      0 & =
      \braket{+_{e}|\,
      (\id_{qe} - Z_{q}Z_{e})
      \,\widetilde{K}\,
      (\id_{qe} + Z_{q}Z_{e})
      \,|+_{e}}.
    \end{align}
    In the computational basis of $q$ and $e$, this equality is equivalent to
    \begin{equation}
    \label{eq:U-eq_ejpp_proc_proof_1}
      0 = \braket{i_{q},(i\oplus 1)_{e}|\widetilde{K}|i_{q},i_{e}} \text{ for } i=0,1.
    \end{equation}
    This proves that a packing process is equivalent to a unitary if and only if the $R_{X}^{(e)}(\varphi)$-calibrated kernel $\widetilde{K}$ has the following form
    \begin{align}
    \label{eq:U-eq_ejpp_proc_proof_2}
      \widetilde{K}
      & =
      \left(
        \begin{array}{cccc}
          \widetilde{K}_{00,00}^{(\bar{q},\bar{e})} & * & * & \widetilde{K}_{00,11}^{(\bar{q},\bar{e})} \\
          0 & * & * & 0 \\
          0 & * & * & 0 \\
          \widetilde{K}_{11,00}^{(\bar{q},\bar{e})} & * & * & \widetilde{K}_{11,11}^{(\bar{q},\bar{e})} \\
        \end{array}
      \right),
    \end{align}
    where $\widetilde{K}_{ii',jj'}^{(\bar{q},\bar{e})}:=\braket{i_{q},i'_{e}|\widetilde{K}|j_{q},j'_{e}}$.
    Since the $R_{X}^{(e)}(\varphi)$ calibration only introduce a global phase to the Kraus operator $\mathcal{K}_{-}$ of a packing process, $\mathcal{P}_{q,e}[K]$ and $\mathcal{P}_{q,e}(\widetilde{K})$ implement the same unitary $U$. The unitary $U$ implemented by $\mathcal{P}_{q,e}[K]$ is therefore given by
    \begin{equation}
      U 
      = \braket{+_{e}|\widetilde{K}(\id + Z_{q}Z_{e})|+_{e}}.
    \end{equation}
    As a result of Eq. \eqref{eq:U-eq_ejpp_proc_proof_1}, the representation of $U$ in the computational basis of $q$ is therefore
    \begin{align}
      \braket{i_{q}|U|j_{q}}
      & =
      \braket{i_{q}, +_{e}|\widetilde{K}(\id + Z_{q}Z_{e})|j_{q},+_{e}}
      \nonumber \\
      & = \braket{i_{q}, i_{e}|\widetilde{K}|j_{q},j_{e}},
    \end{align}
    which leads to
    \begin{align}
      \widetilde{K}
      & =
      \left(
        \begin{array}{cccc}
          \widetilde{U}_{00}^{(\bar{q})} & \widetilde{K}_{00,01}^{(\bar{q},\bar{e})} & \widetilde{K}_{00,10}^{(\bar{q},\bar{e})} & \widetilde{U}_{01}^{(\bar{q})} \\
          0 & \widetilde{K}_{01,01}^{(\bar{q},\bar{e})} & \widetilde{K}_{01,10}^{(\bar{q},\bar{e})} & 0 \\
          0 & \widetilde{K}_{10,01}^{(\bar{q},\bar{e})} & \widetilde{K}_{10,10}^{(\bar{q},\bar{e})} & 0 \\
          \widetilde{U}_{10}^{(\bar{q})} & \widetilde{K}_{11,01}^{(\bar{q},\bar{e})} & \widetilde{K}_{11,10}^{(\bar{q},\bar{e})} & \widetilde{U}_{11}^{(\bar{q})} \\
        \end{array}
      \right).
    \end{align}
    Let $W$ and $V$ be two matrices defined as
    \begin{align}
      W = &
      \left(
        \begin{array}{cc}
          \widetilde{K}_{00,01}^{(\bar{q},\bar{e})} & \widetilde{K}_{00,10}^{(\bar{q},\bar{e})} \\
          \widetilde{K}_{11,01}^{(\bar{q},\bar{e})} & \widetilde{K}_{11,10}^{(\bar{q},\bar{e})} \\
        \end{array}
      \right)
      ,
      \nonumber \\
      V = &
      \left(
        \begin{array}{cc}
          \widetilde{K}_{01,01}^{(\bar{q},\bar{e})} & \widetilde{K}_{01,10}^{(\bar{q},\bar{e})} \\
          \widetilde{K}_{10,01}^{(\bar{q},\bar{e})} & \widetilde{K}_{10,10}^{(\bar{q},\bar{e})} \\
        \end{array}
      \right).
    \end{align}
    Since $\widetilde{K}$ and $U$ are both unitary, it holds then
    \begin{equation}
      V^{\dagger}V = \id, \text{ and } W\,V^{\dagger} = 0,
    \end{equation}
    which leads to
    \begin{equation}
      W = W\,V^{\dagger}\,V = 0.
    \end{equation}
    As a result,
    \begin{align}
    \label{eq:U-eq_ejpp_proc_proof_3}
      \widetilde{K}
      & =
      \left(
        \begin{array}{cccc}
          \widetilde{U}_{00}^{(\bar{q})} & 0 & 0 & \widetilde{U}_{01}^{(\bar{q})} \\
          0 & * & * & 0 \\
          0 & * & * & 0 \\
          \widetilde{U}_{10}^{(\bar{q})} & 0 & 0 & \widetilde{U}_{11}^{(\bar{q})} \\
        \end{array}
      \right).
    \end{align}
    This completes the proof.
  \end{proof}
  \end{lemma}

  \begin{corollary}[Sufficient condition for $U$-equivalent]
  \label{coro:U-eq_suff_cond}
    An entanglement-assisted process with the kernel $C_{q,X_{e}} U C_{q,X_{e}}$ is canonical and equivalent to the unitary $U$,
    \begin{equation}
      U = \mathcal{P}_{q,e}[C_{q,X_{e}} U C_{q,X_{e}}].
    \end{equation}
    The kernel $C_{q,X_{e}} U C_{q,X_{e}}$ is the \emph{primitive kernel} of $U$.
  \begin{proof}
    According to Eq. \eqref{eq:Kraus_op_ejpp_proc}, the Kraus operator of $\mathcal{P}_{q,e}[C_{q,X_{e}} U C_{q,X_{e}}]$ are $\mathcal{K}_{\pm} = \sqrt{2}\braket{\pm_{e}|U|0_{e}} = U$.
  \end{proof}
  \end{corollary}
%

\section{Extended embedding}
\label{sec:destr_hopping}

The embedding introduced in Definition \ref{def:ejpp-embed_U} can merge two non-sequential distributing processes without changing global gates in the kernel.
The packing edges associated with global gates in a packing graph are therefore not affected by an embedding process.
Although an embedding process may introduce conflicts to some other embeddings, one can solve these conflicts through the removal of embeddings, which split their original distributable packets.
It means that the distributability structure of global gates represented by packing edges in the packing graph is not changed by ordinary embeddings.
Since the global gates in a kernel are not changed, the entanglement cost for distributing processes in the kernel does not change.
The merging of two non-sequential distributing processes will then save $1$ ebit.

\begin{figure}[htbp]
  \centering
  \includegraphics[width=0.5\textwidth,height=4.8cm,keepaspectratio]{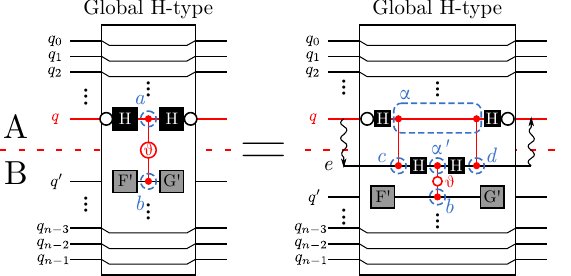}
  \caption{An extended embedding rule for general global $H$-type embeddig unit.}\label{fig:extended_embedding}
\end{figure}

In addition to the ordinary $H$-type embedding rule in Corollary \ref{coro:embed_rules_2q_block} for a two-qubit embedding unit with a control-Z gate, we introduce an extended embedding rule for the $H$-type embedding unit with general controlling phase $\theta\neq\pi$.
\begin{lemma}[Extended embedding]\label{lemma:extended_emb}{\ }\\
  An $H$-type $2$-qubit embedding unit can be packed with the following rule (Fig. \ref{fig:extended_embedding})
  \begin{align}
    & H_{q} U^{(q,q')} H_{q}
    \nonumber \\
    = &
    \mathcal{P}_{q,e}
    \left[H_{q} C_{Z}^{(q,e)} H_{e} U^{(e,q')} H_{e} C_{Z}^{(q,e)} H_{q}\right].
  \end{align}
  \begin{proof}
    Employing the primitive embedding rule in Corollary \ref{coro:primitive_cond_embed}, we know that
    \begin{align}
      & H_{q} U^{(q,q')} H_{q}
      \nonumber \\
      = &
      \mathcal{P}_{q,e}
      \left[ C_{q',X_{e}} H_{q} U^{(q,q')} H_{q} C_{q',X_{e}} \right].
    \end{align}
    One can then show that
    \begin{align}
      & C_{q',X_{e}} H_{q} U^{(q,q')} H_{q} C_{q',X_{e}}
      \nonumber \\
      = &
      H_{q} C_{Z}^{(q,e)} H_{e} U^{(e,q')} H_{e} C_{Z}^{(q,e)} H_{q}.
    \end{align}
  \end{proof}
\end{lemma}%

As it is shown in Fig. \ref{fig:extended_embedding}, this extended embedding of the global $H$-type block localizes the control-phase gate to the $B$ system with two additional global control-Z gates.
On the left-hand side, the original distributable packets $\mathcal{V}_{a}$ and $\mathcal{V}_{b}$ (blue dashed circles) form a packing edge $(\mathcal{V}_{a} \leftrightarrow \mathcal{V}_{b})$.
After the extended embedding, the packet $\mathcal{V}_{a}$ is replaced by $\mathcal{V}_{\alpha}$, and the packing edge $(\mathcal{V}_{a} \leftrightarrow \mathcal{V}_{b})$ is localized to $(\mathcal{V}_{\alpha'} \leftrightarrow b)$ on the right-hand side with two additional packets $\mathcal{V}_{c}$ and $\mathcal{V}_{d}$ globally connected to $\mathcal{V}_{\alpha}$.
\begin{equation}
  (\mathcal{V}_{a} \leftrightarrow \mathcal{V}_{b})
  \underset{
    \substack{\text{extended}\\\text{embedding}}
  }{
    \longrightarrow
  }
  (\mathcal{V}_{c} \leftrightarrow \mathcal{V}_{\alpha} \leftrightarrow \mathcal{V}_{d}, \mathcal{V}_{\alpha'} \leftrightarrow \mathcal{V}_{b}).
\end{equation}
The global packing edges of the kernel are now changed to $\mathcal{V}_{c} \leftrightarrow \mathcal{V}_{\alpha} \leftrightarrow \mathcal{V}_{d}$.
The corresponding graph has the minimum vertex cover $\{\mathcal{V}_{a}\}$.
If $\mathcal{V}_{a}$ is already selected as one of the root packets for the distributing processes, then after the extend embedding, one can replace the original packet $\mathcal{V}_{a}$ in the original vertex cover by $\mathcal{V}_{\alpha}$.
Such a replacement does not change the entanglement cost in the kernel.
It therefore leads to a possible reduction of entanglement by one ebit through the merging of two non-sequential distributing processes hopping over the $H$-type unit.

However, if $\mathcal{V}_{b}$ is selected in the vertex cover of the original packing graph rather than $\mathcal{V}_{a}$, the $1$-ebit entanglement saved by the extended embedding hopping over $\mathcal{V}_{\alpha}$ has to be consumed for the distributing process rooted at $\mathcal{V}_{\alpha}$.
It means that the extended embedding does not bring any reduction of entanglement cost.

It is therefore necessary to select $\mathcal{V}_{a}$ as a root packet for the extended embedding to save one ebit. Suppose that $\mathcal{V}_{a}$ is a root packet. After the extended embedding, the new local control phase gate $C_{V}^{(\alpha',b)}$ may intrinsically prevent the other hopping embeddings. This happens when there is a selected root packet merged by the $q'$-rooted hopping embedding over the packet $\mathcal{V}_{b}$. In this case, one has to remove the $q'$-rooted hopping embedding, and split the corresponding packet, which will reuse the $1$-ebit entanglement saved by the extended embedding.
Such a conflict is the only intrinsic conflict that prevents the reduction of entanglement cost through the extended embedding.
As a result, the necessary condition for reduction of entanglement cost by the extended embedding rule can be summarized as follows.
\begin{corollary}[Entanglement reduction condition]\label{coro:ent_reduction_extended_emb}{\ }\\
  The $q$-rooted extended embedding of an $H$-type embedding unit can save $1$-ebit, only if
  \begin{enumerate}
    \item \label{coro:ent_reduction_extended_emb_cond1}the distributable packet on $q$ in the unit is already selected as a root packet, and
    \item \label{coro:ent_reduction_extended_emb_cond2}the local gate $C_{V}^{(\alpha',b)}$ does not conflict with a selected hopping embedding.
  \end{enumerate}
\end{corollary}
This necessary condition is also sufficient, if there is no extrinsic limits on external resources.

After solving the intrinsic conflicts and identifying the root packets $\mathbb{P}_{i,j}$ in Algorithm \ref{algo:packing_no_ex_limit} (line \ref{algo_line:destr_hopping_1}) and Algorithm \ref{algo:packing_with_ex_limit} (line \ref{algo_line:destr_hopping_2}), one can introduce an addon function ``\textit{extended\_embedding()}'' to further reduce the entanglement cost by checking the two conditions in Corollary \ref{coro:ent_reduction_extended_emb}.
The algorithm for the addon function is given in Algorithm \ref{algo:extended_embedding}.
Note that for Algorithm \ref{algo:packing_with_ex_limit} with extrinsic limits, after the extended embedding, one has to update the extrinsic conflict graph on line \ref{algo_line:required_memory_2}.

\begin{algorithm}[h]
\setcounter{algocf}{\value{theorem}}\addtocounter{theorem}{1}
\caption{Extended embedding}
\label{algo:extended_embedding}
\SetKwProg{Fn}{Function}{:}{}
\setcounter{AlgoLine}{0}
\Fn{\destrHop($\mathbb{R}$)}{
  $\widetilde{\mathbb{H}} \gets \emptyset$ \;
  $\mathbb{H}_{\mathbb{R}} \gets $ Get the hopping embeddings in $\mathbb{R}$\;
  \tcc{Get the neighbouring distributable packets from $\mathbb{R}$}
  $\mathbb{R}_{D} \gets $ Get the neighbouring packets from $\mathbb{R}$ through removal of the hopping embeddings\;
  \ForEach{$\mathcal{V}_{q,T}\in\mathbb{R}_{D}$}{
    \uIf{ $\exists$ sequential $(\mathcal{V}_{q,T_{1}}, \mathcal{V}_{q,T}, \mathcal{V}_{q,T_{2}}) \subseteq \mathbb{R}_{D}$ such that $\nexists$ control phase gates on $q$ between $T_{1}$, $T$, and $T_{2}$}{
      $(t_{1},t_{2}) \gets (\max T_{1}, \min T_{2})$ (Condition \ref{coro:ent_reduction_extended_emb_cond1} in Corollary \ref{coro:ent_reduction_extended_emb})\;
      \uIf{
        \begin{enumerate}
          \item 
            $(q; t_{1},t_{2})\notin\mathbb{H}_{\mathbb{R}}$, and
          \item 
            $W_{q;t_{2},t_{1}}$ is extended embeddable \\ according to Lemma \ref{lemma:extended_emb}, and
          \item 
            $\nexists (q';t_{a},t_{b})\in\mathbb{H}_{\mathbb{R}}$, such that $\exists$a global \\ control phase gate $C_{V}^{(q,q')}(\theta_{t})$ at a \\ depth $t\in(t_{1},t_{2})$ and $t\in(t_{a},t_{b})$ \\
            (Condition \ref{coro:ent_reduction_extended_emb_cond2} in Corollary \ref{coro:ent_reduction_extended_emb})\;
        \end{enumerate}
      }{
        $\widetilde{\mathbb{H}} \gets \widetilde{\mathbb{H}}\cup\{q;(t_{1},t_{2})\}$\;
      }
    }
  }
  $\widetilde{\mathbb{R}} \gets $ merge $\widetilde{\mathbb{R}}$ by $\widetilde{\mathbb{H}}$ according to Lemma \ref{lemma:merging_of_packing}\;
  \KwRet{$\{\widetilde{\mathbb{R}}, \widetilde{\mathbb{H}}\}$}
}
\end{algorithm}

\section{\label{sec:ucc_circuit}UCC circuits}
\begin{figure*}[htpb]
  \centering
  \includegraphics[width=0.8\textwidth]{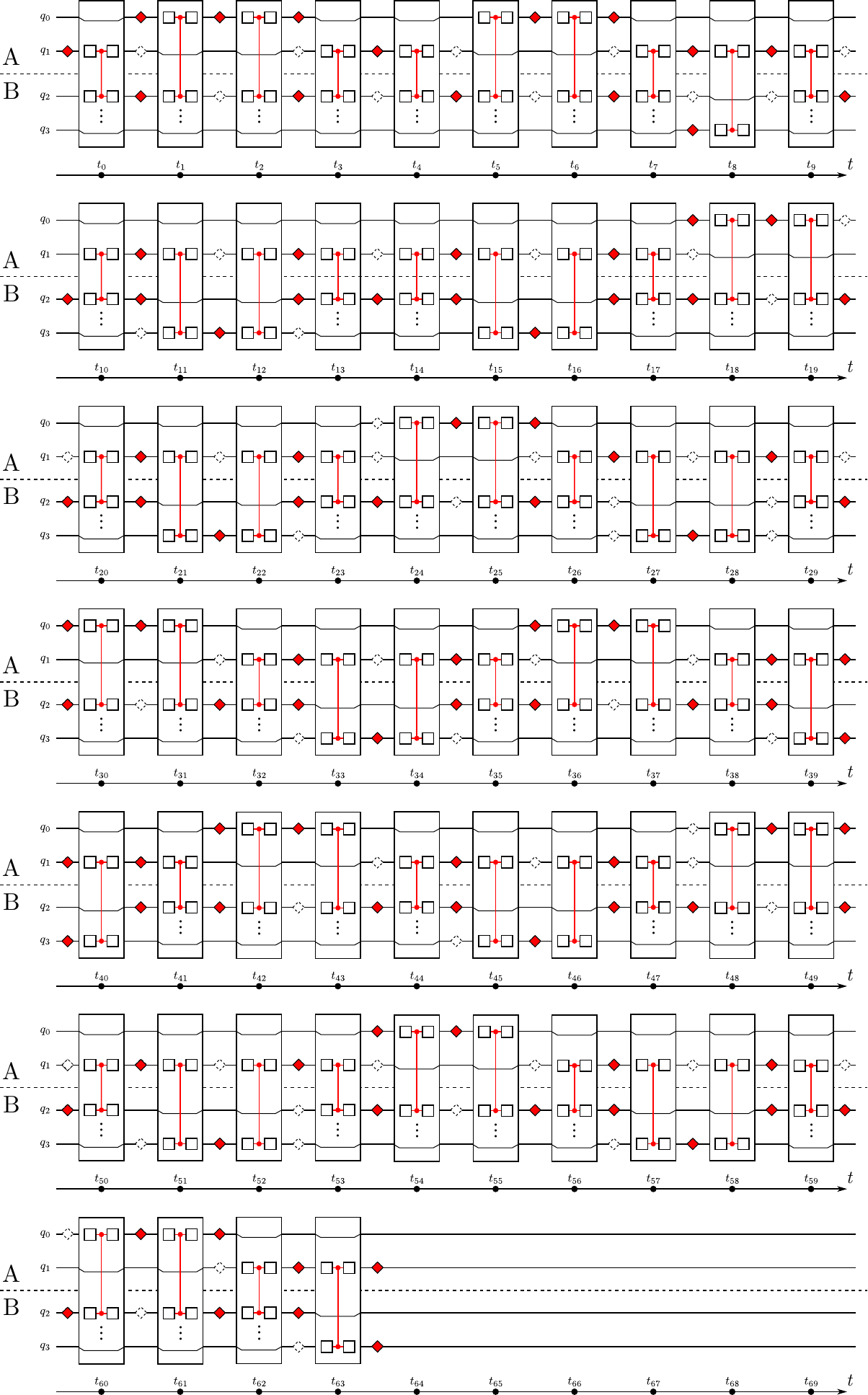}
  \caption{The 2-qubit blocks of a $4$-qubit UCC circuit after the control-phase conversion.}\label{fig:ucc4q}
\end{figure*}

We employ the ``pytket-dqc'' package to generate a 4-qubit UCC circuit.
The 4-qubit circuit is converted according to the control-phase conversion in Eq. \eqref{eq:2q-blocks} and \eqref{eq:2-qubit_block}.
The converted circuit is shown in Fig. \ref{fig:ucc4q}. There are $64$ $2$-qubit blocks. Every block contains one global control-phase gate. After the control-phase conversion, one employs Algorithm \ref{algo:dPacket_identify}, \ref{algo:neighbouring_emb} and \ref{algo:neighbouring_emb} to identify the distributable packets and the corresponding packing graph and conflict graphs, which are shown in Fig. \ref{fig:UCC_appl}.

\section{Proof of theorems and lemmas}
\label{sec:proofs}

\subsection{Proof of Theorem \ref{theorem:EJPP_packing}}
\label{proof:EJPP_packing}
  According to Lemma \ref{lemma:U-eq_ejpp}, the matrices of the kernels $K_{i}$ in the computational basis are
  \begin{align}\label{eq:ejpp_packing_proof_1}
    K_{i}
    & =
    \left(
      \begin{array}{cccc}
        U^{(\bar{q})}_{i,00} & 0 & 0 & U^{(\bar{q})}_{i,01} \\
        0 & * & * & 0 \\
        0 & * & * & 0 \\
        U^{(\bar{q})}_{i,10} & 0 & 0 & U^{(\bar{q})}_{i,11} \\
      \end{array}
    \right),
  \end{align}
  where $U^{(\bar{q})}_{i,jk}:=\braket{j_{q}|U_{i}|k_{q}}$. It holds then
  \begin{align}\label{eq:ejpp_packing_proof_1}
    K_{2}K_{1}
    & =
    \left(
      \begin{array}{cccc}
        \widetilde{U}^{(\bar{q})}_{00} & 0 & 0 & \widetilde{U}^{(\bar{q})}_{01} \\
        0 & * & * & 0 \\
        0 & * & * & 0 \\
        \widetilde{U}^{(\bar{q})}_{10} & 0 & 0 & \widetilde{U}^{(\bar{q})}_{11} \\
      \end{array}
    \right),
  \end{align}
  where
  \begin{equation}
    \widetilde{U}^{(\bar{q})}_{ij}
    =
    \sum_{k} U^{(\bar{q})}_{2,ik}U^{(\bar{q})}_{1,kj}.
  \end{equation}
  As a result of Lemma \ref{lemma:U-eq_ejpp}, one obtains
  \begin{equation}
    U_{2}U_{1}
    =
    \mathcal{P}_{q,e}[K_{2}K_{1}].
  \end{equation}
  This completes the proof.

\subsection{Proof of Theorem \ref{theorem:EJPP_distr}}
\label{proof:EJPP_distr}
  Let $U$ be a $q$-rooted distributable unitary with $q$ being on the local system $A$. According to Lemma \ref{lemma:U-eq_ejpp}, there exists a canonical kernel $K_{A}\otimes K_{B}$ such that
  \begin{align}
    K_{A}\otimes K_{B}
    & =
    \left(
      \begin{array}{cccc}
        \widetilde{U}_{00}^{(\bar{q})} & 0 & 0 & \widetilde{U}_{01}^{(\bar{q})} \\
        0 & * & * & 0 \\
        0 & * & * & 0 \\
        \widetilde{U}_{10}^{(\bar{q})} & 0 & 0 & \widetilde{U}_{11}^{(\bar{q})} \\
      \end{array}
    \right).
  \end{align}
  Since $\braket{i,j|K_{A}\otimes K_{B}|k,l} = \braket{i|K_{A}|k} \otimes \braket{j|K_{B}|l}$, it holds then
  \begin{equation}
    0 = \braket{i|K_{A}|k} \otimes \braket{j|K_{B}|l}, \text{for all } i\oplus k \neq j\oplus l.
  \end{equation}
  It means that $K_{A}$ and $K_{B}$ must be diagonal or anti-diagonal at the same time,
  \begin{align}
  \label{eq:distr_cond_proof_1}
    K_{A} & = \sum_{i,j\in\{0,1\}} \Delta_{ij} \ket{i}\bra{j}_{q}\otimes V_{j}^{(Q_{A}\setminus q)},
    \\
    K_{B} & = \sum_{i,j\in\{0,1\}} \Delta_{ij} \ket{i}\bra{j}_{e}\otimes W_{j}^{(Q_{B})}.
  \end{align}
  where $\Delta_{ij} = \delta^{i}_{j}$ or $\Delta_{ij} = \delta^{i \oplus 1}_{j}$.
  As a result,
  \begin{widetext}
  \begin{equation}
    K_{A}\otimes K_{B}
    =
    \left(
      \begin{array}{cccc}
        \Delta_{00}^{2}V_{0}\otimes W_{0} & 0 & 0 & \Delta_{01}^{2}V_{1}\otimes W_{1} \\
        0 & \Delta_{00}\Delta_{11}V_{0}\otimes W_{1} & \Delta_{11}\Delta_{00}V_{1}\otimes W_{0} & 0 \\
        0 & \Delta_{11}\Delta_{00}V_{1}\otimes W_{0} & \Delta_{00}\Delta_{11}V_{0}\otimes W_{1} & 0 \\
        \Delta_{10}^{2}V_{0}\otimes W_{0} & 0 & 0 & \Delta_{11}^{2}V_{1}\otimes W_{1} \\
      \end{array}
    \right).
  \end{equation}
  \end{widetext}
  It means that the following equality is a necessary condition for a $q$-rooted distributable unitary,
  \begin{equation}
    U = \sum_{i,j} \Delta_{ij}
    \ket{i}\bra{j}_{q}
    \otimes
    V_{j}^{(Q_{A}\setminus q)}\otimes W_{j}^{(Q_{B})}.
  \end{equation}
  This equality is also sufficient for a $q$-rooted distributable unitary, which can be implemented with the kernels given in Eq. \eqref{eq:distr_cond_proof_1}.

\bigskip

\subsection{Proof of Lemma \ref{lemma:cond_recur_compat}}
\label{proof:cond_recur_compat}
  Since $U$ is embeddable on $\mathbb{Q}$ and $\mathbb{Q}'$ with the embedding rules $\mathcal{B}_{\mathbb{Q}}$ and $\mathcal{B}_{\mathbb{Q}'}$, it holds
  \begin{equation}
    U = \mathcal{P}_{\mathbb{Q}',\mathbb{A}'}\left[ \mathcal{B}_{\mathbb{Q}'} \left(\mathcal{P}_{\mathbb{Q},\mathbb{A}}\left[ \mathcal{B}_{\mathbb{Q}(U)}  \right]\right) \right].
  \end{equation}
  According to Eq. \eqref{eq:Kraus_op_ejpp_proc}, the unitary is equivalent to
  \begin{align}
    U = & \sqrt{2} \bra{\pm_{e_{1}},\pm_{e_{2}}, ..., \pm_{e_{k}}}
    C_{\mathbb{Q}',X_{\mathbb{A}'}}
    (K'_{A}\otimes K'_{B})
    \nonumber\\ & \times
    C_{\mathbb{Q},X_{\mathbb{A}}}
    \mathcal{B}_{\mathbb{Q}'}(U)
    C_{\mathbb{Q},X_{\mathbb{A}}}
    \nonumber\\ & \times
    (K_{A} \otimes K_{B})
    C_{\mathbb{Q}',X_{\mathbb{A}'}}
    \ket{0_{e_{1}},0_{e_{2}}, ..., 0_{e_{k}}},
  \end{align}
  where $\mathbb{A}\cup\mathbb{A}'=\{e_{1},...,e_{k}\}$ is the disjoint set union of the auxiliary qubits $\mathbb{A}$ and $\mathbb{A}'$.
  As a result of the commutation relation in  Eq. \eqref{eq:recurs_embed_commut}, the unitary $U$ is then joint embeddable on $\mathbb{Q}\uplus\mathbb{Q}'$ with
  \begin{align}
    U = & \sqrt{2} \bra{\pm_{e_{1}},\pm_{e_{2}}, ..., \pm_{e_{k}}}
    C_{\mathbb{Q}',X_{\mathbb{A}'}}C_{\mathbb{Q},X_{\mathbb{A}}}
    \nonumber\\ & \times
    (K'_{A}\otimes K'_{B})
    \mathcal{B}_{\mathbb{Q}}(U)
    (K_{A} \otimes K_{B})
    \nonumber\\ & \times
    C_{\mathbb{Q},X_{\mathbb{A}}}C_{\mathbb{Q}',X_{\mathbb{A}'}}
    \ket{0_{e_{1}},0_{e_{2}}, ..., 0_{e_{k}}}
    \nonumber \\
    & =
    \mathcal{P}_{\mathbb{Q}\uplus\mathbb{Q}',\mathbb{A}\uplus\mathbb{A}'}
    \left[
      \mathcal{B}_{\mathbb{Q}'}\circ \mathcal{B}_{\mathbb{Q}}(U)
    \right].
  \end{align}

\bigskip

\subsection{Proof of Lemma \ref{lemma:merging_of_packing}}
\label{proof:merging_of_packing}
  First of all, the unitaries $U_{t}$ with $t\in T_{1}\cup T_{2}$ are all distributable by definition.

  Second, for any $t_{1,2}\in T_{1}\cup T_{2}$ with $t_{1}<t_{2}$, we need to prove that the unitary
  \begin{equation}
    W_{t_{2};t_{1}}:=\prod_{t: t_{1}<t<t_{2}}U_{t}
  \end{equation}
  between $t_{1}$ and $t_{2}$ is $q$-rooted embeddable.
  If $(q,t_{1,2})\in \mathcal{V}_{q,T_{i}}$ belong to the same distributable packet, $W_{t_{1};t_{2}}$ is $q$-rooted embeddable by definition.
  For $t_{1}\in T_{1}$ and $t_{2}\in T_{1}$, let $t_{0}\in T_{1}\cap T_{2}$ be a common node in both $\mathcal{V}_{q,T_{1}}$ and $\mathcal{V}_{q,T_{2}}$.
  There are three possible position for $t_{0}$, namely, $t_{0}<t_{1}<t_{2}$, $t_{1}<t_{0}<t_{2}$, and $t_{1}<t_{2}<t_{0}$.
  For $t_{0}<t_{1}<t_{2}$, the unitary $W_{t_{1};t_{0}}$ and $W_{t_{2};t_{0}}$ are both embeddable, since $\{(q,t_{0}), (q,t_{i})\}\subseteq\mathcal{V}_{q,T_{i}}$.
  It holds then
  \begin{equation}
    W_{t_{2},t_{1}} = W_{t_{2},t_{0}}W_{t_{1},t_{0}}^{\dagger}U_{t_{1}}^{\dagger},
  \end{equation}
  which is also $q$-rooted embeddable.
  Analogously, one can prove the $q$-rooted embeddability of $W_{t_{2};t_{1}}$ for $t_{1}<t_{0}<t_{2}$ and $t_{1}<t_{2}<t_{0}$.
  This completes the proof.

\bigskip

\subsection{The time complexity of Algorithm \ref{algo:packing_no_ex_limit} and \ref{algo:packing_with_ex_limit}}
\label{proof:complexity_of_packing}

The time complexity of Algorithm \ref{algo:packing_no_ex_limit} is estimated as follows.
Given a packing graph $G_{\mathbb{B}} = (\mathbb{V},\mathbb{E})$, there are three main steps in the algorithm,
\begin{enumerate}
  \item enumerate all minimum vertex covers $\mathbb{K}_{i}$.
  \item enumerate all minimum vertex covers $\{\mathbb{R}_{i,j}\}_{j}$ of the intrinsice kernel-conflict graph $C_{in}(\mathbb{K}_{i})$.
  \item determine the chromatic number of extrinsic packet-conflict graphs $C_{ex}(\mathbb{R}_{i,j}^{(A,B)})$
\end{enumerate}
The determination of minimum vertex covers is equivalent to the determination of maximum matching according to the K\"oning theorem \cite{DiestelBook-GraphTheory}. The complexity of enumerating all minimum vertex covers of a graph can be therefore estimated as $O(|\mathbb{V}|^{1/2}|\mathbb{E}|+|\mathbb{V}| N_{\mathrm{MVC}})$ according to Theorem 2 in \cite{Uno1997-EnumMtchBiGrph}, where $N_{\mathrm{MVC}}$ is the number of all minimum vertex covers.

Let $m$ be the number of nonlocal control phase gates in the control-phase conversion of a quantum circuit (Eq. \eqref{eq:CV-conversion}). The vertex number $|\mathbb{V}|$ and edge number $|\mathbb{E}|$ of a packing graph are upper bounded by $2m$ and  $m$, respectively.
The complexity of the algorithm is then estimated as
\begin{equation}
  \mathcal{O}(\sqrt{2} m ^{3/2} + 2m N_{\mathrm{MVC}}).
\end{equation}
The complexity of the first two steps is
\begin{equation}
  \mathcal{O} (\sqrt{2} m ^{3/2}(N_{1} + 1)+ 2m (N_{2}+1) N_{1})
\end{equation}
where $N_{1} = |\{\mathbb{K}_{i}\}_{i}|$ is the number of the minimum vertex covers in step 1, while $N_{2}=\max_{i}|\{\mathbb{R}_{i,j}\}_{j}|$ is the maximum cardinality of the set of minimum vertex covers obtained in step 2.

The number of minimum vertex covers can be upper bounded by the number of minimal vertex covers, which is equal to the number of maximal independent sets and upper bounded by $3^{|\mathbb{V}|/3}\le 3^{2m/3}$ according to \cite{MoonMoser1965-CliquesInGraphs}.
In the worst case, the complexity of the first two steps is $O(m\,3^{4m/3})$.

In the last step, one needs to determine the chromatic number of the extrinsic conflict graph $C_{ex}(\mathbb{R}_{i,j}^{(A,B)})$. In general, the determination of the chromatic number has a complexity of $\mathcal{O}(2^{|\mathbb{V}|}|\mathbb{V}|)$ \cite{BjorklunHusfeldtKoivisto2009-GraphPartition}, which is NP-hard.
Overall, to obtain the full information about all the options of packing strategies $\{(\mathbb{R}_{i,j},\chi_{i,j}^{(A)},\chi_{i,j}^{(B)})\}_{i,j}$ with Algorithm \ref{algo:packing_no_ex_limit}, the complexity is roughly upper bounded by
\begin{equation}
\label{eq:packing_algo_complexity}
  \mathcal{O}( m^{2}\,3^{4m/3} 2^{2m} ),
\end{equation}
which is exponentially increasing.

For a heuristic implementation of the algorithm, one can simply find a minimum vertex cover without enumerating all of them. The complexity of finding a minimum vertex cover can be estimated by $O(|\mathbb{V}|^{1/2}|\mathbb{E}|)$ according to the Hopcroft–Karp algorithm \cite{HopcroftKarp1973-MaximumMatching} 
Besides, an upper bound on the chromatic number of a conflict graph can also be efficiently determined by $\chi \le \max_{v\in\mathbb{V}} (d_{v}+1)$, where $d_{v}$ is the degree of a vertex, with a time complexity of $\mathcal{O}(|\mathbb{V}|^2)$.
As a whole, the heuristic implementation of Algorithm \ref{algo:packing_no_ex_limit} has a time complexity of
\begin{equation}
\label{eq:packing_algo_complexity_heuristic}
  \mathcal{O}(m^{7/2}),
\end{equation}
which is polynomial.

\begin{figure}
  \centering
  \includegraphics[width=0.48\textwidth]{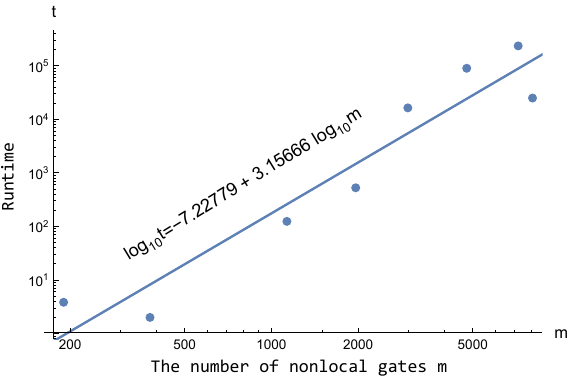}
  \caption{The log-log plot of the runtimes of the packing algorithm (Algorithm \ref{algo:packing_no_ex_limit}) for different UCC circuits.}
  \label{fig:runtime_UCC}
\end{figure}

In Fig. \ref{fig:runtime_UCC}, the runtimes $t$ of Algorithm \ref{algo:packing_no_ex_limit} for different UCC circuits are plotted with respect to the number of the nonlocal gates $m$ in the circuits. It shows a polynomial increasing of the runtime $t$ with respect to the nonlocal gates $m$ in the order of $3.157$.

\bigskip

Algorithm \ref{algo:packing_with_ex_limit} only differs from Algorithm \ref{algo:packing_no_ex_limit} on step 3, where one searches for $\chi_{A}$ and $\chi_{B}$ partitioning on the extrinsic graph $C_{ex}(\mathbb{R}_{i,j}^{(A)})$ and $C_{ex}(\mathbb{R}_{i,j}^{(B)})$, respectively, instead of the determination of their chromatic numbers. The complexity of $\chi_{A,B}$ partitioning is equal to the determination of the chromatic number \cite{BjorklunHusfeldtKoivisto2009-GraphPartition}. The complexity of Algorithm \ref{algo:packing_with_ex_limit} is therefore the same as Algorithm \ref{algo:packing_no_ex_limit}, which is given in Eq. \eqref{eq:packing_algo_complexity}.
The complexity of the heuristic solution for Algorithm \ref{algo:packing_with_ex_limit} is the same as the one given in Eq. \eqref{eq:packing_algo_complexity_heuristic}.

\bigskip

\subsection{Proof of Lemma \ref{lemma:connecting_gate}}
\label{proof::lemma_connecting_gate}
    According to Definition \ref{def:packing-edge}, a connecting gate $g$ is distributable on both $q_{1}$ and $q_{2}$.
    As a result of Theorem \ref{theorem:EJPP_distr}, the gate $g$ can be decomposed as
    \begin{align}
      g
      = &
      \sum_{i,j} u_{ij} \ket{i}\bra{j} \otimes U_{j}
      \label{eq:cph_edge_proof1}\\
      = &
      \sum_{i,j} v_{ij} V_{j} \otimes \ket{i}\bra{j}
      \label{eq:cph_edge_proof2},
    \end{align}
    where $\{u_{ij}\}_{i,j}$ and $\{v_{ij}\}_{i,j}$ are diagonal or antidiagonal.
    Let $D_{1}$ and $D_{2}$ be diagonal or antidiagonal single-qubit gates such that $\{u_{ij}\}_{i,j} = D_{1}.V(\varphi_{1})$ and $\{\tilde{u}_{ij}\}_{i,j} = D_{2}.V(\varphi_{2})$ is diagonal.
    It holds then
    \begin{align}
      &
      (D_{1}^{\dagger} \otimes D_{2}^{\dagger})
      g
      \nonumber\\
      = &
      \projector{0}\otimes D_{2}^{\dagger} U_{0}
      +
      \projector{1} \otimes ( e^{\imI\varphi_{1}} D_{2}^{\dagger} U_{1})
      \\
      = &
      D_{1}^{\dagger} V_{0} \otimes \projector{0}
      +
      (e^{\imI\varphi_{2}} D_{1}^{\dagger} V_{1}) \otimes \projector{1}
    \end{align}
    It follows that $D_{1}^{\dagger}V_{i}$ and $D_{2}^{\dagger}V_{i}$ are diagonal unitary.
    Furthermore, $D_{1}^{\dagger}V_{0}$ and $D_{2}^{\dagger}U_{0}$ fulfill
    $\braket{0|D_{1}^{\dagger}V_{0}|0} = \braket{0|D_{2}^{\dagger}U_{0}|0}$.
    It follows that
    \begin{align}
      &
      (D_{1}^{\dagger} \otimes D_{2}^{\dagger})
      G
      (\tilde{D}_{1}^{\dagger}\otimes\tilde{D}_{2}^{\dagger})
      \nonumber\\ = &
      \projector{0}\otimes\id +
      \\
      &
      \projector{1}\otimes
      (e^{\imI\varphi_{1}}
      \braket{0|D_{1}^{\dagger}V_{0}|0}
      \braket{1|D_{1}^{\dagger}V_{0}|1}^{\ast} D_{2}^{\dagger}U_{1}U_{0}^{\dagger}D_{2})
      \nonumber\\
      = &
      \id\otimes\projector{0} +
      \\
      &
      (e^{\imI\varphi_{2}}\braket{0|D_{2}^{\dagger}U_{0}|0}\braket{1|D_{2}^{\dagger}U_{0}|1}^{\ast} D_{1}^{\dagger}V_{1}V_{0}^{\dagger}D_{1})\otimes\projector{1}
      \nonumber
    \end{align}
    where
    \begin{align}
      \tilde{D}_{1} = &
      \braket{0|D_{1}^{\dagger}V_{0}|0}^{\ast}
      D_{1}^{\dagger}V_{0}
      \\
      \tilde{D}_{2} = &
      D_{2}^{\dagger}U_{0}.
    \end{align}
    Furthermore, one can then show the following equality
    \begin{align}
      &
      e^{\imI\varphi_{1}}
      \braket{0|D_{1}^{\dagger}V_{0}|0}
      \braket{1|D_{1}^{\dagger}V_{0}|1}^{\ast} D_{2}^{\dagger}U_{1}U_{0}^{\dagger}D_{2}
      \nonumber\\
      =  & e^{\imI\varphi_{2}}
      \braket{0|D_{2}^{\dagger}U_{0}|0}
      \braket{1|D_{2}^{\dagger}U_{0}|1}^{\ast} D_{1}^{\dagger}V_{1}V_{0}^{\dagger}D_{1}
      \nonumber\\
      = & \projector{0} + e^{\imI \theta}\projector{1} = V(\theta).
    \end{align}
    As a result,
    \begin{equation}
      (D_{1}^{\dagger} \otimes D_{2}^{\dagger})
      g
      (\tilde{D}_{1}^{\dagger}\otimes\tilde{D}_{2}^{\dagger})
      =
      \projector{0}\otimes\id +
      \projector{1}\otimes V(\theta).
    \end{equation}

\subsection{Proof of Lemma \ref{lemma:indecomp_hopping_emb}}
\label{proof:indecomp_hopping_emb}
  Two remote control-phase nodes $\{(q,t_{i}), (q,t_{j})\}$ with $i\neq j-2$ form a $\mathbb{B}$-distributable packet $\mathcal{V}_{q,\{t_{i},t_{j}\}}$, if and only if the unitary $W_{t_{j};t_{i}}$ between them is $\mathbb{B}$-embeddable through a hopping embedding consisting of several embedding units.
  According to the embedding rules $\mathbb{B}$ in Theorem \ref{theorem:distr_and_embed} and Corollary \ref{coro:embed_rules_2q_block}, all embedding rules for embedding units are the global D-type, local D-type and global H-type and as shown in Fig. \ref{fig:embedding_rules} (a,b,c).
  The unitary $W_{t_{j};t_{i}}$ between $\{(q,t_{i}), (q,t_{j})\}$ is $\mathbb{B}$-embeddable, if and only if it can be decomposed as
  \begin{equation}
  \label{eq:uninsertable_pSet_proof_1}
    W_{t_{j};t_{i}}
    =
    F_{q,t_{j}}
    \left(\prod_{k:i<k<j}  \widetilde{U}_{t_{k}}\right)
    G_{q,t_{i}},
  \end{equation}
  where $\widetilde{U}_{t_{k}}$ are D-type or H-type embedding units.

  Suppose there exists a D-type embedding unit $U_{t_{l}}$, one can then decompose $W_{t_{j};t_{i}}$ into two embeddings
  \begin{equation}
    W_{t_{j};t_{i}}
    =
    W_{t_{j};t_{l}}
    C_{V}^{q,q'_{l}}(\theta_{l})
    W_{t_{l};t_{i}},
  \end{equation}
  where $W_{t_{j};t_{l}}$ and $W_{t_{l};t_{i}}$ are both $\mathbb{B}$-embeddable given by
  \begin{align}
    W_{t_{j};t_{l}}
    & =
    F_{q,t_{j}}
    \left(\prod_{k:i<k<l}  \widetilde{U}_{t_{k}}\right)
    G_{q,t_{l}},
    \\
    W_{t_{l};t_{i}}
    & =
    F_{q,t_{l}}
    \left(\prod_{k:l<k<i}  \widetilde{U}_{t_{k}}\right)
    G_{q,t_{i}}.
  \end{align}
  As a result, the distributable packet $\mathcal{V}_{q,\{t_{i},t_{j}\}}$ is indecomposable, only if the embedding units in Eq. \eqref{eq:uninsertable_pSet_proof_1} are all H-type.
  This completes the proof.

\bigskip

\subsection{The time complexity of Algorithm \ref{algo:dPacket_identify}}
\label{proof:id_dPacket_complexity}
Up to line \ref{algo_line:dPacket_id_neighbour}, the complexity is equal to $\mathcal{O}(m_{1}+m_{2})$, where $m_{1}$ and $m_{2}$ are the numbers of single-qubit and two-qubit gates, respectively.
For each qubit, the identification of neighboring distributable packets has the complexity $\mathcal{O}(d)$, where $d$ is the depth of a circuit.
For the identification of hopping distributable packets in Algorithm \ref{algo:hopping_emb}, the complexity is determined by the examination of the embeddability of $W_{t_{j};t_{i}}$ on line \ref{algo_line:hopping_emb_check}, which is upper bounded by $\mathcal{O}(d(d-1))$.
As a result, the complexity of Algorithm \ref{algo:dPacket_identify} is estimated by $\mathcal{O}(m_{1}+m_{2}+n\times d(d-1))$, where $n$ is the number of total qubits from all parties. Since $m_{1,2}\le nd$, the complexity is upper bounded by $\mathcal{O}( n d^{2})$.

\bigskip

\subsection{Proof of line \ref{algo_line:hopping_emb_phase_cond} in Algorithm \ref{algo:hopping_emb}}
\label{proof:phase_cond_for_hopping_emb}
We need to prove that a unitary $W_{t_{j};t_{i}}$ given in the following form is $q$-rooted embeddable according to the embedding rules $\mathbb{B}$, if and only if $\alpha_{k+1}+\gamma_{k} = n\pi$ for all $k$,
\begin{align}
  W_{t_{j};t_{i}}
  & =
  V(\gamma_{j-1}) \,H\, V(\alpha_{j-1}) \; C_{V}^{(q,q_{j-1})}
  \nonumber\\
  & \times V(\gamma_{j-2}) \,H\, V(\beta_{j-2}) \,H\, V(\alpha_{j-2}) \; C_{V}^{(q,q_{j-2})}
  \nonumber\\
  & \times \cdots
  \nonumber\\
  & \times V(\gamma_{i+1}) \,H\, V(\beta_{i+1}) \,H\, V(\alpha_{i+1})  \; C_{V}^{(q,q_{i+1})}
  \nonumber\\
  & \times V(\gamma_{i}) \,H\, V(\alpha_{i}).
\end{align}
Since the phase gates in unitary $W_{t_{j};t_{i}}$ commute with $C_{V}$ gates, one can merge them together as follows
\begin{align}
  W_{t_{j};t_{i}}
  & =
  V(\gamma_{j-1}) \,H\, C_{V}^{(q,q_{j-1})}
  \nonumber\\
  & \times V(\alpha_{j-1}+\gamma_{j-2}) \; H\, V(\beta_{j-2}) \,H\, C_{V}^{(q,q_{j-2})}
  \nonumber\\
  & \times \cdots
  \nonumber\\
  & \times V(\alpha_{i+1}+\gamma_{i+1}) \; H\, V(\beta_{i+1}) \,H\, C_{V}^{(q,q_{i+1})}
  \nonumber\\
  & \times V(\alpha_{i+1}+\gamma_{i})\; H\, V(\alpha_{i}).
\end{align}
According to the embedding rules $\mathbb{B}_{2}$ in Corollary \ref{coro:embed_rules_2q_block}, $W_{t_{j};t_{i}}$ is $q$-embeddable only if the gate $C_{V}$ must form a $H$-type embedding unit.
To satisfy this necessary condition, we insert $H\,H$ into $W_{t_{j};t_{i}}$ to construct $H$-type embedding units $H C_{V} H$,
\begin{align}
  W_{t_{j};t_{i}}
  & =
  V(\gamma_{j-1}) \,H\, C_{V}^{(q,q_{j-1})} H
  \nonumber\\
  & \times H V(\alpha_{j-1}+\gamma_{j-2}) \; H\, V(\beta_{j-2}) \,H\, C_{V}^{(q,q_{j-2})} H
  \nonumber\\
  & \times \cdots
  \nonumber\\
  & \times H V(\alpha_{i+1}+\gamma_{i+1}) \; H\, V(\beta_{i+1}) \,H\, C_{V}^{(q,q_{i+1})} H
  \nonumber\\
  & \times H V(\alpha_{i+1}+\gamma_{i})\; H\, V(\alpha_{i}).
\end{align}
This unitary is $q$-embeddable according to $\mathbb{B}_{1}$ in Theorem \ref{theorem:distr_and_embed} and $\mathbb{B}_{2}$ in Corollary \ref{coro:embed_rules_2q_block}, if and only if $H V(\alpha_{k+1}+\gamma_{k}) H V (\beta_{k})$ is embeddable for all $k$. It is equivalent to the following condition
\begin{equation}
  \alpha_{k+1} + \gamma_{k} = n\pi.
\end{equation}
This completes the proof.


\bigskip

\subsection{Proof of Theorem \ref{theorem:incompt_2qubit_circ}}\label{proof:incompt_2qubit_circ}
  In this section, we prove the compatibility of the primitive embedding rules in a circuit consisting of single-qubit and two-qubit gates.
  Fig. \ref{fig:nested_embedding} shows all the three possible cases of nested distributable packets.
  In Fig. \ref{fig:nested_embedding} (a,b), the two distributable packets are on the same qubit $q$.

  For (a), it shows the distributable packets $\mathcal{V}_{q,\{t_{0},t_{5}\}}$ and $\mathcal{V}_{q,\{t_{1},t_{6}\}}$ identified by two nested embeddings.
  We first implement the primitive embedding of $W_{t_{5};t_{0}}$.
  According to the embedding rules $\mathbb{B}$ in Eq. \eqref{eq:emb_rules_B}, the additional kernels are all gates acting on system B, which are irrelevant for the embedding of $W_{t_{6};t_{2}}^{(q)}$ on $q$.
  For the embedding of $W_{t_{6};t_{2}}^{(q)}$ on $q$, the only additional gate is the control-X gate $C_{q,X_{e}}$ in the ending process of embedding of $W_{t_{5};t_{0}}^{(q)}$, which acts before $(q,t_{5})$.
  One can shift the control-X gate $C_{q,X_{e}}$ into the block $\widetilde{U}_{t_{5}}$
  \begin{equation}
    \widetilde{U}_{t_{5}}
    =
    (\widetilde{D}_{q}\otimes G_{q_{n-1}})
    \,
    C_{q,X_{e}} C_{V}(\theta_{5})
    \,
    (D_{q}\otimes F_{q_{n-1}})
  \end{equation}
  For the indecomposable embedding of $W_{t_{6};t_{2}}^{(q)}$, one can find a decomposition of local gates on $q$ that forms a $H$-type embedding unit for $\widetilde{U}_{t_{5}}$.
  \begin{equation}
    \widetilde{U}_{t_{5}}
    =
    (\widetilde{D}_{q}H_{q} \otimes G_{q_{n-1}})
    \,
    C_{q,X_{e}} C_{V}(\theta_{5})
    \,
    (H_{q} D_{q}\otimes F_{q_{n-1}}).
  \end{equation}
  The unitary $\widetilde{U}_{t_{5}}$ can be decomposed into two $H$-type embedding units by inserting $H_{q}^{2}\otimes\id$ between the two control gates $C_{q,X_{e}}$ and $C_{V}(\theta_{5})$,
   as follow
  \begin{align}
    \widetilde{U}_{t_{5}}
    =
    (\widetilde{D}_{q}H_{q} \otimes G_{q_{n-1}})
    \,
    C_{q,X_{e}}
    (H_{q}\otimes\id_{B})
    \nonumber\\
    \times (H_{q}\otimes\id_{B})
    C_{V}(\theta_{5})
    \,
    (H_{q} D_{q}\otimes F_{q_{n-1}}).
  \end{align}
  It means that one can implement the embedding of $W_{t_{6};t_{2}}$ with an additional local gate $(H_{e}\otimes H_{e'}) C_{Z}^{(e,e')} (H_{e}\otimes H_{e'})$ acting on the two auxiliary qubit $\{e,e'\}$.
  Removing duplicated Hadamard gates between two control-Z gate on the auxiliary qubits $\{e,e'\}$, the implementation of embedding is shown in Fig. \ref{fig:nested_embedding} (b), where the additional local gate $C_{Z}^{(e,e')}$ is highlighted in green.
  This proves the compatibility of two nested embedding shown in Fig. \ref{fig:nested_embedding} (a).

  For the case shown in Fig. \ref{fig:nested_embedding} (c), one of the embedding $W_{t_{5};t_{2}}^{(q)}$ is included in the embedding of $W_{t_{6};t_{0}}^{(q)}$. One can first implement the embedding $W_{t_{6};t_{0}}^{(q)}$, of which the local kernels of embedding all acts on $B$ and are all irrelevant with the embedding of $W_{t_{5};t_{2}}$. It means that the embedding of $W_{t_{5};t_{2}}^{(q)}$ is not affected by the embedding of $W_{t_{6};t_{0}}^{(q)}$. These two embeddings are therefore compatible. The implementation of the two embeddings are shown in Fig. \ref{fig:nested_embedding} (d).

  For the case shown in Fig. \ref{fig:nested_embedding} (e), the two nested embeddings are rooted on two qubits $\{q,q'\}$, which are on two different local systems. The embeddings of $W_{t_{5};t_{0}}^{(q)}$ can be decomposed into two embedding units $\widetilde{U}_{t_{2}}$ and $\widetilde{U}_{t_{3}}$, while $W_{t_{6};t_{1}}^{(q')}$ can be decomposed into $\widetilde{U}_{t_{3}}$ and $\widetilde{U}_{t_{4}}$. The two hopping embeddings share the same embedding unit $\widetilde{U}_{t_{3}}$. If one wants to implement the two hopping embedding together, one has to implement the joint embedding of $\widetilde{U}_{t_{3}}$ on $\{q,q'\}$. The primitive recursive embedding of $\widetilde{U}_{t_{3}}$ will lead to a global control-Z gate $
    C_{Z}^{(e,e')}$ acting on the two auxiliary $\{e,e'\}$,
  \begin{align}
    & C_{q, X_{e}}C_{q', X_{e'}}
    \left(H_{q}\otimes H_{q'}\right)
    C_{V}(\theta)
    \left(H_{q}\otimes H_{q'}\right)
    C_{q', X_{e'}}C_{q, X_{e}}
    \nonumber \\
    = &
    \left(H_{e}\otimes H_{e'}\right)
    C_{Z}^{(q,e')}C_{Z}^{(q',e)}
    C_{V}(\theta)
    C_{Z}^{(e,e')}
    \left(H_{e}\otimes H_{e'}\right).
  \end{align}
  The incompatibility of the nested embeddings is caused by the incompatibility of the recursive embedding of the gate embedding unit $\widetilde{U}_{t_{3}}$.

  The distributable packets $\mathcal{V}_{q,\{t_{i}, t_{j}\}}$ and $\mathcal{V}_{q',\{t_{k}, t_{l}\}}$ are incompatible, if and only if $\{q,q'\}$ belong to two local systems, and there is an embedding unit $\left(H_{q}\otimes H_{q'}\right) C_{V}^{(q,q')}(\theta) \left(H_{q}\otimes H_{q'}\right)$ located on $t$ with $t_{i}<t<t_{j}$ and $t_{k}<t<t_{l}$. Existence of such an embedding unit for two distributable packets is equivalent to the existence of a control-phase gate acting on $\{q,q'\}$ at a depth of $t$ with $t_{i}<t<t_{j}$ and $t_{k}<t<t_{l}$.
  This completes the proof.


\myprintglossary

\myprintbibliography

\end{document}